\documentstyle[mnras_cite,psfig,ifthen,color]{mn}
\definecolor{lgrey}{gray}{0.9}

\def\gsim{~\rlap{$>$}{\lower 1.0ex\hbox{$\sim$}}}
\def\lsim{~\rlap{$<$}{\lower 1.0ex\hbox{$\sim$}}}
\def\G{{\rm G}}

\def\d{{\rm d}}

\newboolean{extras}
\setboolean{extras}{false}

\begin{document}

\title[Heating of Galactic Disks by Infalling Satellites]{Heating of Galactic Disks by Infalling Satellites}
\author[A.~J.~Benson, C.~G.~Lacey, C.~S.~Frenk, C.~M.~Baugh, \&
  S.~Cole ]{A.~J.~Benson$^{1,3}$, C.~G.~Lacey$^2$, C.~S.~Frenk$^2$,
  C.~M.~Baugh$^2$ \& S.~Cole$^2$\\
1. California Institute of Technology, MC 105-24, Pasadena, CA 91125,
U.S.A. \\
2. Physics Department, University of Durham, Durham, DH1 3LE, England \\
3. Current address: University of Oxford, Keble Road, Oxford, OX1 3RH (e-mail: abenson@astro.ox.ac.uk)
}

\maketitle

\begin{abstract}
We develop an analytic model to calculate the rate at which galaxy
disks are heated by dark matter substructures orbiting in their
halos. The model takes into account the internal structure, mass
function and accretion rate of satellites expected in the $\Lambda$CDM
cosmology, as well as the growth of the disk by accretion and mergers,
but it ignores resonant heating of the disk and the dynamical effects
of spiral arms and bars. We calibrate this model against N-body
simulations and demonstrate that it is able to reproduce the N-body
heating rates to within a factor of 3 in the majority of cases. Our
model gives the distribution of disk scale-heights for galaxies of
different luminosities.  For $L_*$ spiral galaxies, it predicts a
median disk thickness of only 5\% of the radial scale-length if
substructure is the only source of heating. The median disk thickness
increases to nearly 20\% of the radial scale-length when heating due
to gravitational scattering of stars by molecular clouds is also
included. The latter value is close to the thickness estimated
observationally for the disk of the Milky Way galaxy. The distribution
of disk thickness predicted by the model is also consistent with a
recent observational determination for sub-$L_*$ galaxies by
\scite{bizyaev02}. Thus, the observed thickness of the stellar disks
of spiral galaxies seems to be entirely compatible with the abundance
of substructure in dark matter halos predicted by the standard
$\Lambda$-dominated cold dark matter model of structure formation. In
an $\Omega_0=1$ universe, our best model of galaxy formation produces
similar scale-heights, a consequence of the fact that similar amounts
of substructure are accreted by halos during the lifetime of the disk
in $\Omega_0=1$ and $\Omega_0=0.3$, $\Lambda_0=0.7$ cold dark matter
cosmologies.
\end{abstract}

\section{Introduction}

A generic prediction of hierarchical models of structure formation,
such as the cold dark matter (CDM) model, is that the dark matter
halos of galaxies and clusters should contain large amounts of
substructure, in the form of small, gravitationally bound subhalos
orbiting within the larger potential. This substructure arises because
large halos are built up by mergers of smaller halos whose
tidally-stripped remnants can survive in favourable
conditions. Recently, it has been claimed that the CDM model predicts
an order of magnitude too many subhalos around the Milky Way galaxy,
compared to what is inferred from the number of satellite galaxies
\cite{klypin99,moore99}. Several authors have now pointed out that
this apparent discrepancy is readily explained if some process (such
as the heating of the intergalactic medium (IGM) during reionization)
is efficient at suppressing the formation of galaxies in most of these
subhalos \cite{bullock00,benson02b,somerville02}. In this picture,
galaxy halos should be filled with many small subhalos containing
negligible amounts of luminous material. A strong test of this idea is
possible by searching for gravitational signatures of subhalos, thus
bypassing the problem of relating subhalos to the visible material in
satellite galaxies.

The most direct probe of substructure in dark matter halos is
gravitational microlensing. Its properties are reasonably well
understood theoretically
\cite{mao98,metmad01,chiba01,dal01,dal02}. Although the interpretation of
the current datasets remains controversial in some cases, the observed
microlensing rates appear to be consistent with the abundance of
substructure predicted by CDM.

An alternative constraint on the amount of substructure in halos may
be obtained by considering the thickness of the stellar disks of
galaxies. Subhalos on orbits that pass through or near to a galactic
disk perturb it gravitationally and deposit energy into it, gradually
heating the disk and increasing its scale-height. Since there are
other mechanisms that also heat stellar disks (but with uncertain
efficiency), the observed thickness of galactic disks sets an upper
limit on the abundance of such substructure. The heating of galactic
disks by infalling satellites was invoked as a constraint on models of
structure formation by \ncite{TO}~(1992; hereafter TO). They
calculated this effect semi-analytically, and concluded that the
thinness of the Milky Way's disk is inconsistent with the hierarchical
build-up of galaxies in a high density ($\Omega_0=1$) CDM
universe. This conclusion was disputed by \scite{NFW94} whose
cosmological simulations showed that many of the satellites that are
incorporated into a galactic dark halo do not actually merge with the
central galaxy. Subsequent numerical simulations of mergers of
single satellites with larger disk galaxies (e.g. \pcite{huang97,VW})
indicated that TO's analytical estimates of the heating rate were
somewhat too high, weakening their constraint on structure formation
models. More recently, \scite{font01} have numerically simulated the
heating of disks by the ensemble of subhalos predicted to exist within
dark halos in the CDM model. Their simulations of Milky Way-like
galaxies only set an upper limit to the rate of disk heating by
satellites, because of numerical effects, but they conclude that this
is less than the total disk heating rate that is inferred
observationally for the Solar neighbourhood. They argue that the
heating rates are low because the most massive satellites, which are
the ones that cause the most heating, are few in number and because
few satellites penetrate the inner regions of the disk. Although their
conclusions agree with those of \scite{NFW94}, they are limited by the
fact that they simulated only two realizations of the halo
substructure.

In this paper, we develop a new semi-analytical model of disk heating
by halo substructure. Our calculation builds upon earlier
semi-analytical modeling of galaxy formation within the framework of
CDM cosmology \cite{cole00}, and on recently developed analytical
models of the evolution of satellites within dark matter halos
\cite{taybab,benson02a,taff03,taybab03}. The rate at which satellite
halos of different masses are incorporated into the main halo is given
by the galaxy formation model. The satellite model then predicts how
the masses, radii and orbits of subhalos evolve due to dynamical
friction and tidal stripping by the halo, disk and bulge of the host
galaxy. In this paper, we add a calculation of how much of the orbital
energy of the satellites that is lost by dynamical friction goes into
increasing the thickness and vertical motions of the galactic
disk. The interaction between the satellite and the disk is modeled in a
simplified way, ignoring details such as resonant interactions and the
possible role of spiral features and bars. We test and calibrate our
analytical model of satellite evolution against a new set of
high-resolution N-body simulations of single satellites merging with
disk galaxies. We find (as has also been shown by
\pcite{taybab} and \pcite{taff03}) that such an analytical model is
able to reproduce well the behaviour seen in the N-body
simulations. We measure the disk heating in the same simulations, and
find that it is quite well reproduced by our analytical model. We then
apply this model of heating by satellites within the framework of our
semi-analytical model of galaxy formation, in order to predict the
distribution of scale-heights for disk galaxies of different
luminosities.

Both the N-body and semi-analytical approaches have advantages and
disadvantages when applied to this problem. N-body simulations fully
account for the non-linear interaction of substructure and disk
(e.g. for the excitation of global modes such as warps and bars in the
disk). However, they are limited by resolution and artificial
numerical heating and, because of computational cost, they are limited
to few (two, in the case of the best cosmological simulations of disk
heating to date, by \ncite{font01}). The semi-analytical approach has
the advantage that it is not limited by resolution or artificial
heating, and it allows the calculation of a large number of
realizations. Since heating by substructure is a highly stochastic
process, it is important to account for the galaxy-to-galaxy variation
in the heating rate by calculating a large number of realizations. At
present, this is only possible with the semi-analytical approach. 

The remainder of this paper is arranged as follows. In
\S\ref{sec:model} we describe our analytical model for disk heating by
subhalos and for the evolution of the disk scale-height. In
\S\ref{sec:Nbody} we calibrate and test our analytical model against
numerical simulations of single satellite-disk mergers. In
\S\ref{sec:results} we present our predictions for the distribution of
scale-heights of disk galaxies in the CDM model, and compare with
observational data for the Milky Way and for other galaxies. Finally,
in \S\ref{sec:disc} we present our conclusions. Appendices detail
derivations of various formulae related to dynamical friction and disk
energies and present convergence tests for the N-body simulations.

\section{Model}
\label{sec:model}

\subsection{Evolution of Satellites and their Orbits}

We calculate the evolution of the masses, radii and orbits of
satellites using a development of the model presented in
\ncite{benson02a}~(2002a; hereafter Paper~I). That work, in turn, was
based on the satellite evolution model of \scite{taybab}. We summarize
here the main features of our model. The growth of the main halo is
described by a merger history tree which is calculated by a Monte
Carlo method \cite{cole00}. When smaller halos (in general containing
one or more visible galaxies) merge with the main halo, they become
satellite halos. The satellite halos are given initial orbits which
start close to the virial radius but have a range of eccentricities
consistent with the distribution seen in the N-body simulations of
\scite{ghigna98}. The satellite orbits are followed in the potential
of the host system; they evolve due to dynamical friction against the
dark halo, disk and bulge of the main galaxy. At the same time, the
satellites lose mass by tidal stripping, both ``static'' tidal
limitation and tidal shocking. As a satellite is tidally stripped, its
radius and internal structure also change.

We have made a few improvements to our satellite orbit model
from that presented in Paper~I. These are described in
Appendix~\ref{sec:improve}. 

\subsection{Disk Heating}

\subsubsection{Rate of Heating}

We now wish to calculate the rate at which a satellite halo heats the
disk of the galaxy in its host halo. The satellite experiences
dynamical friction against the disk, and the energy lost from the
orbital motion of the satellite by this mechanism goes into increasing
the energy of the disk. Working in the frame in which the centre of
mass of the central galaxy and its halo are at rest, the satellite
injects energy into the disk at a rate
\begin{equation}
P = -{\bf F}_{\rm df,disk} \cdot {\bf v}_{\rm sat},
\end{equation}
where ${\bf F}_{\rm df,disk}$ is the dynamical friction force exerted
by the disk and ${\bf v}_{\rm sat}$ is the velocity of
satellite. (Note that while we typically expect the satellite to lose
energy to the disk, it is possible for the satellite to gain energy
from the ordered motions of the disk if ${\bf F}_{\rm df,disk} \cdot
{\bf v}_{\rm sat}>0$. This occurs because the dynamical friction force
depends, in our approximation, on the relative velocity vector of the
satellite and the local disk stars. If the local disk velocity is
sufficiently large, the relative velocity vector may point in the
opposite direction to the satellite velocity vector, resulting in a
transfer of energy from the disk to the satellite.) This energy is
initially injected in the form of kinetic energy, but it is
subsequently mixed between kinetic and potential energies by the
motions of the stars. We are interested in the increase in the
vertical energy of the disk, which is given by
\begin{equation}
\dot{E}_{\rm z} = - \epsilon_{\rm z} {\bf F}_{\rm df,disk} \cdot {\bf v}_{\rm sat},
\label{eq:heatz}
\end{equation}
We derive an expression for the efficiency factor $\epsilon_{\rm z}
\leq 1$ in Appendix~\ref{app:zeffic} by considering the increases in
the vertical and horizontal components of the velocity dispersion of
the stars responsible for dynamical friction during scattering
events. This expression (\ref{eqn:epsz}) depends only on the Coulomb
logarithm, $\ln \Lambda$, and the angle $\theta_{{\rm V}_0}$ between
the disk-satellite relative velocity and the $z$-axis. We then simply
integrate $\dot{E}_{\rm z}$ along the satellite orbit to determine the
net increase in the disk's vertical energy.

\begin{figure}
\psfig{file=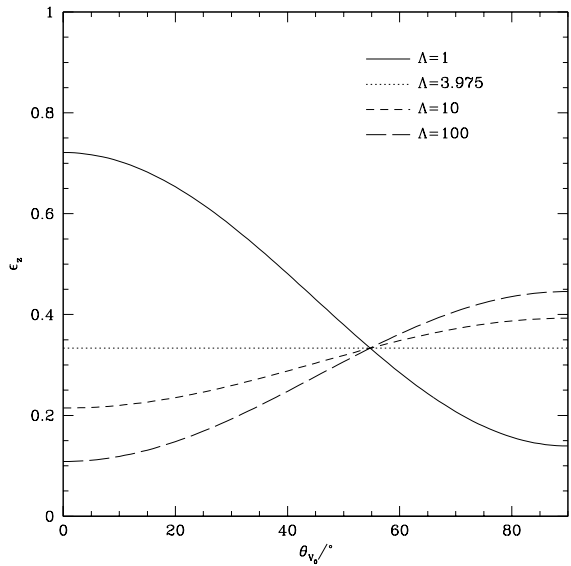,width=80mm}
\caption{The efficiency of energy transfer to vertical motions in the
disk as a function of the angle between the disk-satellite relative
velocity and the $z$-axis, $\theta_{{\rm V}_0}$. Results are plotted
for several values of $\Lambda$ as indicated in the figure.}
\label{fig:epsz}
\end{figure}

Figure~\ref{fig:epsz} shows how $\epsilon_{\rm z}$ depends on the
angle $\theta_{\rm V_0}$ for a few representative values of
$\Lambda$. Note that $\epsilon_{\rm z}=\frac{1}{3}$ when $\cos
\theta_{{\rm V}_0}=\frac{1}{\sqrt{3}}$, independently of $\Lambda$. For
small $\Lambda$ the efficiency is greatest when $\theta_{{\rm
V}_0}=0^\circ$ (approaching unity as $\Lambda$ approaches zero) and
smallest for $\theta_{{\rm V}_0}=90^\circ$ (approaching zero as
$\Lambda$ approaches zero). For large $\Lambda$ the trend is reversed,
with $\epsilon_{\rm z}$ being smallest at $\theta_{{\rm V}_0}=0^\circ$
(approaching zero as $\Lambda$ approaches infinity) and largest for
$\theta_{{\rm V}_0}=90^\circ$ (approaching $\frac{1}{2}$ as $\Lambda$
approaches infinity). The transition between these two regimes occurs
for $\Lambda \approx 3.975$, for which $\epsilon_{\rm z}$ is
independent of $\theta_{{\rm V}_0}$.

We can understand the behaviour of $\epsilon_{\rm z}$ in simple
terms. For example, for $\theta_{{\rm V}_0}=0$, the efficiency drops
to zero as $\Lambda$ becomes large. In this case, vertical motions in
the disk are parallel to the relative velocity vector of the satellite
and the disk stars. Consequently, only the $\Delta V_{{\rm m}||}$ term
(see eqn.~\ref{eq:deltavmperp}) contributes to increasing the energy
in these vertical motions. As $\Lambda$, and hence the maximum impact
parameter, $b$, increases, energy transfer from the satellite becomes
dominated by large $b$ scatterings. For large impact parameters, the
increase in velocities (and hence energies) perpendicular to the
satellite motion dominates over that parallel to the motion, since
$\Delta V_{{\rm m}||} \propto b^{-2}$ while $\Delta V_{{\rm m}\perp}
\propto b^{-1}$ (see eqn.~\ref{eq:deltavmpara}). Consequently, the
efficiency of transfer to vertical motions in the disk drops to zero
as $\Lambda$ becomes large.

The reversal of the trend of $\epsilon_{\rm z}$ with $\theta_{{\rm
V}_0}$ at $\Lambda\approx 3.975$ is also simple to understand. For
larger $\Lambda$, energy transfer is predominantly into motions
perpendicular to the motion of the satellite (as discussed
above). Thus, the efficiency of energy transfer to motions in the vertical
direction is greatest when the satellite moves perpendicular to that
direction ($\theta_{{\rm V}_0}=90^\circ$). For smaller $\Lambda$,
energy transfer occurs mostly into the parallel direction, and so
$\epsilon_{\rm z}$ is maximized for $\theta_{{\rm V}_0}=0^\circ$. For
$\Lambda\approx 3.975$ energy transfers into perpendicular and
parallel directions are equal and so $\epsilon_{\rm z}$ is constant.

When $\cos \theta_{{\rm V}_0}=\frac{1}{\sqrt{3}}$, the energy
transferred to vertical motions is always one third of the increase in
energy parallel to the satellite velocity, plus one third of the
increase in the energies in the two directions perpendicular to the
satellite velocity. Thus, this energy is always exactly one third
of the total energy transferred from the satellite and hence
$\epsilon_{\rm z}=\frac{1}{3}$ independently of $\Lambda$.

\begin{table*}
\caption{Properties of the galaxy and satellite models used in the N-body
simulations. The first column specifies the component in question. The
second column gives the density profile, either in spherical
coordinates ($r$), or cylindrical polar coordinates ($R,z$). The
values of the parameters of each profile are listed in column 3. The
satellites are all described by King models \protect\cite{king66}. For
these, we specify the core radius, $r_{\rm c}$, and the concentration
$c=\log _{10}r_{\rm t}/r_{\rm c}$ where $r_{\rm t}$ is the tidal
radius of the satellite. The final column lists the number of
particles used to represent each component in the standard case.}
\label{tb:simpars}
\begin{tabular}{lllc}
\hline
Component & Density Profile & Parameters & Number of Particles \\
\hline
Halo & $\rho_{\rm h}(r) = {M_{\rm h} \alpha \over 2 \pi^{3/2}r_{\rm cut}} {\exp(-r^2/r_{\rm cut}^2) \over r^2 + \gamma^2}$ & $M_{\rm h}=7.84\times 10^{11}M_\odot$ & 687008 \\
 & & $\gamma = 3.5$kpc & \\
 & & $r_{\rm cut} = 84$kpc & \\
 & & $\alpha = 1.076$ & \\
Disk & $\rho_{\rm d}(R,z)={M_{\rm d} \over 4 \pi R_{\rm d}^2 H_{\rm d}}\exp(-R/R_{\rm d}) {\rm sech} ^2(z/H_{\rm d})$ & $M_{\rm d}=5.6\times 10^{10}M_\odot$ & 163840 \\
 & & $R_{\rm d}=3.5$kpc & \\
 & & $H_{\rm d}=700$pc & \\
Bulge & $\rho_{\rm b}(r) = {M_{\rm b} \over 2\pi} {a \over r(a+r)^3}$ & $M_{\rm b}=1.87\times 10^{10}M_\odot$ & 16384 \\
 & & $a=525$pc & \\
Satellite S1 & King Model & $M_{\rm s} = 5.60\times 10^9M_\odot$ & 32768 \\
 & & $r_{\rm c}=1$kpc & \\
 & & $c=0.8$ & \\
Satellite S2 & King Model & $M_{\rm s} = 5.60\times 10^9M_\odot$ & 32768 \\
 & & $r_{\rm c}=500$pc & \\
 & & $c=1.1$ & \\
Satellite S3 & King Model & $M_{\rm s} = 1.12\times 10^{10}M_\odot$ & 32768 \\
 & & $r_{\rm c}=875$pc & \\
 & & $c=1.0$ & \\
\hline
\end{tabular}
\end{table*}

\begin{table}
\begin{center}
\caption{Properties and initial orbital parameters of the satellites
in the N-body simulations. Column~2 specifies the satellite model used
(as defined in Table~\protect\ref{tb:simpars}). Column~3 lists $\theta_{\rm
i}$, the angle between the initial angular momentum vector of the
satellite and that of the disk. Column~4 lists the circularity of the
satellite's initial orbit, $\epsilon_{\rm J}$, while column~5 lists
the initial radial position of the satellite (which is the apocentre
of its orbit), $r_{\rm a}$. Column~6 specifies whether the simulation
contains a disk or not (note that $\theta_{\rm i}$ is undefined for
diskless simulations G2Sxx).}
\label{tb:simorbs}
\begin{tabular}{lcrcccc}
\hline
Model & Satellite & $\theta_{\rm i}$ & $\epsilon_{\rm J}$ & $r_{\rm a}$/kpc & Disk? \\
\hline
G1S1 & S1 & $45^\circ$ & 0.33 & 59.0 & $\surd$ \\
G1S2 & S1 & $0^\circ$ & 0.55 & 55.0 & $\surd$ \\
G1S3 & S1 & $45^\circ$ & 0.55 & 55.0 & $\surd$ \\
G1S4 & S1 & $90^\circ$ & 0.55 & 55.0 & $\surd$ \\
G1S5 & S1 & $135^\circ$ & 0.55 & 55.0 & $\surd$ \\
G1S6 & S1 & $180^\circ$ & 0.55 & 55.0 & $\surd$ \\
G1S7 & S1 & $0^\circ$ & 0.82 & 46.5 & $\surd$ \\
G1S8 & S1 & $45^\circ$ & 0.82 & 46.5 & $\surd$ \\
G1S9 & S2 & $0^\circ$ & 0.55 & 55.0 & $\surd$ \\
G1S10 & S2 & $45^\circ$ & 0.55 & 55.0 & $\surd$ \\
G1S11 & S2 & $90^\circ$ & 0.55 & 55.0 & $\surd$ \\
G1S12 & S2 & $135^\circ$ & 0.55 & 55.0 & $\surd$ \\
G1S13 & S2 & $180^\circ$ & 0.55 & 55.0 & $\surd$ \\
G1S14 & S3 & $45^\circ$ & 0.55 & 55.0 & $\surd$ \\
G1S15 & S3 & $135^\circ$ & 0.55 & 55.0 & $\surd$ \\
G2S1 & S1 & N/A & 0.33 & 59.0 & $\times$ \\
G2S2 & S1 & N/A & 0.55 & 55.0 & $\times$ \\
G2S7 & S1 & N/A & 0.82 & 46.5 & $\times$ \\
G2S9 & S2 & N/A & 0.55 & 55.0 & $\times$ \\
G2S14 & S3 & N/A & 0.55 & 55.0 & $\times$ \\
\hline
\end{tabular}
\end{center}
\end{table}

\begin{table}
\caption{Comparison of results from the analytic and N-body
calculations of satellite evolution. Column~1 lists the model number;
column~2 lists the type of calculation (analytic or N-body); column~3
gives the change in the disk vertical kinetic energy generated by the
satellite at the end of the simulation, both in absolute units and as
a percentage of the initial disk vertical kinetic energy (values in
parentheses). Where the analytic and N-body estimates of the disk
energy differ by more than a factor of two, we show the values in bold
type. Columns~4 and 5 list the times at which the satellite reaches
50\% and 10\% of its initial mass respectively.}
\label{tb:comparison}
\begin{center}
\rule{74mm}{0.1mm}\newline
\makebox[1cm][c]{Model}\mbox{ }\makebox[1.25cm][c]{Type}\mbox{ }\makebox[2cm][c]{$\frac{\Delta T_{\rm z}(4\mbox{\scriptsize Gyr})}{10^{14}M_\odot \mbox{\scriptsize km}^2 \mbox{\scriptsize s}^{-2}}$}\mbox{ }\makebox[1.25cm][c]{$t_{50}$/Gyr}\mbox{ }\makebox[1.25cm][c]{$t_{10}$/Gyr} \newline
\rule{74mm}{0.1mm}\newline

\vspace{-0.75mm}\colorbox{white}{\makebox[1cm][l]{G1S1}\mbox{ }\makebox[1.25cm][l]{Analytic}\mbox{ }\makebox[1cm][c]{\rm 0.133}\mbox{ }\makebox[1cm][c]{(25\%)}\mbox{ }\makebox[1.25cm][c]{1.24}\mbox{ }\makebox[1.25cm][c]{2.01}} \newline
\vspace{-0.75mm}\colorbox{white}{\makebox[1cm][l]{G1S1}\mbox{ }\makebox[1.25cm][l]{N-body}\mbox{ }\makebox[1cm][c]{\rm 0.101}\mbox{ }\makebox[1cm][c]{(18\%)}\mbox{ }\makebox[1.25cm][c]{1.38}\mbox{ }\makebox[1.25cm][c]{2.16}} \newline
\vspace{-0.75mm}\colorbox{lgrey}{\makebox[1cm][l]{G1S2}\mbox{ }\makebox[1.25cm][l]{Analytic}\mbox{ }\makebox[1cm][c]{\rm 0.257}\mbox{ }\makebox[1cm][c]{(49\%)}\mbox{ }\makebox[1.25cm][c]{1.51}\mbox{ }\makebox[1.25cm][c]{2.89}} \newline
\vspace{-0.75mm}\colorbox{lgrey}{\makebox[1cm][l]{G1S2}\mbox{ }\makebox[1.25cm][l]{N-body}\mbox{ }\makebox[1cm][c]{\rm 0.342}\mbox{ }\makebox[1cm][c]{(60\%)}\mbox{ }\makebox[1.25cm][c]{1.69}\mbox{ }\makebox[1.25cm][c]{2.08}} \newline
\vspace{-0.75mm}\colorbox{white}{\makebox[1cm][l]{G1S3}\mbox{ }\makebox[1.25cm][l]{Analytic}\mbox{ }\makebox[1cm][c]{\rm 0.184}\mbox{ }\makebox[1cm][c]{(35\%)}\mbox{ }\makebox[1.25cm][c]{2.20}\mbox{ }\makebox[1.25cm][c]{3.19}} \newline
\vspace{-0.75mm}\colorbox{white}{\makebox[1cm][l]{G1S3}\mbox{ }\makebox[1.25cm][l]{N-body}\mbox{ }\makebox[1cm][c]{\rm 0.238}\mbox{ }\makebox[1cm][c]{(42\%)}\mbox{ }\makebox[1.25cm][c]{2.25}\mbox{ }\makebox[1.25cm][c]{3.13}} \newline
\vspace{-0.75mm}\colorbox{lgrey}{\makebox[1cm][l]{G1S4}\mbox{ }\makebox[1.25cm][l]{Analytic}\mbox{ }\makebox[1cm][c]{\bf 0.029}\mbox{ }\makebox[1cm][c]{(5\%)}\mbox{ }\makebox[1.25cm][c]{2.32}\mbox{ }\makebox[1.25cm][c]{3.73}} \newline
\vspace{-0.75mm}\colorbox{lgrey}{\makebox[1cm][l]{G1S4}\mbox{ }\makebox[1.25cm][l]{N-body}\mbox{ }\makebox[1cm][c]{\bf 0.101}\mbox{ }\makebox[1cm][c]{(18\%)}\mbox{ }\makebox[1.25cm][c]{2.54}\mbox{ }\makebox[1.25cm][c]{3.53}} \newline
\vspace{-0.75mm}\colorbox{white}{\makebox[1cm][l]{G1S5}\mbox{ }\makebox[1.25cm][l]{Analytic}\mbox{ }\makebox[1cm][c]{\rm 0.032}\mbox{ }\makebox[1cm][c]{(6\%)}\mbox{ }\makebox[1.25cm][c]{2.24}\mbox{ }\makebox[1.25cm][c]{3.77}} \newline
\vspace{-0.75mm}\colorbox{white}{\makebox[1cm][l]{G1S5}\mbox{ }\makebox[1.25cm][l]{N-body}\mbox{ }\makebox[1cm][c]{\rm 0.057}\mbox{ }\makebox[1cm][c]{(10\%)}\mbox{ }\makebox[1.25cm][c]{2.46}\mbox{ }\makebox[1.25cm][c]{3.55}} \newline
\vspace{-0.75mm}\colorbox{lgrey}{\makebox[1cm][l]{G1S6}\mbox{ }\makebox[1.25cm][l]{Analytic}\mbox{ }\makebox[1cm][c]{\rm 0.177}\mbox{ }\makebox[1cm][c]{(33\%)}\mbox{ }\makebox[1.25cm][c]{1.94}\mbox{ }\makebox[1.25cm][c]{3.00}} \newline
\vspace{-0.75mm}\colorbox{lgrey}{\makebox[1cm][l]{G1S6}\mbox{ }\makebox[1.25cm][l]{N-body}\mbox{ }\makebox[1cm][c]{\rm 0.090}\mbox{ }\makebox[1cm][c]{(16\%)}\mbox{ }\makebox[1.25cm][c]{2.21}\mbox{ }\makebox[1.25cm][c]{3.14}} \newline
\vspace{-0.75mm}\colorbox{white}{\makebox[1cm][l]{G1S7}\mbox{ }\makebox[1.25cm][l]{Analytic}\mbox{ }\makebox[1cm][c]{\bf 0.056}\mbox{ }\makebox[1cm][c]{(11\%)}\mbox{ }\makebox[1.25cm][c]{2.70}\mbox{ }\makebox[1.25cm][c]{$>4.00$}} \newline
\vspace{-0.75mm}\colorbox{white}{\makebox[1cm][l]{G1S7}\mbox{ }\makebox[1.25cm][l]{N-body}\mbox{ }\makebox[1cm][c]{\bf 0.443}\mbox{ }\makebox[1cm][c]{(77\%)}\mbox{ }\makebox[1.25cm][c]{2.22}\mbox{ }\makebox[1.25cm][c]{2.28}} \newline
\vspace{-0.75mm}\colorbox{lgrey}{\makebox[1cm][l]{G1S8}\mbox{ }\makebox[1.25cm][l]{Analytic}\mbox{ }\makebox[1cm][c]{\bf 0.096}\mbox{ }\makebox[1cm][c]{(18\%)}\mbox{ }\makebox[1.25cm][c]{3.23}\mbox{ }\makebox[1.25cm][c]{$>4.00$}} \newline
\vspace{-0.75mm}\colorbox{lgrey}{\makebox[1cm][l]{G1S8}\mbox{ }\makebox[1.25cm][l]{N-body}\mbox{ }\makebox[1cm][c]{\bf 0.324}\mbox{ }\makebox[1cm][c]{(57\%)}\mbox{ }\makebox[1.25cm][c]{3.72}\mbox{ }\makebox[1.25cm][c]{4.03}} \newline
\vspace{-0.75mm}\colorbox{white}{\makebox[1cm][l]{G1S9}\mbox{ }\makebox[1.25cm][l]{Analytic}\mbox{ }\makebox[1cm][c]{\rm 0.272}\mbox{ }\makebox[1cm][c]{(51\%)}\mbox{ }\makebox[1.25cm][c]{1.92}\mbox{ }\makebox[1.25cm][c]{3.30}} \newline
\vspace{-0.75mm}\colorbox{white}{\makebox[1cm][l]{G1S9}\mbox{ }\makebox[1.25cm][l]{N-body}\mbox{ }\makebox[1cm][c]{\rm 0.307}\mbox{ }\makebox[1cm][c]{(54\%)}\mbox{ }\makebox[1.25cm][c]{1.82}\mbox{ }\makebox[1.25cm][c]{1.88}} \newline
\vspace{-0.75mm}\colorbox{lgrey}{\makebox[1cm][l]{G1S10}\mbox{ }\makebox[1.25cm][l]{Analytic}\mbox{ }\makebox[1cm][c]{\bf 0.244}\mbox{ }\makebox[1cm][c]{(46\%)}\mbox{ }\makebox[1.25cm][c]{2.53}\mbox{ }\makebox[1.25cm][c]{3.28}} \newline
\vspace{-0.75mm}\colorbox{lgrey}{\makebox[1cm][l]{G1S10}\mbox{ }\makebox[1.25cm][l]{N-body}\mbox{ }\makebox[1cm][c]{\bf 0.588}\mbox{ }\makebox[1cm][c]{(103\%)}\mbox{ }\makebox[1.25cm][c]{3.01}\mbox{ }\makebox[1.25cm][c]{3.35}} \newline
\vspace{-0.75mm}\colorbox{white}{\makebox[1cm][l]{G1S11}\mbox{ }\makebox[1.25cm][l]{Analytic}\mbox{ }\makebox[1cm][c]{\bf 0.114}\mbox{ }\makebox[1cm][c]{(22\%)}\mbox{ }\makebox[1.25cm][c]{2.86}\mbox{ }\makebox[1.25cm][c]{3.59}} \newline
\vspace{-0.75mm}\colorbox{white}{\makebox[1cm][l]{G1S11}\mbox{ }\makebox[1.25cm][l]{N-body}\mbox{ }\makebox[1cm][c]{\bf 0.363}\mbox{ }\makebox[1cm][c]{(63\%)}\mbox{ }\makebox[1.25cm][c]{3.16}\mbox{ }\makebox[1.25cm][c]{3.80}} \newline
\vspace{-0.75mm}\colorbox{lgrey}{\makebox[1cm][l]{G1S12}\mbox{ }\makebox[1.25cm][l]{Analytic}\mbox{ }\makebox[1cm][c]{\rm 0.131}\mbox{ }\makebox[1cm][c]{(25\%)}\mbox{ }\makebox[1.25cm][c]{2.89}\mbox{ }\makebox[1.25cm][c]{3.58}} \newline
\vspace{-0.75mm}\colorbox{lgrey}{\makebox[1cm][l]{G1S12}\mbox{ }\makebox[1.25cm][l]{N-body}\mbox{ }\makebox[1cm][c]{\rm 0.229}\mbox{ }\makebox[1cm][c]{(40\%)}\mbox{ }\makebox[1.25cm][c]{3.26}\mbox{ }\makebox[1.25cm][c]{4.06}} \newline
\vspace{-0.75mm}\colorbox{white}{\makebox[1cm][l]{G1S13}\mbox{ }\makebox[1.25cm][l]{Analytic}\mbox{ }\makebox[1cm][c]{\rm 0.507}\mbox{ }\makebox[1cm][c]{(96\%)}\mbox{ }\makebox[1.25cm][c]{2.32}\mbox{ }\makebox[1.25cm][c]{2.62}} \newline
\vspace{-0.75mm}\colorbox{white}{\makebox[1cm][l]{G1S13}\mbox{ }\makebox[1.25cm][l]{N-body}\mbox{ }\makebox[1cm][c]{\rm 0.350}\mbox{ }\makebox[1cm][c]{(61\%)}\mbox{ }\makebox[1.25cm][c]{2.87}\mbox{ }\makebox[1.25cm][c]{3.28}} \newline
\vspace{-0.75mm}\colorbox{lgrey}{\makebox[1cm][l]{G1S14}\mbox{ }\makebox[1.25cm][l]{Analytic}\mbox{ }\makebox[1cm][c]{\rm 0.521}\mbox{ }\makebox[1cm][c]{(98\%)}\mbox{ }\makebox[1.25cm][c]{1.58}\mbox{ }\makebox[1.25cm][c]{2.52}} \newline
\vspace{-0.75mm}\colorbox{lgrey}{\makebox[1cm][l]{G1S14}\mbox{ }\makebox[1.25cm][l]{N-body}\mbox{ }\makebox[1cm][c]{\rm 0.873}\mbox{ }\makebox[1cm][c]{(153\%)}\mbox{ }\makebox[1.25cm][c]{1.62}\mbox{ }\makebox[1.25cm][c]{1.90}} \newline
\vspace{-0.75mm}\colorbox{white}{\makebox[1cm][l]{G1S15}\mbox{ }\makebox[1.25cm][l]{Analytic}\mbox{ }\makebox[1cm][c]{\rm 0.438}\mbox{ }\makebox[1cm][c]{(83\%)}\mbox{ }\makebox[1.25cm][c]{1.78}\mbox{ }\makebox[1.25cm][c]{2.09}} \newline
\vspace{-0.75mm}\colorbox{white}{\makebox[1cm][l]{G1S15}\mbox{ }\makebox[1.25cm][l]{N-body}\mbox{ }\makebox[1cm][c]{\rm 0.374}\mbox{ }\makebox[1cm][c]{(65\%)}\mbox{ }\makebox[1.25cm][c]{1.80}\mbox{ }\makebox[1.25cm][c]{2.18}} \newline

\rule{74mm}{0.1mm}\newline
\end{center}
\end{table}

It is worth considering at this point some of the simplifications
which go into our dynamical model of disk heating. Dynamical friction
is treated using Chandrasekhar's approximation which is clearly not
strictly applicable to our halo-plus-disk system. While this
approximation has been shown to be a reasonable one for dark matter
halos \cite{weinberg86,bont87,cora97,vdb99,VW,colpi99}, its validity
when disks are included is less clear. Importantly, this
approximation ignores any possible resonant interaction between the
satellite and the disk.

A further simplification of our model is that the phase space
distributions of halo dark matter and disk stars are assumed to be
fixed, with the exception that the disk vertical velocity dispersion
and density profile are allowed to change with time. (We further
assume that the vertical motions of stars in the disk do not couple to
radial and azimuthal motions, which will be approximately true
provided that the disk remains thin.) In reality, all three components
of the disk velocity dispersion will be affected by substructure
heating. However, the changes in the radial and azimuthal velocity
dispersions have only a small effect on the overall structure of the
disk in the majority of cases. Thus, our approach should be a
reasonable first approximation.

A final, important simplification is that we ignore some possible
interactions between the disk and the dark matter halo, e.g. those
driven by non-axisymmetric structures such as bars or warps in the
disk. This complex set of interactions could, in principle, result in
energy initially transferred from the satellite to the disk finding
its way into the halo dark matter. The efficiency with which this
happens will clearly depend upon the frequency with which substructure
excites bars and other global modes in the disk and is therefore
beyond the scope of our current calculations.

Given these simplifications it is important to test our analytic
calculations against N-body simulations of the disk heating
process. We perform such tests in \S\ref{sec:Nbody}.

\subsubsection{Disk Scale-Height and Vertical Energy}
\label{sec:sch}

Having calculated the energy deposited into vertical motions of disk
stars, we now wish to calculate the resulting scale-height of the
disk. We work throughout in the {\em thin disk approximation}, in
which the vertical extent of the disk is always assumed to be small compared
to its radial extent, and the non-circular velocities are assumed to
be small compared to the circular velocity. In this approximation, the
disk can be treated as being locally plane-parallel, with the
consequence that the vertical motions separate from the motions in the
plane, and there is a well-defined vertical energy which (in the
absence of perturbations by satellites or other objects) is conserved
both for individual stars and for the disk as a whole. The vertical
energy given to a star by an encounter with a satellite is initially
in the form of vertical kinetic energy, but the orbital motion of the
star subsequently mixes this between vertical kinetic and potential
energies, while keeping the sum of the kinetic and potential energies
constant. In the thin disk approximation, the total vertical energy
per unit area of the disk, $e_{\rm z}$, can be written as (TO)
\begin{equation}
e_{\rm z} = t_{\rm z} + w_{\rm dd} + w_{\rm dh},
\label{eq:toten}
\end{equation}
where all quantities are surface energy densities, $t_{\rm z}$ is the
disk vertical kinetic energy, $w_{\rm dd}$ is the disk
self-gravitational energy and $w_{\rm dh}$ is the gravitational energy
due to the disk/halo interaction. The vertical energy $e_{\rm z}$ is
defined to be zero in a state where the disk has zero thickness and
zero vertical velocities. Expressions for $t_{\rm z}$, $w_{\rm
dd}$ and $w_{\rm dh}$ are derived in
Appendix~\ref{app:enrels}. Following TO, we assume virial equilibrium
and find
\begin{equation}
2t_{\rm z}-w_{\rm dd}-2w_{\rm dh}=0,
\label{eq:virial}
\end{equation}
and so
\begin{equation}
e_{\rm z}={3 \over 2}w_{\rm dd} + 2w_{\rm dh}.
\label{eq:ezvirial}
\end{equation}
The density of our model disks in the vertical direction is
proportional to $\hbox{sech}^2 z/H_{\rm d}$. For this density profile
we find from eqn.(\ref{eq:ezvirial}) (TO)
\begin{equation}
e_{\rm z}={3\over 2} \pi \G \Sigma_{\rm d}^2(R) H_{\rm d} 
+ \frac{\pi^2}{12}\Sigma_{\rm d}(R)H_{\rm d}^2 {\G M_{\rm h}(R)\over R^3},
\label{eq:ez}
\end{equation}
where $R$ is radius in the disk plane, $\Sigma_{\rm d}(R)$ is the
disk surface mass density, and $M_{\rm h}(R)$ is the mass in the
(spherical) halo plus bulge within radius $R$. Since the vertical
kinetic energy per unit area is $t_{\rm z}=\frac{1}{2}\Sigma_{\rm
d}\sigma_{\rm z}^2$, we also find from eqn.(\ref{eq:virial})
\begin{equation}
\sigma_{\rm z}^2=\pi \G \Sigma_{\rm d}(R)H_{\rm d} + \frac{\pi^2}{12}\G M_{\rm
h}(R)H_{\rm d}^2/R^3
\label{eq:sigmaz}
\end{equation}
This expression is used to calculate the vertical velocity dispersion
at each radius from the scale-height $H_{\rm d}$. \footnote{Note that here we
differ slightly from VW by including the contribution of the halo
gravity to the disk vertical velocity dispersion. This is typically a
small, although not negligible, contribution over the bulk of the
disk.}.

To relate the radially-dependent vertical energy per unit area to the
global total vertical energy, we make the assumption that the disk
scale-height is constant with radius, since this is observed to be a
good approximation for real galaxies (e.g. \pcite{reynier}). We can
then integrate eqn.(\ref{eq:ez}) over the whole disk to find the total
vertical energy. Using $\Sigma_{\rm d} = (M_{\rm d}/2\pi R_{\rm d}^2)
\exp(-R/R_{\rm d})$ for an exponential disk of radial scale-length
$R_{\rm d}$ we find
\begin{equation}
E_{\rm z} = {3 \over 16} M_{\rm d} V_{\rm d}^2 h + \frac{\pi^2}{12}
M_{\rm d} V_{\rm d}^2 h^2 \int_0^\infty \left[{V_{\rm h}\over
    V_{\rm d}} \right]^2 {\exp(-x) \over x} \d x, 
\label{eq:eztotal}
\end{equation}
where the fractional scale-height $h=H_{\rm d}/R_{\rm d}$, $V^2_{\rm d}=\G
M_{\rm d}/R_{\rm d}$ and $V^2_{\rm h}=\G M_{\rm h}(R)/R$. Integrating
eqn.(\ref{eq:sigmaz}) gives a similar expression for the total
vertical kinetic energy $T_z$. Once the total vertical energy $E_{\rm
z}$ is known, the above equation is easily solved for $h$ and hence
$H_{\rm d}$.

\subsubsection{Local vs. Global Heating}

In \S\ref{sec:sch} we made the assumption that the energy deposited in
the disk by satellites was distributed throughout the disk in such a
way as to produce a scale-height that was independent of
radius. However, the increase in energy per unit mass caused by a
satellite passing through or near the disk will initially be greatest
close to the point of impact. Since satellite encounters frequently
trigger global modes of the disk it is not implausible that this
energy quickly becomes redistributed throughout the disk. However, it
is interesting to consider the opposite extreme in which energy is
deposited at the position of the satellite and remains there. We refer
to these two extremes as ``global'' and ``local'' heating. To study
local heating we accumulate the energy deposited by satellites in a
narrow annulus of the disk (in practice we use a Gaussian window
function), centred on the disk half-mass radius.  We then assume that
the specific energy of disk material is proportional to the same
window function and use the relations of \S\ref{sec:sch} to compute
the resulting scale-height at the half-mass radius.

Observations of real galactic disks \cite{reynier} indicate that the
scale-height is reasonably constant with radius, at least for late-type
galaxies. For this reason we prefer the global heating assumption, but
consider local heating also as an interesting comparison.

\subsubsection{Further Aspects}

Below we detail how we deal with energy\footnote{For convenience, we
use the expression ``energy'' to imply ``disk vertical energy'' from
here on, unless explicitly stated otherwise.} deposited in a gaseous
disk and how we treat galaxy mergers, gas accretion and star
formation.

\emph{Gas in Galaxy Disks:} Disks in our model in general consist of
both stars and gas. The gas is assumed to be in an infinitely thin
layer with zero velocity dispersion in the disk midplane. We include
the contribution of the gas to the disk gravitational potential and
when computing the disk scale-height. With our choice of zero-points
for the energy, the vertical kinetic energy of the gas and also its
self-gravitational energy are both zero (because it is at $z=0$), but
the gas contributes to the total energy per unit area of the disk
$e_z$ through the gravitational interaction energy between the gas and
stars (see TO for more details).  We assume that gas and stars in the
disk are heated at the same rate per unit mass, but that the gas
dissipates this energy rapidly, so that energy deposited in the gas
is effectively lost.

\emph{Adiabatic Heating due to Gas Accretion:} Gas accreted onto the
disk is assumed to initially have zero energy. However, the growth of
the disk surface density causes gravitational compression in the
vertical direction, which tends to increase the vertical energies of
disk stars. We follow TO and assume that gas infall occurs
adiabatically, adopting their equation (3.12) to describe the change
in energy of the disk stars due to adiabatic heating. In our model, gas
can also be lost from the disk due to feedback processes, resulting in
a decrease in the energies of stars. We account for this process in
the same way as for the adiabatic heating, simply changing the sign of
the effect. We find that these are minor effects that have little
impact on the predicted scale-heights of disk galaxies.

\emph{Star Formation:} When gas turns into stars, we assume that these
stars start out with zero energy, but then rapidly mix with the
pre-existing stellar population, conserving the total disk vertical
energy.

\emph{Galaxy Mergers:} In a major merger all disks are destroyed, and
so we zero the energy of the resulting galaxy. In minor mergers, stars
from the satellite galaxy disk and bulge are added to the bulge of the
central galaxy. In the merger, the energy of the satellite disk is
lost, while that of the central galaxy disk is unchanged, unless the
infalling satellite contains gaseous material, which will contribute
to the adiabatic heating of the central galaxy disk.

\subsubsection{Heating of Disks by Scattering by Clouds}
\label{sec:cloudheat}

Substructure in the halo is not the only source of heating for
disks. Two other plausible mechanisms for disk heating are
gravitational scattering of stars by massive gas clouds
\cite{spitschwar53,lacey84} and scattering of stars by spiral arms
\cite{carl85}. The latter mechanism is inefficient at producing any
heating in the vertical direction, so we will focus on the first
mechanism. \scite{lacey84} derived analytical expressions for the rate
at which scattering by clouds increases the vertical and horizontal
epicyclic energies of the stars.  In general, these expressions depend
on the radial and vertical disk velocity dispersions, $\sigma_{\rm R}$
and $\sigma_{\rm z}$, but, acting by themselves, the clouds tend to
drive the ratio $\sigma_{\rm z}/\sigma_{\rm R}$ to an equilibrium
value. We calculate the rate of increase of vertical energy per unit
mass for the stars, $\varepsilon_z$, using Lacey's eqn.(39), evaluated
for the equilibrium $\sigma_{\rm z}/\sigma_{\rm R}$ and in the limit
in which the scale-height of the stars is larger than that of the
clouds. This gives
\begin{equation}
\left(\frac{\d\varepsilon_{\rm z}}{\d t}\right)_{\rm clouds} = \frac{2 \G^2 \Sigma_{\rm c} M_{\rm C}
  \ln\Lambda_{\rm c}\, \nu}{\sigma_{\rm z}^2} \alpha^3_{\rm s}(\beta) K_{\rm s}(\beta)
\label{eq:cloudheat}
\end{equation}
where $\Sigma_{\rm c}$ is the surface density in clouds, $M_{\rm c}$
is the cloud mass, $\ln\Lambda_{\rm c}$ is the Coulomb logarithm for
scattering of stars by clouds and $\nu$ is the vertical epicyclic
frequency.  $\alpha_{\rm s}(\beta)$ and $K_{\rm s}(\beta)$ are
functions of $\beta=2\Omega/\kappa$ which are tabulated by Lacey,
$\Omega$ being the angular velocity for circular orbits and $\kappa$
the radial epicyclic frequency. We obtain the total contribution of
scattering by clouds to the increase of vertical energy by integrating
eqn.(\ref{eq:cloudheat}) over radius:
\begin{equation}
\dot{E}_{\rm z,clouds} = \int_0^\infty \Sigma_{\rm d}
\left(\frac{\d\varepsilon_{\rm z}}{\d t}\right)_{\rm clouds} 2\pi R\, \d R
\label{eq:cloudheat_tot}
\end{equation}
Numerical simulations of heating by clouds agree fairly well with the
velocity dependence predicted analytically, ($\d\sigma^2/\d t \propto
\sigma^{-2}$), but have given somewhat conflicting results about the
amplitude of the effect; \scite{villumsen} found heating rates
$\d\sigma^2/\d t$ at a given $\sigma$ about 6 times lower than the
analytical prediction, while \scite{hanninen} found rates 3--8 times
higher. 

Our galaxy formation model predicts the total mass of gas in the disk
of each galaxy as a function of time. We assume that the gas is
distributed radially in the same way as the stars, with a constant
fraction being in the form of giant molecular clouds. For our standard
case we will assume that 25\% of the gaseous mass of the disk is in
clouds \cite{granato00}, that they have mass-weighted mean mass of
$M_c=6.6\times 10^5M_\odot$ \cite{lacey84} and typical radius
$a_c=16$pc \cite{granato00}, and that $\beta=1.5$. For each model
galaxy, we integrate the heating due to scattering from molecular
clouds over each timestep in the calculations, and add this energy
change to that which arises from interactions with satellites.

\section{Calibration Using N-body Simulations}
\label{sec:Nbody}

As has been noted by several authors, the amount of heating caused by
a satellite is difficult to determine analytically since some of
the energy may drive global perturbations (e.g. warps) in the disk,
and satellites may trigger bar instabilities leading to an enhanced
heating rate. Furthermore, our approach to dynamical friction in the
disk follows the methods of Chandrasekhar (e.g. \pcite{bintrem},
section~7.1), which assume that each particle interacts with the
satellite only once. If the satellite orbital period is close to the
rotation period of the disk  (or to some other resonance of the
disk orbits), this assumption fails. Instead, a single particle may
interact with the satellite multiple times on consecutive orbits.
This problem should therefore ideally be approached in terms of
resonant interactions between satellite and disk
\cite{goldtre79,donner93,wahde96,weinberg02}.  We retain the
Chandrasekhar methods for their simplicity, and  show that they
provide a reasonable approximation to the dynamical friction due to
disks in the regimes of interest.

\subsection{N-body Simulations}

We begin by testing and calibrating our analytic calculations against
numerical simulations of disk heating. In principle, the simulations
of VW are ideal for this purpose. However, the central densities and
velocity dispersions of the King model satellites given by VW are too
low to be consistent with their assumed concentration
parameters. Thus, the satellites seem to more weakly bound than the
authors intended. It is unclear a priori how this would affect the
results and we have therefore decided to repeat their
calculations. This has two other advantages:
\begin{itemize}
\item We can repeat each simulation without the disk component,
allowing us to constrain separately the contributions of the halo and
disk to the dynamical friction experienced by the satellite.
\item We can perform convergence tests by increasing the number of
particles in the simulation in order to ensure that disk heating is
being estimated accurately.
\end{itemize}
We carry out the same set of simulations as VW. Briefly, each
simulation consists of a galaxy containing a bulge, disk and dark
matter halo, plus a satellite object. Density profiles and the number
of particles used for each component are listed in
Table~\ref{tb:simpars}, while other details of each simulation (type
of satellite used, initial satellite orbital parameters and whether or
not a disk is included) are listed in Table~\ref{tb:simorbs}. Initial
conditions are created using the techniques of
\scite{hernquist93}. The galaxy and satellite are then evolved
separately, as described by VW, using the {\sc gadget} code
\cite{springel01} to allow them to reach equilibrium. We employ {\sc
gadget}'s new cell-opening criterion for tree walks ({\tt
TypeOfOpeningCriteria}$=1$) with an accuracy of {\tt
ErrTolForceAcc}$=0.001$, together with {\tt
TypeOfTimestepCriterion}$=1$ with {\tt ErrTolVelScale}$=10.0$. {\sc
gadget} uses adaptive time-stepping. We impose no minimum timestep
size, but impose a maximum size of {\tt MaxSizeTimestep}$=0.01$ (in
{\sc gadget}'s default internal units). All particles in the
simulation are given a softening length of $0.110$kpc. With these
choices, energy is conserved to better than 1\% throughout the
simulations. The two sets of initial conditions are then superimposed
and evolved for 4Gyr.

The simulations are labelled G1S1 to G1S15 as in VW. We also perform a
simulation with no satellite, G1S0, to measure the two-body heating
rate in the disk. We repeat each simulation without a disk component,
labelling these G2S1 to G2S15 (note that in the absence of the disk,
only models G2S1, G2S2, G2S7, G2S9 and G2S14 are different). We also
repeated all of these calculations with one half and one quarter the
number of particles, in order to test how well the results have
converged. The convergence tests are described in
Appendix~\ref{sec:convergence}. They indicate that the convergence is
good for the evolution of the mass and orbit of the satellite, and
adequate for the increase in the vertical energy of the disk. Unless
otherwise noted, we show results from the highest resolution
simulations.

Each simulation output is analyzed in order to determine the position,
velocity and mass of the satellite (computed for those particles which
remain bound to the satellite), and the vertical kinetic energy of the
disk. We determine which particles are bound to the satellite using
the following algorithm:
\begin{enumerate}
\item Begin by considering all the satellite particles that were bound
to the satellite at the previous timestep (or simply  all satellite
particles for the first timestep).
\item Compute the mean position and velocity, and the mass of the
satellite from these particles.
\item For each particle in this set, determine if it is
gravitationally bound to the other particles in the set.
\item Retain only those particles which are bound and go back to step
(ii). Repeat until the mass of the satellite has converged.
\end{enumerate}

To determine the vertical kinetic energy of the disk, $T_{\rm z}$, at
each output time, we first locate the centre of mass of the disk and
its mean velocity. (Since the satellite mass is comparable to that of
the disk, the disk moves around significantly as the satellite passes
by.) We then rotate the system to the frame defined by the principal
axes of the disk inertia tensor, and sum the kinetic energies of
particles in the direction defined by the shortest axis (which
corresponds to the $z$-axis for an untilted disk). This rotation is
necessary because the disk can become tilted through its interaction
with the satellite (as also noted by VW). In the original frame (i.e.
without rotation), purely circular motions in a tilted disk appear as
vertical energy. 

One final step is necessary in order to obtain the {\em increase} in
the disk vertical energy due to the interaction with the
satellite. Even in the absence of a satellite, the vertical kinetic
energy of the disk increases as the simulation proceeds due to
numerical relaxation (mainly two-body scattering), from $0.57\times
10^{14}M_\odot$km$^2$s$^{-2}$ at $t=0$ to $0.64\times
10^{14}M_\odot$km$^2$s$^{-2}$ at $t=4$Gyr, for our standard particle
number. This increase of $0.07\times 10^{14}M_\odot$km$^2$s$^{-2}$ due
to two-body relaxation is comparable to the heating by the satellite
in many of the cases considered. Therefore, to obtain the increase in
vertical energy due to the satellite at time $t$, which we denote as
$\Delta T_{\rm z}(t)$, we subtract off the energy of the unperturbed
disk (from model G1S0) at the same time $t$. Based on runs of model
G1S0 with different random number seeds but the same number of
particles, the increase in $T_{\rm z}$ due to numerical relaxation is
determined to an accuracy of better than $0.005 \times
10^{14}M_\odot$km$^2$s$^{-2}$ in the standard case, so the uncertainty
in $\Delta T_{\rm z}$ introduced by the subtraction is small compared
to $\Delta T_{\rm z}$ itself.

\subsection{Comparison with Analytic Calculations}
\label{sec:VWcomp}

To test our analytical model of satellite orbital evolution and
constrain its parameters, we adapt the analytical model so as to mimic
the set-up of each N-body simulation.  Thus, we assume density
profiles for host and satellite systems identical to those of the
N-body simulations. \scite{taybab} compared their model of satellite
galaxy orbital evolution to the orbital radii and satellite masses as
a function of time in the simulations of VW, finding generally good
agreement. We repeat their analysis here, using our own model of
satellite dynamics, extended to include the calculation of disk
heating. We will use this comparison to fix the four free parameters
of our satellite orbit model, $f_{\rm orb}$, $f_{\Lambda, \rm h}$,
$f_{\Lambda, \rm d}$ and $\epsilon_{\rm h}$. As described in
Appendix~\ref{sec:improve}, $f_{\rm orb}$ controls the timescale on
which tidally stripped mass is lost from the satellite, while
$f_{\Lambda, \rm h}$ and $f_{\Lambda, \rm d}$ are the factors that
appear in the Coulomb logarithms, $\Lambda_{\rm h}$ and $\Lambda_{\rm
d}$, for the dynamical friction force due to the halo and disk
respectively. The parameter $\epsilon_{\rm h}$ controls the strength
of gravitational shock-heating and is defined in \scite{benson02a}.

\begin{figure*}
\begin{tabular}{cc}
\multicolumn{2}{c}{Model G2S2 (S1, $\theta_{\rm i}=0^\circ$, $\epsilon_{\rm J}=0.55$, $r_{\rm a}=55$kpc, no disk)} \\
\vspace{-5mm} \psfig{file=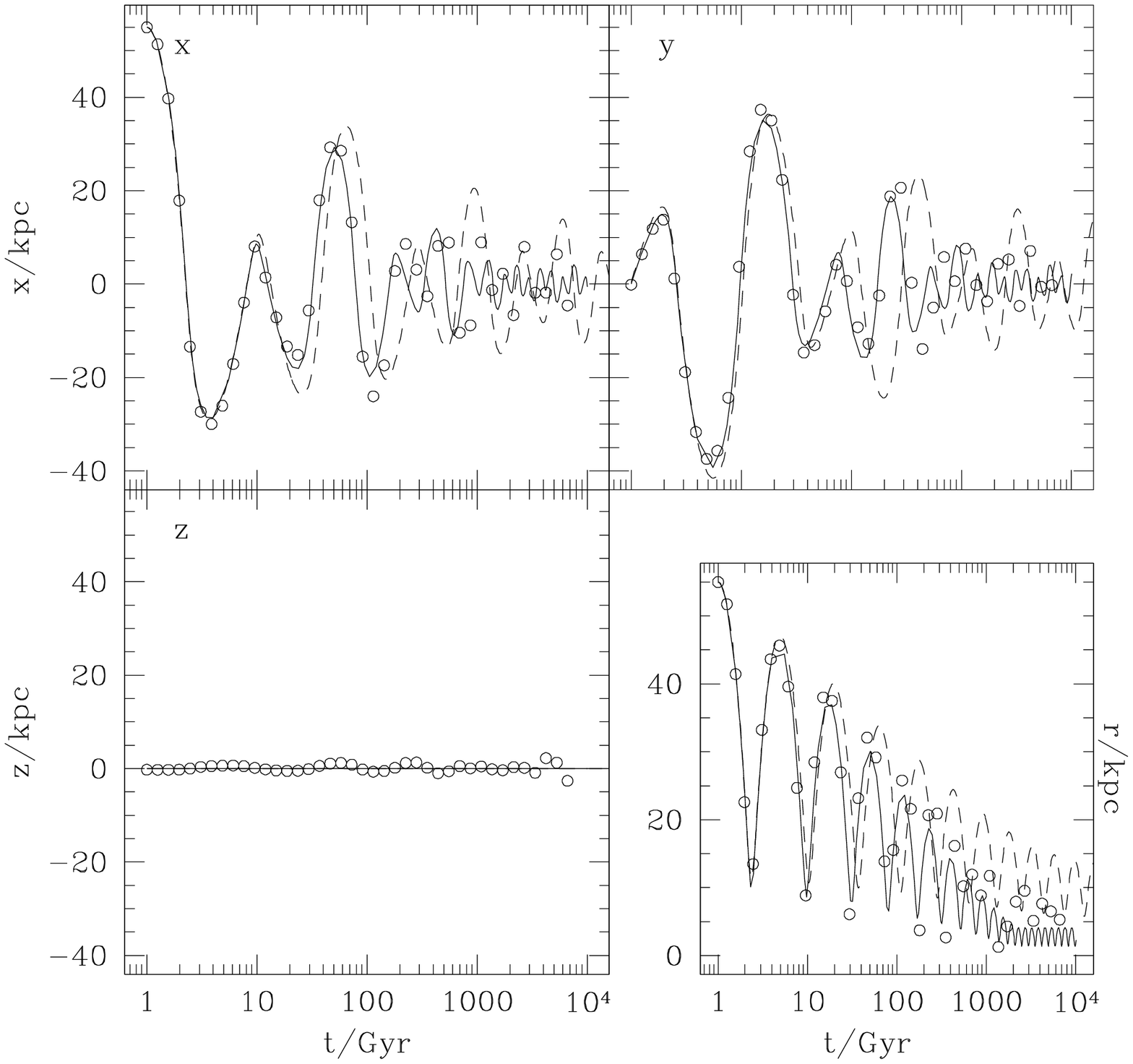,width=75mm} & \psfig{file=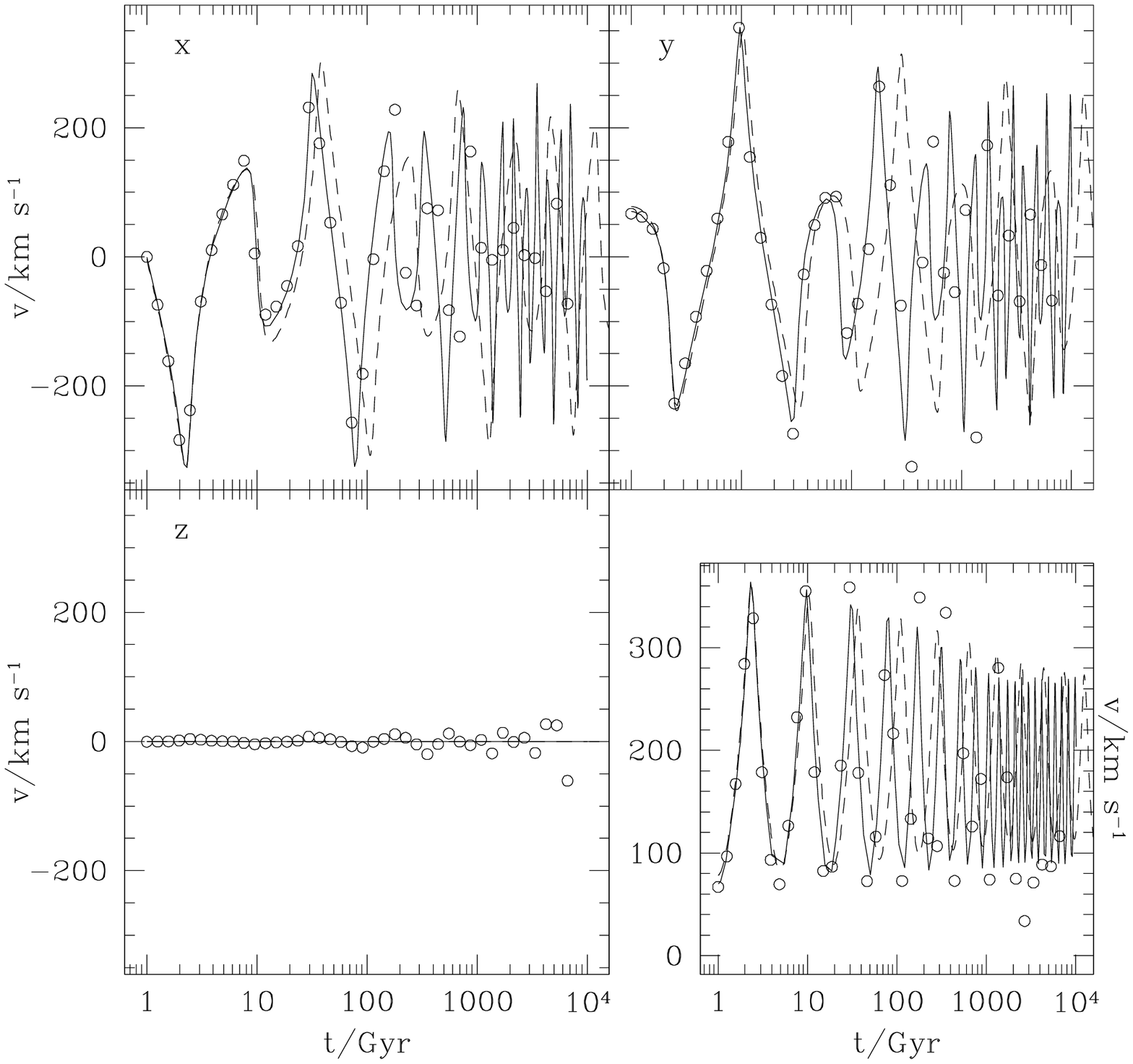,width=75mm} \\
\vspace{-5mm} \psfig{file=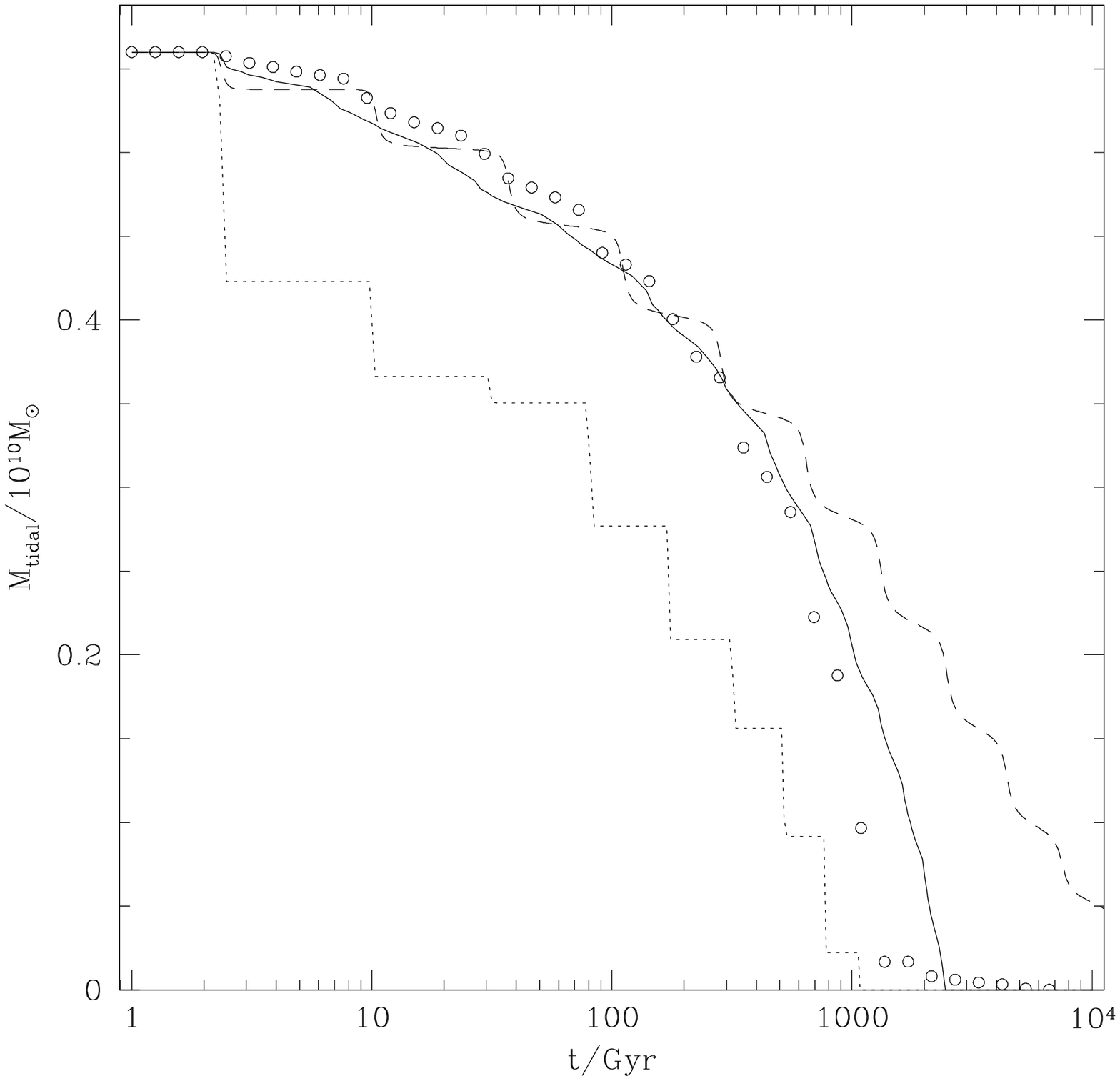,width=75mm} &  \psfig{file=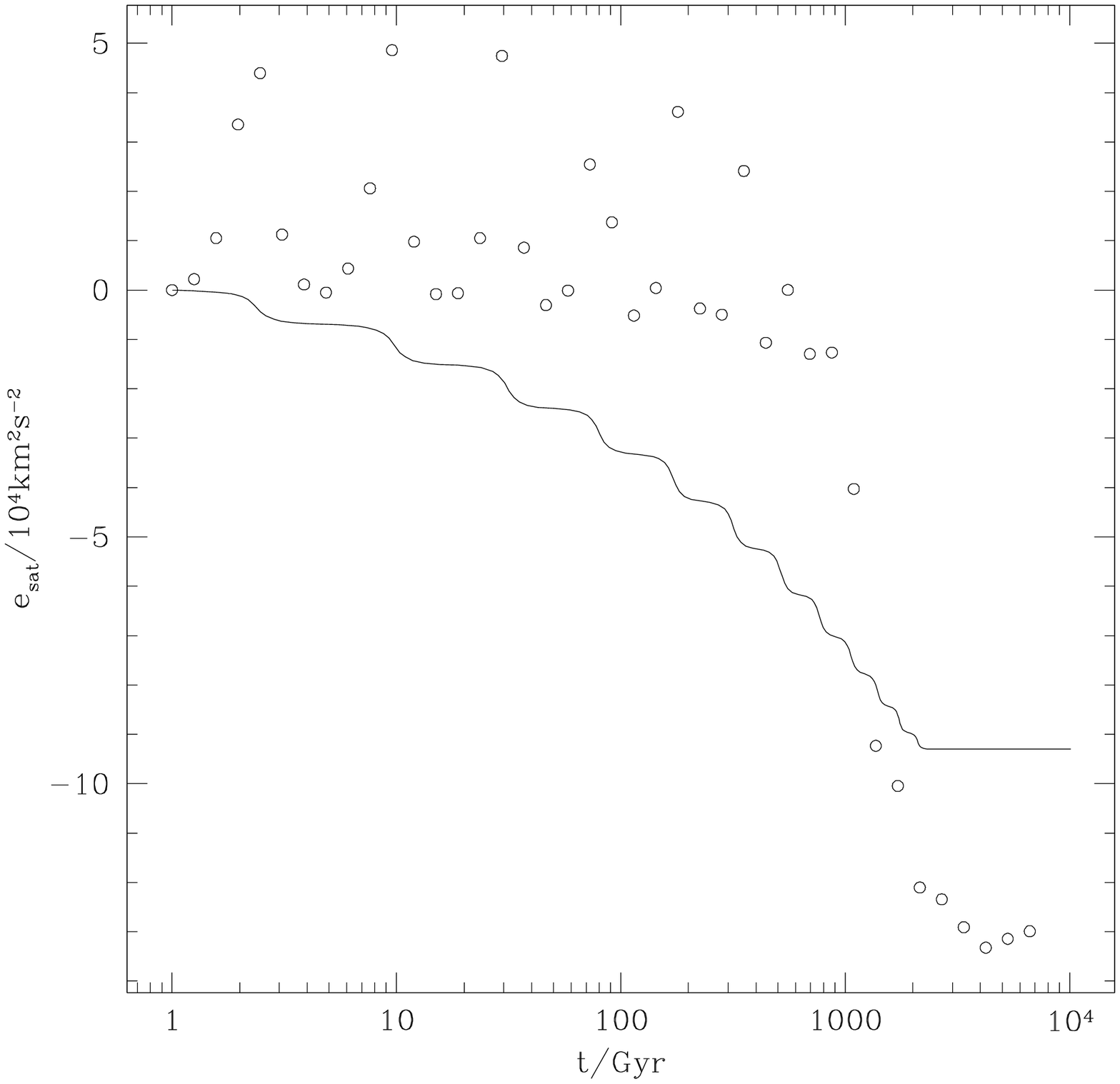,width=75mm} \\
\end{tabular}
\caption{Evolution of the satellite and its orbit in the diskless
  model G2S2. We compare the results from our analytical model (solid
  lines) with the analytical model of \protect\scite{taybab} (dashed
  lines) and with our N-body simulation (circles). \emph{Top left-hand
  panel:} The orbital position and radius of the satellite as a
  function of time. \emph{Top right-hand panel:} The orbital velocity
  of the satellite and its components as a function of
  time. \emph{Lower left-hand panel:} The remaining bound mass of the
  satellite as a function of time. The dotted line shows the mass of
  the satellite if mass loss beyond the tidal radius is assumed to
  occur instantaneously (i.e. $f_{\rm orb}=0$). \emph{Lower right-hand
  panel:} The change in the specific orbital energy of the satellite
  as a function of time.}
\label{fig:VW_G2S2}
\end{figure*}

Using our satellite orbit model, each orbit is integrated for
4~Gyr. Figures~\ref{fig:VW_G2S2} and \ref{fig:VW_G1S3} show the
orbital position and velocity and the remaining bound mass and orbital
energy of the satellite for models G2S2 and G1S3 respectively, with
our N-body results shown as open circles. Figure~\ref{fig:VW_G1S3}
also shows the energy deposited in the disk in model G1S3. This is
given by the vertical kinetic energy of the simulated disk minus the
vertical kinetic energy of the disk in model G1S0 which contains no
satellite. The subtraction removes both the initial energy of the
disk, and the energy gained by two-body relaxation during the
simulation. We indicate at the top of each figure the label of the
satellite model, the initial inclination of the orbit with respect to
the galaxy disk ($\theta_{\rm i}$), the initial circularity
($\epsilon_{\rm J}$; the angular momentum of the satellite divided by
the angular momentum of a circular orbit with the same energy) and the
initial apocentric distance of the orbit ($r_{\rm a}$). Where they are
available, we show the results of VW as triangles. Note that in the
simulation of VW the satellite loses mass more rapidly, due to the
incorrect density profile used. For comparison, we show, as dashed
lines, the orbital radius and remaining bound mass derived from the
analytical calculations of \scite{taybab} for the same model.

The results in Figs.~\ref{fig:VW_G2S2} and \ref{fig:VW_G1S3} are for
the parameter combination $(f_{\rm orb}, f_{\Lambda, \rm h},
f_{\Lambda, \rm d}, \epsilon_{\rm h})=(2.5,1.5,3.0,1.0)$. The values
of $f_{\rm orb}$ and $\epsilon_{\rm h}$ are fixed by matching the mass
loss rates found in the simulations with no disk component. The value
of $f_{\Lambda, \rm h}$, which controls the strength of the dynamical
friction force due to the halo, is fixed by matching the rate of decay
of the orbital radius in models with no disk (so that the orbital
decay is caused entirely by the halo plus bulge system). Finally,
$f_{\Lambda, \rm d}$ is fixed by matching the rate of orbital decay in
the models which include a disk. In these models, the disk is the
dominant source of dynamical friction throughout a substantial
fraction of the orbital evolution.

The parameter values that we have selected produce the best agreement
with the set of fifteen models that were simulated. Generally, we find 
quite good agreement with the numerical results, comparable to that
achieved by
\scite{taybab}\footnote{It should be noted that \protect\scite{taybab}
were attempting to match the simulations of VW, rather than our
simulations, so that one should not expect exact agreement of their
results with ours.}. Our model uses more general expressions for
$\Lambda_{\rm h}$ and $\Lambda_{\rm d}$ than that of
\scite{taybab}. If we treat those numbers as free parameters (instead
of $f_{\Lambda, \rm h}$ and $f_{\Lambda, \rm d}$) we are able to
achieve even better agreement with the numerical simulations. However,
our approach has the advantage that $\Lambda_{\rm d}$ and
$\Lambda_{\rm h}$ scale in a physically reasonable way when we apply
our model to very different satellite/host systems. In any case,
orbital positions and velocities are typically matched accurately
until the final merging of the satellite (where it becomes difficult
to determine these quantities precisely in the N-body
simulations). The satellite mass as a function of time is typically
matched to within about 30--40\%. Table~\ref{tb:comparison} lists
several quantities -- the final change in the disk energy and the time
at which the satellite reaches 50\% and 10\% of its original mass --
from both analytic and N-body calculations for comparison.

\begin{figure*}
\begin{tabular}{cc}
\multicolumn{2}{c}{Model G1S3 (S1, $\theta_{\rm i}=45^\circ$, $\epsilon_{\rm J}=0.55$, $r_{\rm a}=55$kpc)} \\
\vspace{-5mm} \psfig{file=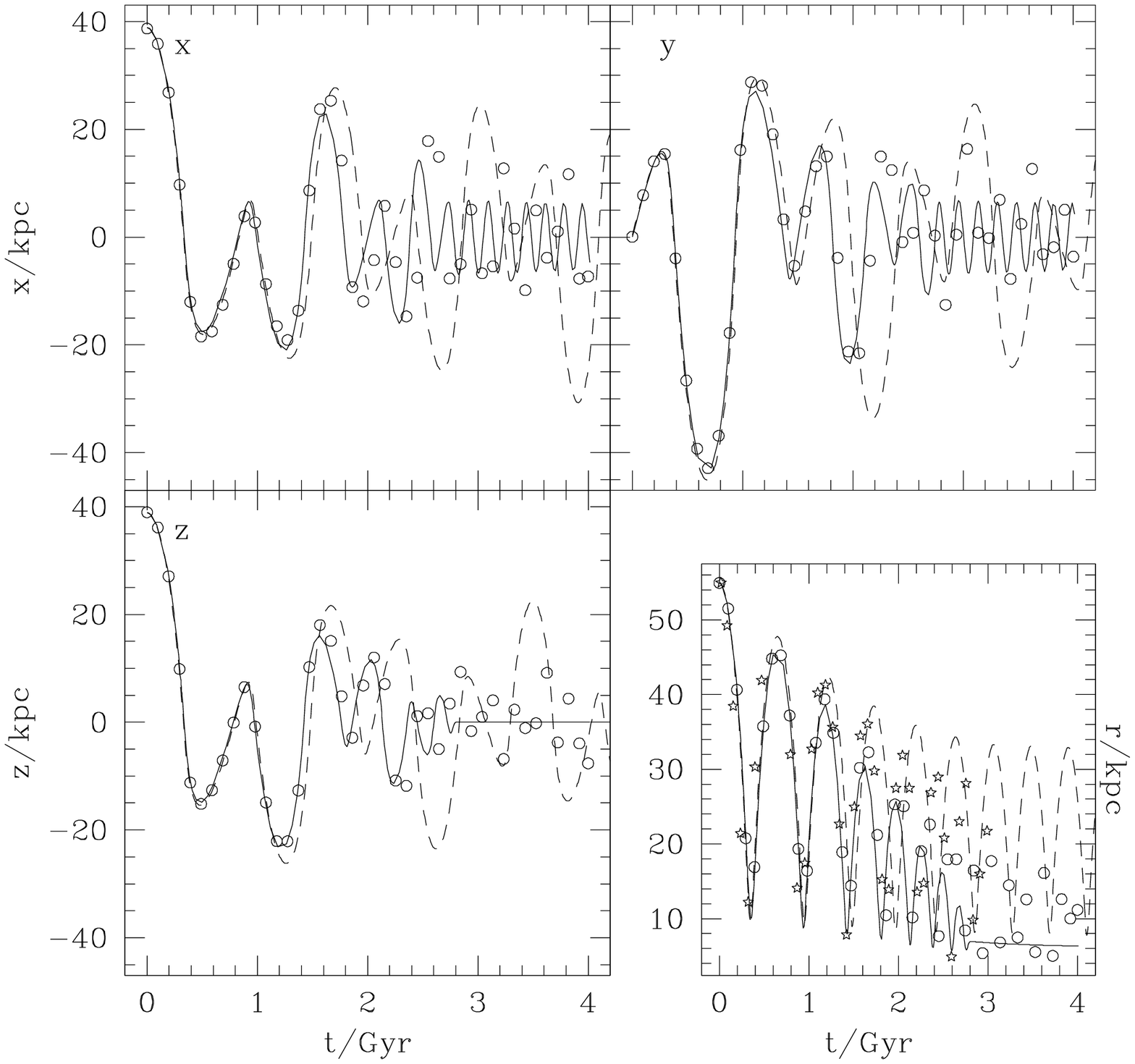,width=75mm} & \psfig{file=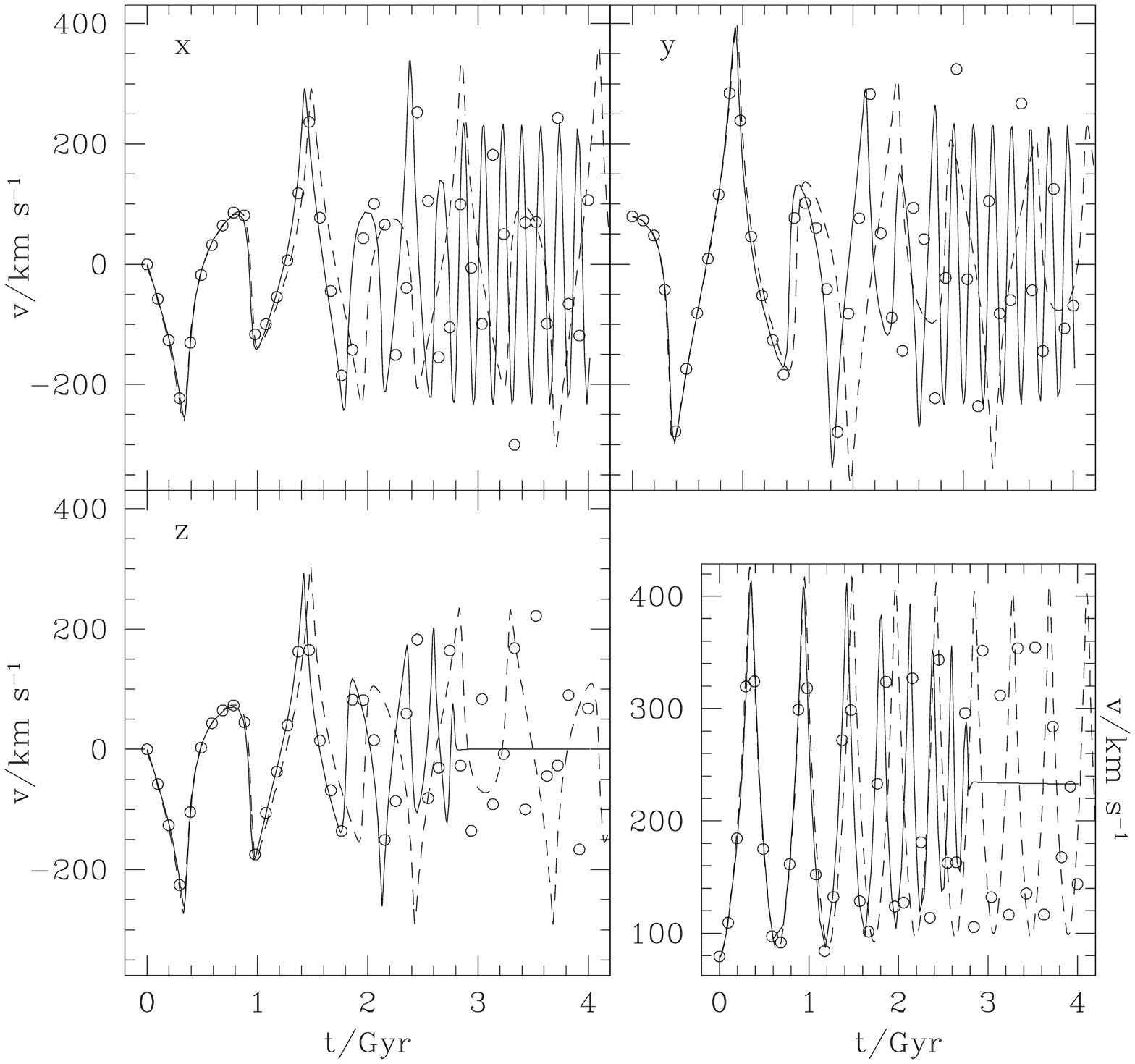,width=75mm} \\
\vspace{-5mm} \psfig{file=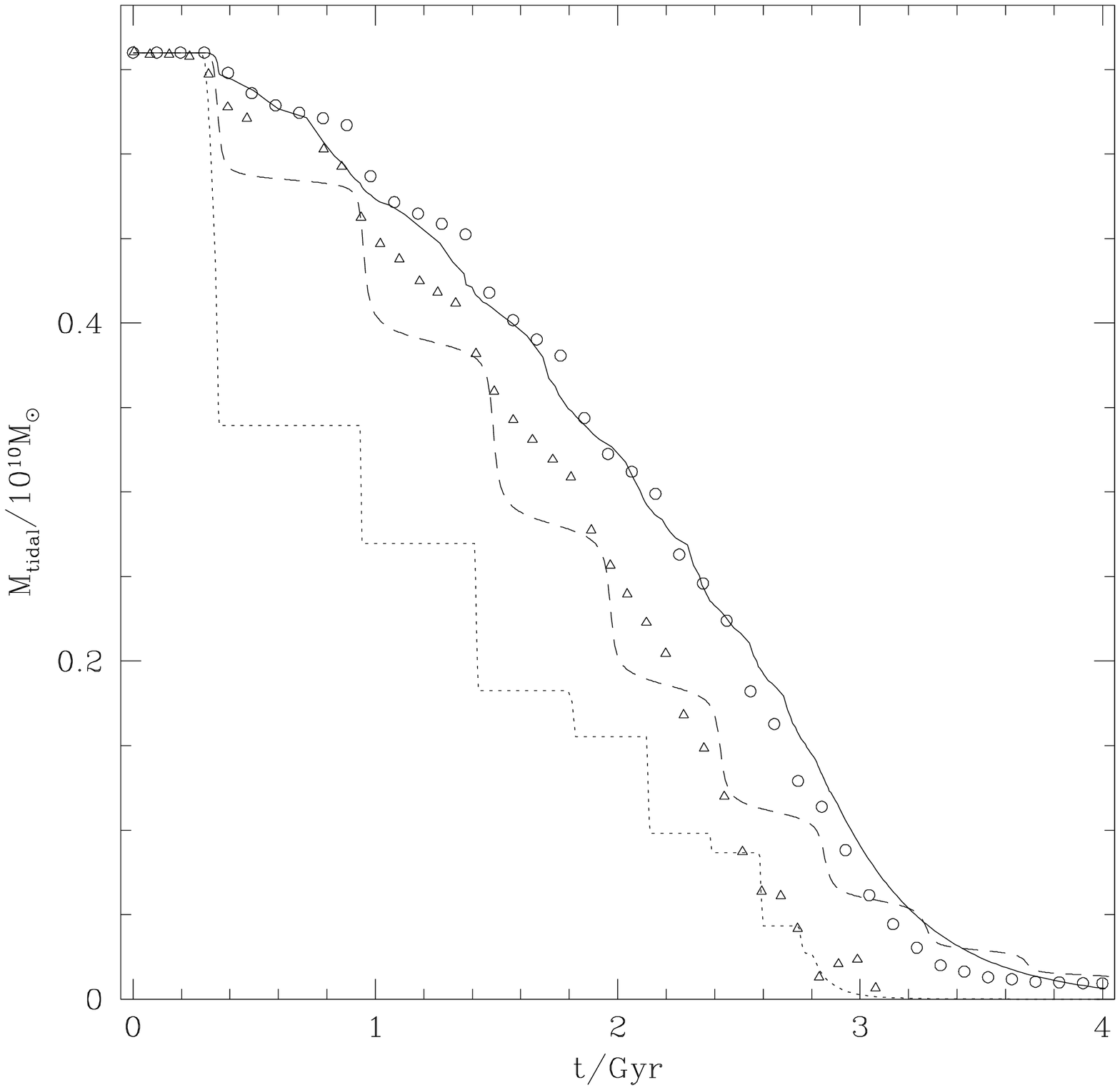,width=75mm} & \psfig{file=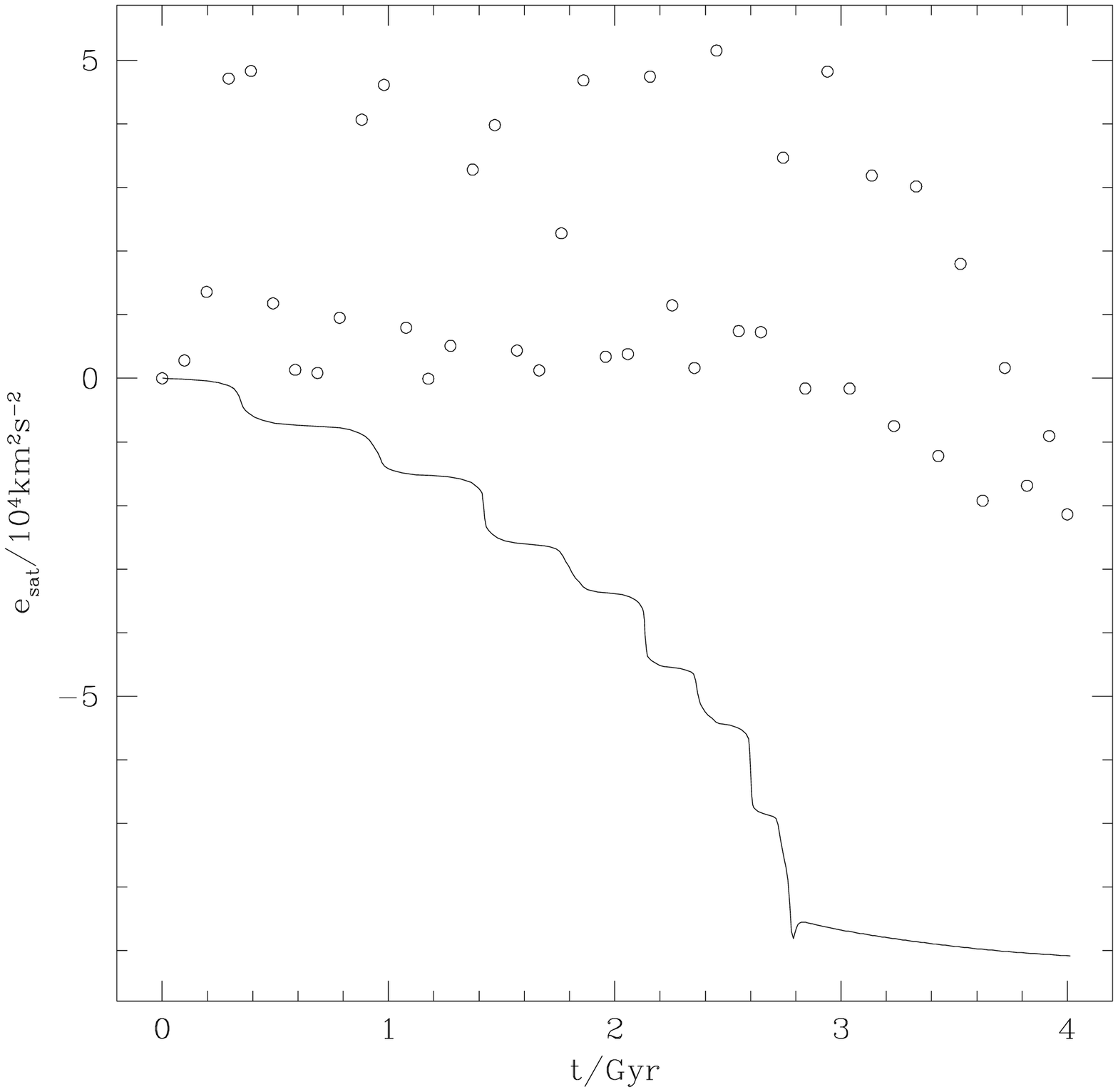,width=75mm} \\
\vspace{-5mm}  & \psfig{file=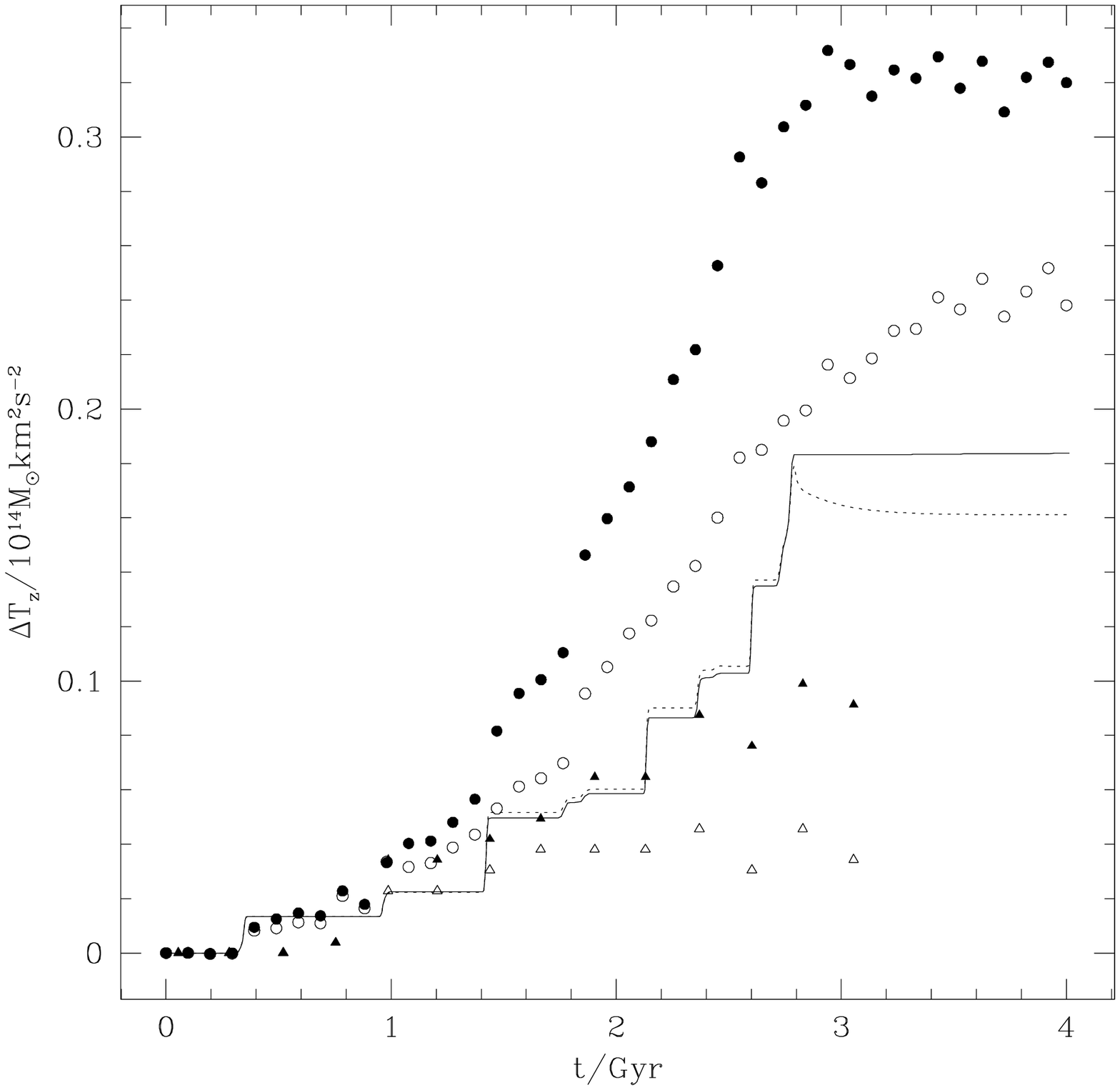,width=75mm}
\end{tabular}
\caption{}
\label{fig:VW_G1S3}
\end{figure*}

\begin{figure*}
\addtocounter{figure}{-1}
\caption{\emph{(cont.)} Properties of the orbiting satellite and host
halo galaxy disk in model G1S3. We compare the results from our
analytic calculations (solid lines) and those of
\protect\scite{taybab} (dashed lines), with those from our N-body simulation
(circles) and those of \protect\scite{VW} (triangles). \emph{Top
left-hand panel:} The orbital position and radius of the satellite as
a function of time. \emph{Top right-hand panel:} The orbital velocity
and speed of the satellite as a function of time. \emph{Centre
left-hand panel:} The remaining bound mass of the satellite as a
function of time. The dotted line shows the mass of the satellite if
mass loss beyond the tidal radius is assumed to occur
instantaneously. \emph{Centre right-hand panel:} The change in
specific orbital energy of the satellite with time. \emph{Lower
right-hand panel:} The vertical kinetic energy of the central galaxy
disk. Filled symbols show the energy measured in the original
coordinate frame of the disk, whereas the open symbols show the energy
measured in a frame that coincides with the principal axes of the
inertia tensor of the disk at each epoch. The dotted line shows the
result obtained if the energies of the disk in each direction
($R,\phi,z$) are assumed to reach equipartition.}
\end{figure*}

It is worth noting that the free parameters of our orbit model are set
without reference to the disk heating rate seen in the numerical
simulations. Thus, the heating rates we predict are entirely specified
by other considerations. The lower right-hand panel in
Fig.~\ref{fig:VW_G1S3} shows the change in disk vertical kinetic
energy from our analytic model calculated as described in
\S\ref{sec:sch} and from the N-body simulation. We find that our
analytic model reproduces the final disk energy in the numerical
simulations to better than a factor of two in ten out of the fifteen
simulations (see Table~\ref{tb:comparison}) but, in extreme cases, the
difference can be a factor of three or more.  Of the five models which
do not agree to within a factor of two, one (G1S7) has a prograde
satellite orbit in the disk plane ($\theta_{\rm i}=0^\circ$), two
(G1S4 and G1S11) have polar orbits ($\theta_{\rm i}=90^\circ$), and
two (G1S8 and G1S10) are on inclined prograde orbits ($\theta_{\rm
i}=45^\circ$).

For all five of the most discrepant cases, the analytical calculation
predicts less heating than the N-body simulation.  The largest
disagreement occurs for model G1S7 which has a prograde orbit in the
disk plane. Here, the analytical determination overestimates the
dynamical friction force in the disk as measured in the N-body
simulation. The satellite then becomes trapped in an orbit rotating
with the disk and there is no further energy transfer to the disk,
resulting in and underestimate of the heating in the analytic model by
a factor of 8. For the polar orbits (G1S4 and G1S11), mass loss in the
analytic model is too rapid and this again reduces the heating rate
compared to the N-body calculation. These two models underpredict the
N-body heating by a factor of approximately 3. For the inclined orbits
(G1S8 and G1S10), it is possible that the disk is no longer well
described by a single inclination (for example, it may have become
warped), leading to an overestimate of the energy in the N-body
simulations. (With the number of particles employed in our simulated
disks, the inclination of an un-warped disk can be determined to very
high precision, so there is very little inaccuracy in the
determination of the disk energy).

\begin{figure}
\psfig{file=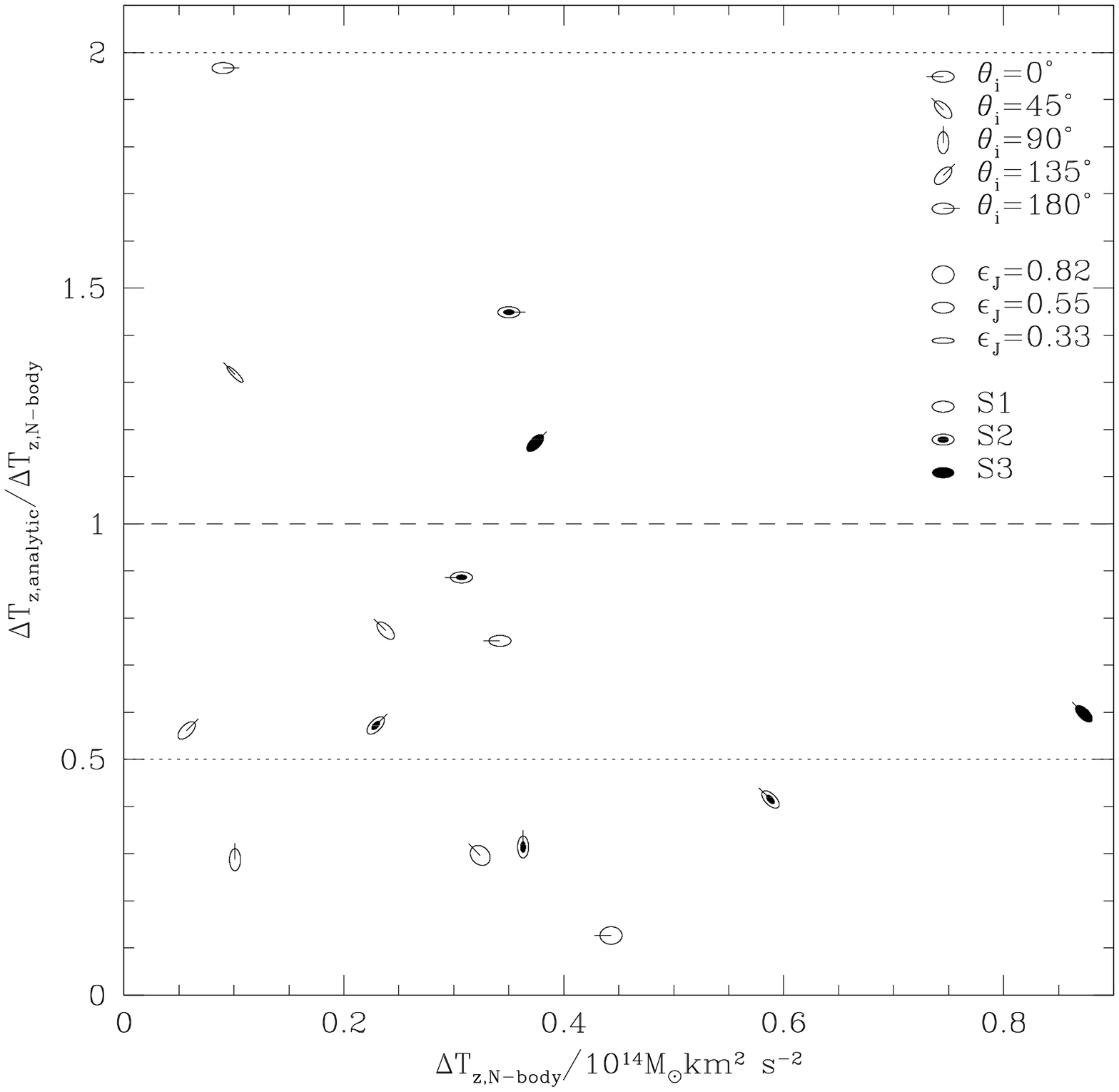,width=80mm}
\caption{A comparison between the analytic and N-body results for the 
change in the vertical component of the disk kinetic energy, $\Delta
T_{\rm z}$.  The dashed line is the locus of perfect agreement, with
the two dotted lines indicating factor of two discrepancies. The
symbols, one for each of the fifteen simulations, G1S1 to G1S15,
indicate through their orientation, shape and shading the orbital
inclination, the orbital circularity and the satellite model
respectively.}
\label{fig:accuracy}
\end{figure}

Figure~\ref{fig:accuracy} compares the N-body and analytic results for
the change, $\Delta T_z$, in disk kinetic energy.  The dashed line is
the locus of perfect agreement between the two calculations, and the
two dotted lines indicate a factor of two discrepancy. The symbols,
one for each of the fifteen simulations, G1S1 to G1S15, indicate
through their orientation, shape and shading the orbital inclination,
the orbital circularity and the satellite model respectively (as
indicated by the key in the figure).

Several of the N-body simulations show evidence of bar formation in
the central regions of the disk. This is particularly evident when the 
satellite is on a prograde orbit in the disk plane. Bars may be expected
to enhance the transfer of energy to the disk, and may be part of the
reason why the analytic model (which does not allow for bar formation)
substantially underpredicts the amount of heating in some cases
(e.g. G1S5 and G1S7, of which the latter shows a particularly strong
bar in the N-body simulation).

The efficiency of vertical heating, $\epsilon_{\rm z}$, is an
important component of our calculations. If we did not include this
efficiency factor, the predicted heating rates would be up to 4 times
higher (depending on the orbit ---the
effect is largest for near-circular prograde orbits in the disk plane
and polar orbits), with a factor of 3 being typical.

The inclusion of the $\theta$ dependence in the expression for
$\Lambda_{\rm d}$ (see Appendix~\ref{app:couldisk}) tends to reduce
the heating rate slightly. The effect is small for most orbits, but it
is of greater importance for orbits in the disk plane, helping to
improve the agreement with the simulations in those cases. The use of
an anisotropic disk velocity dispersion in the dynamical friction
force generally has an even smaller effect, typically increasing the
disk heating rate by a few percent (although in some cases the rate is
decreased by an equally small amount). Prograde orbits in the disk
plane are, once again, most strongly affected, with heating rates
reduced by 20--40\%
 
The galaxy in the N-body model contains a bulge of mass $1/3$ that of the
disk. VW also performed simulations with bulges of mass $1/5$ and
$2/3$ that of the disk to examine the influence of the bulge on the
heating rate, finding that a more massive bulge reduced the amount of
disk heating. Our analytical model typically reproduces this trend, with
approximately the same strength. 

To summarize, we are able to reproduce the rates of disk heating seen
in numerical simulations for the majority of the cases
considered. Where the analytic approach ``fails'' (we say ``fails'',
since the N-body techniques have their own inadequacies and do not
necessarily represent the correct solution), it underestimates the
heating by a factor of 3 on average. In many of the discrepant cases,
the incorrect heating rate is a consequence of an incorrect estimate
of the disk dynamical friction force or tidal mass loss rate, but in
some of the other discrepant cases, the reason is less obvious. It is
worth emphasizing that our analytic calculation reproduces several
important trends observed in the N-body heating rates. For example:
\begin{itemize}
\item heating is greatly suppressed for satellites on polar orbits;
\item differences between heating rates for prograde and retrograde orbits
(which are not always in the same sense, depending on the  satellite
type) are reproduced; 
\item differences due to the concentration or initial mass of the
satellite are clearly reproduced.
\end{itemize}
The trend of increased heating for more circular orbits, seen in
the N-body simulations, is not reproduced, however.

While it is clear that the analytic model does not match the N-body
heating rates perfectly, in the majority of cases, the differences are
compatible with the accuracy of the simulations themselves, as judged
by the convergence tests. We conclude that, in general, the analytic
model provides a reasonable approximation to the simulation results. 

\section{Results}
\label{sec:results}

\subsection{Scale-Height Distribution for Disk Galaxies}
\label{sec:field}

Having demonstrated that our model can be used to calculate disk
heating rates with reasonable accuracy, we now proceed to apply these
calculations to galaxy formation in a cosmological
setting. Specifically, we implement this model of disk heating in the
{\sc galform} semi-analytic model of galaxy formation described by
\scite{cole00} and \scite{benson02a}, based on a standard $\Lambda$CDM
cosmology with $\Omega_0=0.3$ and
$\Lambda_0=0.7$\footnote{\protect\scite{benson02c} describe small
changes in the parameters of this model, relative to those of
\protect\scite{benson02a}, which we retain here.}. This model follows
the growth of galactic disks in a merging hierarchy of dark matter
halos. At each time the model predicts the mass and radial size of the
galactic disk forming at the centre of each halo. It also gives the
rate at which subhalos are merging into each halo, which we take as
input for our calculations of satellite evolution and disk heating. We
assume that only direct progenitors of the halo cause heating
(i.e. subhalos can heat the disk, but sub-subhalos are not
considered). This is to avoid double-counting of mass. We will
consider briefly below the effect of allowing all progenitors to heat
disks.

\begin{figure*}
\psfig{file=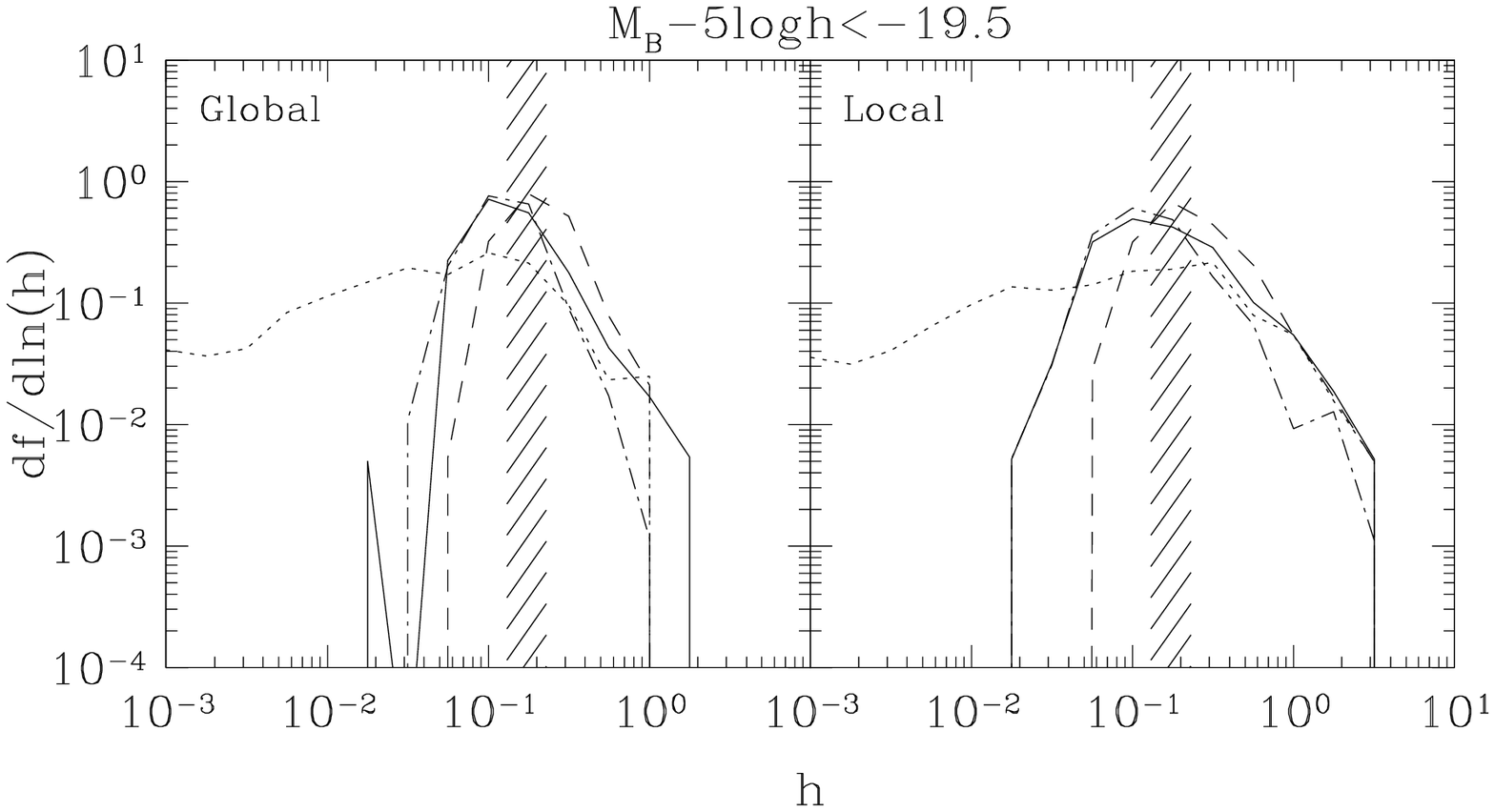,width=180mm,bbllx=0mm,bblly=165mm,bburx=190mm,bbury=270mm,clip=}
\caption{Normalized disk scale-height distributions for spiral
galaxies with $M_{\rm B}-5\log h\leq-19.5$. Left and right-hand panels
show results for global and local heating respectively. Solid lines
show results from our full calculation, including heating from
substructure and from scattering by molecular clouds. The dotted line
corresponds to ignoring the molecular cloud heating, while the dashed
line corresponds to increasing 
the masses of individual clouds and the total mass in clouds by a factor
of two over our standard values. These results correspond to satellite
orbits which are integrated until they reach a radius $(R_{\rm
\frac{1}{2}sat}+R_{\rm \frac{1}{2}host})/8$. For comparison, the
dot-dashed line shows the result when the integration is stopped when
a radius $(R_{\rm \frac{1}{2}sat}+R_{\rm \frac{1}{2}host})$ is
reached. The vertical shaded strip shows the observationally allowed
range for the scale-height of the Milky Way galaxy, discussed in
\S\protect\ref{sec:MW}.}
\label{fig:dhfieldbright}
\end{figure*}

\begin{figure*}
\psfig{file=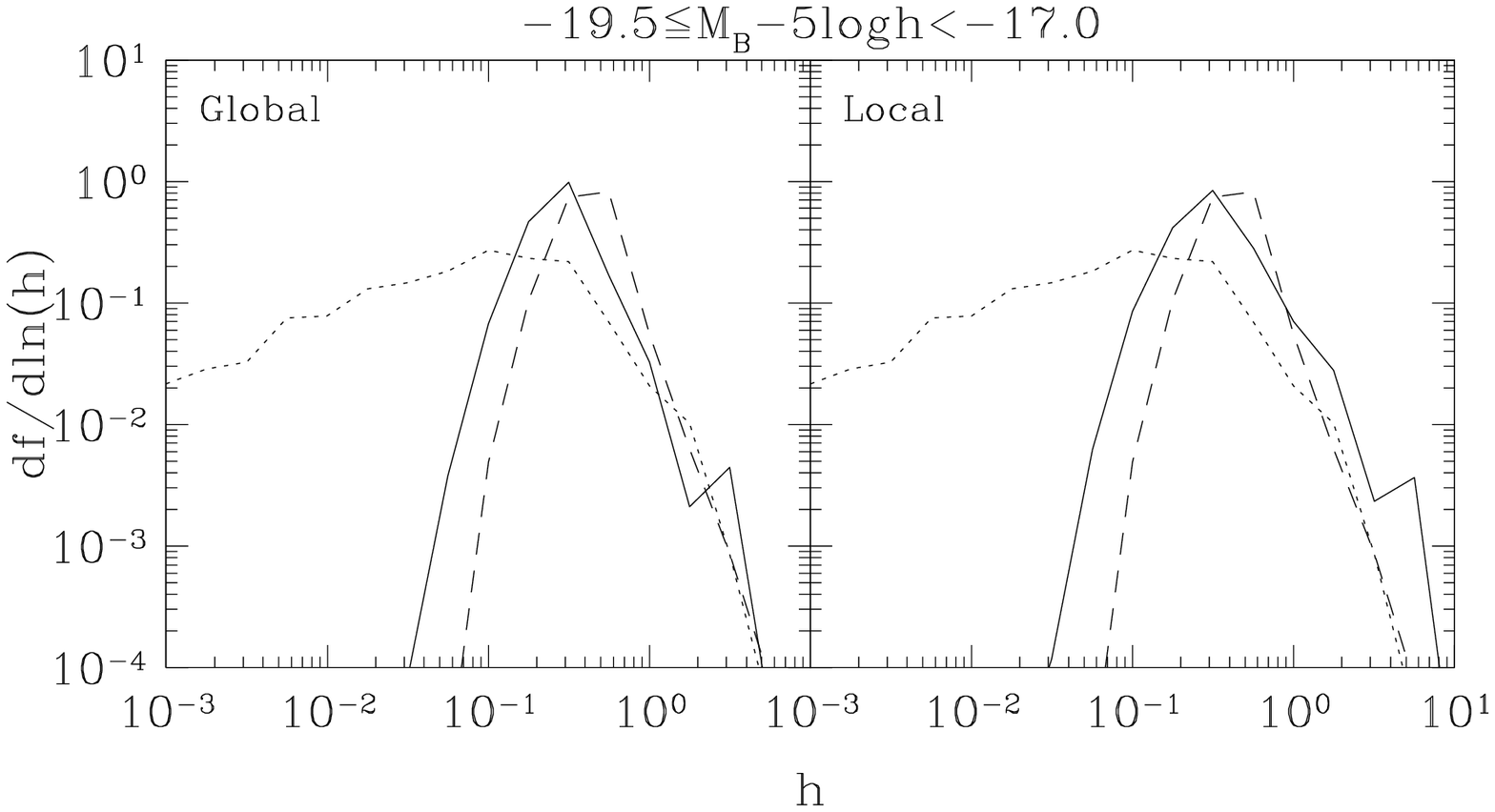,width=180mm,bbllx=0mm,bblly=165mm,bburx=190mm,bbury=270mm,clip=}
\caption{Normalized disk scale-height distributions for spiral
galaxies with $-19.5<M_{\rm B}-5\log h\leq-17.0$. Left and right-hand
panels show results for global and local heating respectively. Solid
lines show results from our full calculation, including heating from
substructure and from scattering by molecular clouds. The dotted line
corresponds to ignoring the molecular cloud heating, while the dashed
line corresponds to increasing the masses of individual clouds and the
total mass in clouds by a factor of two over our standard values.  }
\label{fig:dhfieldfaint}
\end{figure*}

Using this model, we generate a representative sample of galaxies
living in dark matter halos spanning a wide range of masses. For each
galaxy, the model computes the usual properties predicted by this type
of modeling (masses, luminosities, etc. ---see \pcite{cole00}), and
now also the vertical scale-height of the galactic
disk. Figures~\ref{fig:dhfieldbright} and \ref{fig:dhfieldfaint} show
the resulting distribution of disk scale-heights, expressed in units
of the disk radial scale-length, for galaxies with $M_{\rm B}-5\log
h\leq-19.5$ (approximately $L_*$ and brighter
galaxies) and $-19.5<M_{\rm B}-5\log h\leq-17.0$ respectively. We
include only spiral galaxies (which we define as
any galaxy with a bulge-to-total ratio measured in dust-extinguished
B-band light less than 0.4). 

We remind the reader that we define the dimensionless scale-height, 
$h=H_{\rm d}/R_{\rm d}$, in terms of the thickness parameter $H_{\rm
d}$ in the ${\rm sech}^2$ vertical density law and the radial
exponential scale-length, $R_{\rm d}$. The disk thickness can be
equivalently defined as $H_{\rm d} = \Sigma_{\rm d}/(2\rho_0)$, where 
$\Sigma_{\rm d}$ is the disk surface density and $\rho_0$ the
density at the midplane. However, many authors prefer to use the {\em
exponential scale-height} as the measure of disk thickness. Since
${\rm sech}^2(z/H_{\rm d}) \propto \exp(-2z/H_{\rm d})$ for $z \gg
H_{\rm d}$, the exponential scale-height that would be measured for our
assumed vertical profile is $H_{\rm d,exp} = H_{\rm d}/2$.

The left and right-hand panels in Figures~\ref{fig:dhfieldbright} and
\ref{fig:dhfieldfaint} show the scale-height distributions for the
global and local heating assumptions, from which we see that the
results are not very sensitive to this choice. The figures also show
the sensitivity of the results to two other parameters, one numerical
and the other physical.

The numerical parameter characterizes the galactocentric radius at which
the satellite is assumed to merge with the main galaxy and stop
heating the disk. In \scite{cole00} and in Paper~I, we assumed that
two galaxies merge at the time when the separation of their centres, 
$R$, equals the sum of their half-mass radii, $R_{\rm
\frac{1}{2}sat}+R_{\rm \frac{1}{2}host}$. However, once 
tidal stripping is taken into account, it would seem reasonable to allow the satellite to
sink down to $R=0$ and continue heating the disk while it does
so. However, for numerical reasons we cannot integrate the satellite
orbits down to $R=0$. We have therefore calculated the disk heating
when the satellite orbit is followed down to $R=f_{\rm heat}(R_{\rm
\frac{1}{2}sat}+R_{\rm \frac{1}{2}host})$ for $f_{\rm heat}=1$,
$\frac{1}{2}$, $\frac{1}{4}$ and $\frac{1}{8}$. We find that the
distribution of scale-heights has converged for $f_{\rm
heat}=\frac{1}{8}$, and use this as our standard value in what
follows. We show in Fig.~\ref{fig:dhfieldbright} results for $f_{\rm
heat}=1$ (dot-dashed lines) and $f_{\rm heat}=\frac{1}{8}$ (solid
lines) in both cases with disk heating by molecular clouds also
included. The differences in the scale-height distributions are fairly
small (they are somewhat more significant if we do not include disk
heating by molecular clouds).

The physical parameter concerns the heating of the disk that results
from scattering of stars by giant molecular clouds, computed using
eqn.~(\ref{eq:cloudheat}). Our standard calculation 
(solid lines in Figs.~\ref{fig:dhfieldbright} and
\ref{fig:dhfieldfaint}) includes heating by clouds with the
parameters described in \S\ref{sec:cloudheat}. The figures also show
the results when no clouds are present (dotted lines) and when the
masses of individual clouds and the fraction of gas in clouds are both
doubled (dashed lines). Removing the clouds entirely results in a tail to
very low $h$ in the height distribution. These galaxies experienced
very little heating by substructures, and so their thickness is
almost entirely due to heating by molecular cloud scattering. The peak
of the distribution is little changed, but the median scale-height is
significantly reduced (see Figure~\ref{fig:scmag}). Doubling the cloud
mass (dashed lines) results in a shift towards somewhat thicker disks
without changing the shape of the distribution.

Global and local heating are found to produce rather similar
distributions of scale-heights. Note that we have shown the results
for local heating at the disk half-mass radius. Our local heating
model, in fact, predicts a trend of increasing scale-height with disk
radius; we defer a detailed study of this radial variation to a future
paper.

It should be noted that the tails of the distributions extend to
$h>1$, which is clearly unphysical. Our analytical calculation is
based on the thin disk approximation, $h\ll1$, and so breaks down when
$h \sim 1$. We interpret these objects as disks that have been heated
so much that they are no longer disks, and must instead resemble a
spheroidal or irregular galaxy. For these galaxies, our calculations
break down, but we can safely assume that they are no longer
recognizable disk galaxies.

\subsection{Other Effects on the Scale-Height Distribution}

We now discuss tests of various potential systematic effects in our
calculations.

\emph{Merger Tree Resolution:} Our calculations typically resolve dark
matter substructures with mass greater than $5\times
10^9h^{-1}M_\odot$ in every merger tree. Thus, we ignore the heating
due to lower mass halos. Increasing the resolution of to
$10^9h^{-1}M_\odot$ results in no significant increase in the amount
of heating experienced by galaxies, indicating that our resolution is
sufficient to estimate the total heating rate. (Note that the heating
produced by a satellite of mass $M$ should scale approximately as
$M^2$, making it relatively easy to achieve convergence here provided
the number of satellites, $\d N/\d\ln M$, varies with mass less
steeply than $M^{-2}$ at small mass. In fact, numerical simulations
indicate $\d N/\d\ln M \sim M^{-0.7}$ for subhalos
\cite{springelclus}.)

\emph{Effects of Sub-subhalos:} In our standard calculation,
sub-subhalos (i.e. halos which reside inside a larger halo which
subsequently fell into a yet larger halo) do not contribute separately
to the heating of disks. (Note that this is different from our
treatment of galaxy mergers; the merging times of sub-subhalos are
computed from their own properties, not those of the subhalo in which
they reside.) An alternative approach would be to treat sub-subhalos
(and higher levels of the merging hierarchy) on an equal basis as
subhalos. To avoid double-counting of mass in this case, we must
remove the mass bound to sub-subhalos when determining the mass of a
subhalo. We do this by scaling down the density profile of the subhalo
so as to remove this amount of mass before computing heating rates.

If we adopt this approach, we find that the distribution of scale
heights is shifted to slightly larger values. Heating by sub-subhalos,
however, would only be important if these sub-subhalos survived after
their host had been tidally destroyed. While it is unlikely that this
would occur to any great extent, numerical simulations could, in
principle, answer this question.

\emph{Effect of Cosmological Model:} Finally, we have repeated our
calculations in an $\Omega_0=1$ cosmology, using the $\tau$CDM
parameter set used by \scite{cluster1}, but including the effects of
photoionization suppression\footnote{Note that
\protect\scite{cluster1} adopted an artificially high merger rate in
order to obtain a good match to the galaxy luminosity functions. With
our more detailed model of merging, we no longer have the freedom to
adjust the merger rate in this way. We find that, in this cosmology,
our revised merger model produces somewhat too few elliptical
galaxies.}. This model is not as successful at matching the properties
of $z=0$ galaxies as our standard $\Lambda$CDM model. In particular,
galaxies are somewhat too faint to match the observed luminosity
function (by about 0.75 magnitudes in the B-band), forcing us to adopt
an unphysical value of the mass-to-light ratio normalization
parameter, $\Upsilon$ of 0.7. We find that the median scale-height of
$L_*$ disk galaxies is slightly smaller in this cosmology than in our
$\Lambda$CDM model. At first sight, this seems surprising, since, as
noted by TO, there is more infall of substructure at late times in an
$\Omega_0=1$ cosmology, which would result in a larger rate of heating
at the present day. However, our model galaxies in this cosmology are
younger than their $\Lambda$CDM counterparts (due to the later growth
of structure and to the stronger feedback required in this model), and
so they have less time in which to be heated. These two effects
counteract each other.

The age of our Galactic disk has been estimated using studies of white
dwarfs. For example,
\scite{fontaine01} find an age of 11~Gyr for the Galactic disk. Since
they assume a constant star formation rate, this implies a mean stellar
age of 5.5~Gyr for the disk stars. Our $\tau$CDM disks typically have a
mean stellar age of 4~Gyr, somewhat less than the true
value. Consequently, our model underestimates the amount of heating
experienced by disks in this cosmology, albeit only by a small factor.

We can understand the similarity of the disk scale-heights in the two
cosmologies in more detail by examining the growth histories of the
dark matter halos hosting $L_*$ spiral galaxies. In our model, halos
of present-day mass $2\times 10^{12}h^{-1}M_\odot$ have, on average,
assembled half of their mass by redshifts of $0.45$ and $0.91$
respectively in the $\tau$CDM and $\Lambda$CDM cosmologies.  The mean
stellar ages of $L_*$ disk galaxies --- $4.0$ and $5.5$Gyr for
$\tau$CDM and $\Lambda$CDM respectively --- reflect this difference in
halo assembly epoch. We find that the host halos on average accrete
close to 25\% of their total mass over these galaxy lifetimes in both
cosmologies. Therefore, the number of substructures infalling onto a
galaxy over its lifetime is roughly the same in both cosmologies,
consistent with their similar disk scale-height distributions.  It
should be kept in mind that the disks in our $\Omega_0=1$ model are
somewhat unrealistic (e.g. they are too faint for a reasonable
$\Upsilon$ and, more importantly, too young). An $\Omega_0=1$ model
which produced realistic disks might predict larger (or smaller)
scale-heights. The important lesson to derive from these
considerations is that disk scale-heights depend on the details of
galaxy formation as well as on the cosmological model.

\subsection{Scale-Heights as a Function of Luminosity}

\begin{figure}
\psfig{file=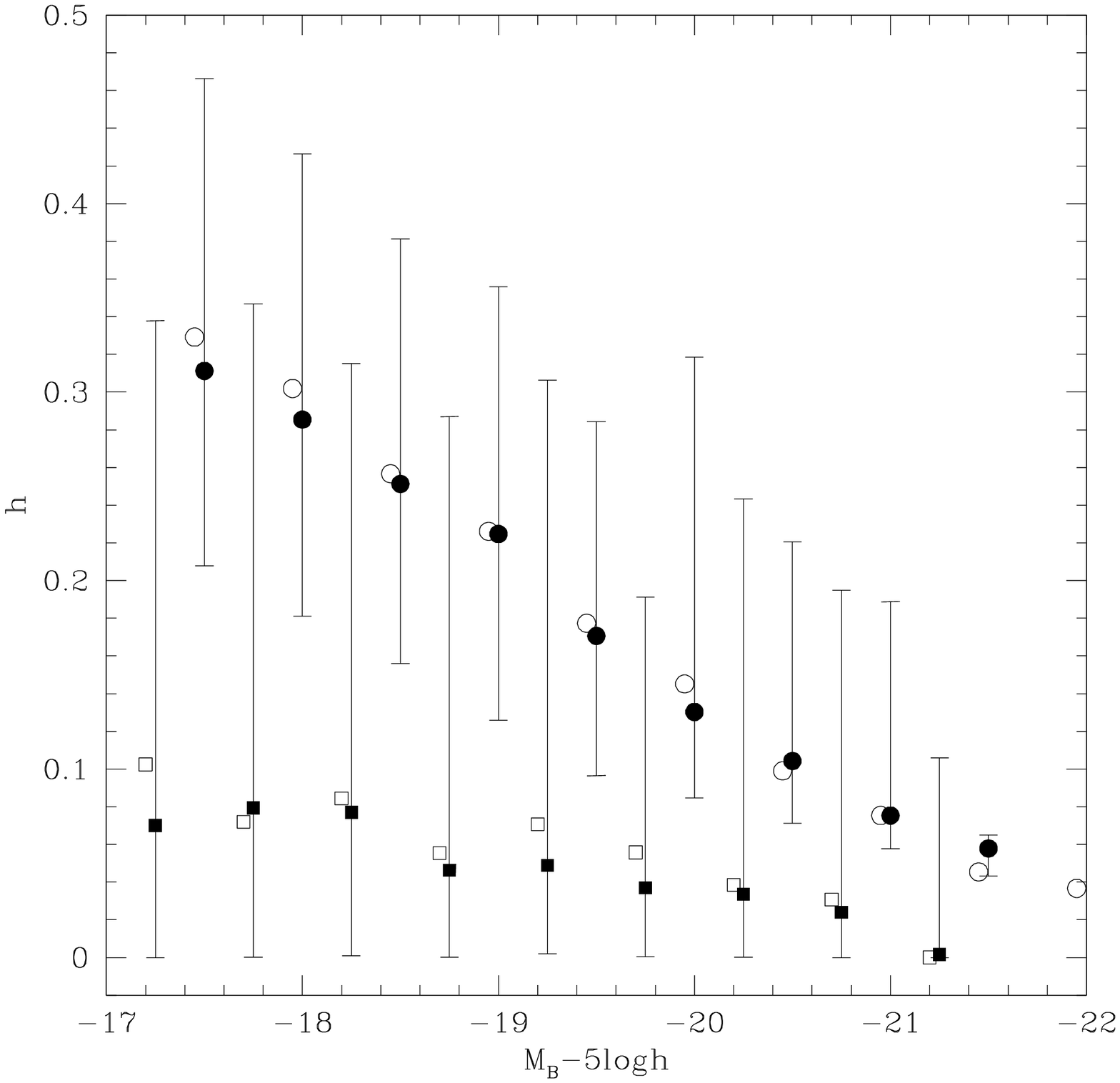,width=80mm}
\caption{The median fractional scale-heights, $h=H_{\rm d}/R_{\rm d}$, of
spiral galaxies as a function of absolute magnitude. The squares show the
results for heating by substructures only, and the circles 
for heating by substructures and clouds together. In each case, the
filled symbols are for global heating, and the open symbols (offset
slightly for clarity) for local heating. The errorbars indicate
the 10\% and 90\% intervals of the distribution of scale-heights. For
clarity, the errorbars are suppressed for the local heating case, but
are similar to those for global heating.}
\label{fig:scmag}
\end{figure}

In Fig.~\ref{fig:scmag}, we show the median value of $h$ as a function
of absolute magnitude for spiral galaxies in our standard model. The
squares show the results for heating by substructure alone, and the
circles for heating by substructure and clouds together. The median
scale-height at all luminosities is much larger when heating by clouds
is included. However, the scatter in scale-height at a given luminosity
is extremely large for the case of heating by substructures only, reflecting the
strongly stochastic nature of this process.  Our calculations predict
that brighter galaxies should host thinner disks than fainter galaxies
(when measured in terms of the fractional disk thickness, $h=H_{\rm d}/R_{\rm
d}$). This trend is apparent in calculations with and without
molecular clouds, and reflects a similar trend in the fractional
vertical energy, $E_{\rm z}/M_{\rm disk}V_{\rm disk}^2$.

\subsection{Comparison with the Milky Way galaxy}
\label{sec:MW}

It has been conventional to compare predictions for disk scale-heights
with the observed value for the Milky Way galaxy. As a way of testing
models against the real universe, this comparison has significant
drawbacks, since (i) the global parameters of the Milky Way (such as
the disk radial scale-length, total luminosity and bulge-to-disk
ratio) are, in fact, quite difficult to determine observationally and
(ii) the models predict a {\em distribution} of scale-heights at a
given luminosity, and this cannot be constrained well from a single
measured point. Therefore, we will make only a brief comparison with
the Milky Way here, before comparing with the distribution of
scale-heights measured for external galaxies. 

The vertical scale-height of the galactic disk in the Solar
neighbourhood has been measured from star counts. We use the recent
determination by \scite{guzman} which, for a ${\rm sech}^2(z/H_{\rm
d})$ vertical profile, gives $H_{\rm d}=0.50\pm0.08$kpc (corresponding
to an exponential scale-height of $0.25\pm0.04$kpc), somewhat smaller
than earlier determinations. The measurement of the radial exponential
scale-length of the galactic disk has, in the past, been a matter of more
disagreement. We use the models of the galactic mass distribution by
\scite{dehnen}, which imply $R_{\rm d}=3.0\pm0.4$kpc. Combining these,
we find the fractional scale-height of the Milky Way stellar disk, 
$h=H_{\rm d}/R_{\rm d} = 0.18\pm0.05$. This range in $h$ is indicated as
a shaded region in Figure~\ref{fig:dhfieldbright}, from which one can
see that the scale-height of the Milky Way is entirely typical of $L_*$
disk galaxies in the model (with 35\% of galaxies predicted to have
$h>0.18$). We have repeated the comparison using the same definition
of ``Milky Way-like'' galaxies as in
\scite{benson02b}, namely a circular velocity at the disk half-mass
radius between 210--230km/s and a bulge-to-total ratio by mass between
5--20\%. We again find that the observed scale-height of the Milky Way
lies well within the distribution of $h$ predicted by the model (with
80\% of such galaxies predicted to have $h>0.18$).

\subsection{Comparison with the Observed Scale-Height Distribution for
  Other Galaxies}
\label{sec:obsdist}

The best way to test models of disk heating is by comparing with the
observed distribution of scale-heights for external galaxies. This
distribution has recently been measured in a complete sample of disk
galaxies, for the first time, by
\scite{bizyaev02}.  They estimated the vertical and radial
scale-lengths of a statistically complete sample of 60 edge-on
galaxies using K-band photometry from the 2MASS survey. 

\begin{figure}
\psfig{file=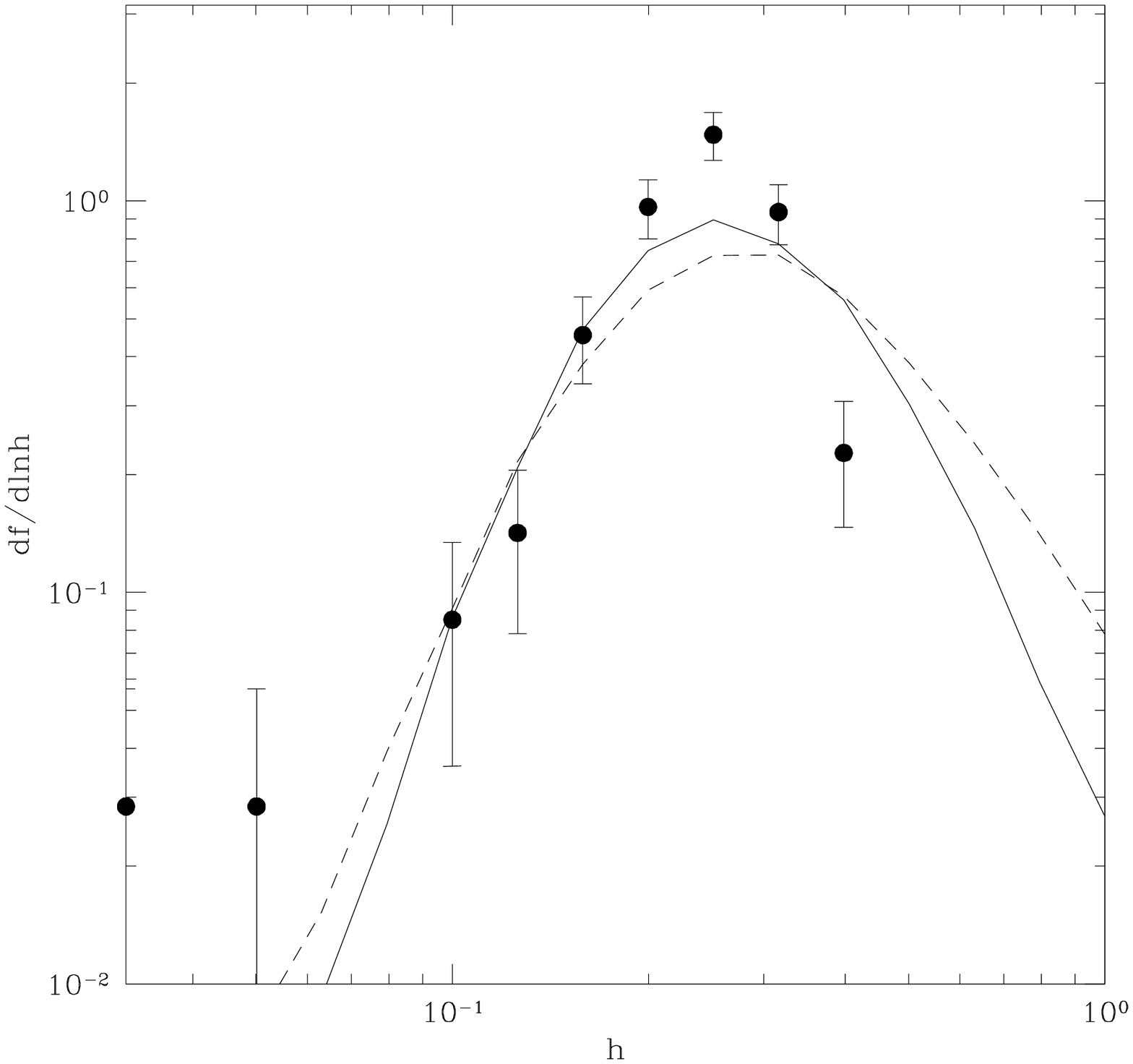,width=80mm}
\caption{The normalized distribution of scale-heights, $h$, in the
observational sample of \protect\scite{bizyaev02} compared to the
prediction of our model. Error bars on the observational datapoints
indicate Poisson errors. The model predictions are shown by solid
and dashed lines for global and local heating respectively. The model
galaxies have been weighted to match the distribution of absolute
magnitudes and morphological types found in the observational sample.}
\label{fig:obsh}
\end{figure}

We compare the scale-height distribution in \scite{bizyaev02} sample
with our model predictions in Figure~\ref{fig:obsh}. Since the
selection criteria for the observational sample are somewhat complex,
we weight model galaxies so as to match the distribution of absolute
magnitudes in the observational sample (which peaks in the range
$-19<M_{\rm B}-5\log h\leq -18$), and select only those galaxies with
bulge-to-total luminosity ratios typical of the morphological types
found in the observational sample (which are mostly Sc spirals). We
see that the model provides quite a good match to the observed
distribution, with both distributions peaking around $h=0.2$. The only
significant discrepancy is that the model predicts too many systems
with large $h\gsim 0.4$. However, it is not clear that such thick
galaxies would be recognized as disk galaxies. The conclusions that
can be drawn at present are limited by the relatively small size of
the current observational sample. However, this situation should soon
improve with the availability of data from large CCD-based sky
surveys, which will allow much more thorough tests of the theoretical
predictions.

\section{Discussion}
\label{sec:disc}

We have developed a model to calculate the rate of heating of galactic
disks by substructures orbiting in their halos. To calibrate the
model, we performed N-body simulations of disk heating which we tested
for convergence. We find that the analytical model reproduces the
heating rates in the N-body simulations to within a factor 3 in most
cases. One could perhaps improve the accuracy of the analytical model
by treating the satellite--disk interaction in terms of a sum of
interactions with resonances in the disk
(e.g. \pcite{donner93,weinberg02}).  It is unclear, however, whether
such a calculation would ever be worth performing semi-analytically,
i.e. whether its computational cost would be any less than that of a
full N-body simulation. Nevertheless, it is clear from the
calculations presented here that N-body estimates of disk heating
rates have their own problems (e.g. very large numbers of particles
are required in the disk to determine the heating rate accurately),
and so it may yet prove worthwhile to pursue analytical estimates of
disk heating.

We find that for galaxy formation in the standard $\Lambda$CDM
cosmology, heating by substructure alone produces a distribution of
disk scale-heights which is very broad and skewed to low values, with
median fractional scale-height, $h=H_{\rm d}/R_{\rm d}$, around 0.05
for $L_*$ spiral galaxies. The width of the distribution reflects the
stochastic nature of the heating process, which is, in turn, related
to the distribution of orbital parameters of the satellites. Including
the additional heating generated by stars scattering from gas clouds
in the disk increases the median value of $h$ significantly, to around
0.2. The distribution is considerably less broad once the contribution
from gas cloud heating is included. Heating by clouds is treated as a
deterministic process here, with variations in the amount of heating
for a given type of galaxy reflecting the distribution of ages of
galactic disks. The fractional scale-height for the Milky Way galaxy,
estimated observationally to be around 0.2, is then entirely
consistent with our model expectations for a typical $L_*$ spiral
galaxy. We find that the predicted distribution of scale-heights for
slightly sub-$L_*$ spiral galaxies agrees remarkably well with a
recent observational determination by \scite{bizyaev02} based on data
from the 2MASS survey.

It is intriguing that for the $L_*$ galaxies considered here,
satellites and gas clouds give rise to comparable amounts of disk
heating. A simple order of magnitude estimate of the scale-heights
produced by these two processes illustrates why this is so. Using the
expressions given in this paper, we find (for parameter values typical
of Milky Way-like galaxies) that the fractional scale-height generated by
scattering from giant molecular clouds is 
\begin{eqnarray}
h & = & 7.2 \times 10^{-3} \left[ {f \over 0.025} \right]^{1/2} \left[ {M_{\rm c}/M_{\rm d} \over 3 \times 10^{-5}} \right]^{1/2} \left[ {\ln \Lambda \over 3} \right]^{1/2} \nonumber \\
 & & \times \left[ {\nu \over 90\hbox{ Gyr}^{-1}} \right]^{1/2} \left[ {\alpha_{\rm S}(\beta) \over 0.7} \right]^{3/2} \left[ {K_{\rm S}(\beta) \over 0.15} \right]^{1/2} \left[ {t \over \hbox{Gyr}} \right]^{1/2},
\end{eqnarray}
while that generated  by dark matter substructures is 
\begin{eqnarray}
h & = & 0.16 \left[ {f_{\rm mass} \over 0.1} \right] \left[ {f_{\rm max} \over 0.01} \right] \left[ {\epsilon_{\rm z} \over 0.3} \right] \left[ {M_{\rm halo}\over 10^{12}M_\odot}\right]^2 \left[ {\ln \Lambda \over 3}\right] \nonumber \\
 & & \times \left[ {V_{\rm s} \over 200\hbox{ km/s}} \right]^{-1} \left[ {R_{\rm d}  \over 3.5\hbox{ kpc}} \right]^{-1}  \left[ {r_{\rm orb} \over 200\hbox{ kpc}} \right]^{-1} \nonumber \\
 & & \times \left[ {M_{\rm d} \over 5\times 10^{10}M_\odot} \right]^{-1} \left[ { t \over \hbox{Gyr}} \right]
\end{eqnarray}
In the first equation, $f$ is the fraction of the total disk mass in
the form of giant molecular clouds. In the second equation, $V_{\rm
s}$ is the typical orbital velocity of satellites, $r_{\rm orb}$ their
typical orbital radius, $f_{\rm mass}$ is the fraction of the total
halo mass in the form of substructures, $f_{\rm max}$ is the mass of
the largest substructure in units of the total halo mass and we have
assumed a distribution of substructure masses ${\rm d}N/{\rm d}M
\propto M^{-1.7}$. In both cases $t$ is the time for which heating has
occurred. To derive the second expression, we have assumed that
substructures heat the disk only over a fraction of their orbit
approximately equal to $H_{\rm d}/r_{\rm orb}$. Taking $t\approx
10$~Gyr, these estimates imply $h\sim 0.1--1$ for both heating
mechanisms, confirming the coincidence that the two contribute
approximately equally to the scale-height of Milky Way-like disks
(given the crude approximations made above and the fact that we have
ignored the stochastic nature of heating by satellites). However,
these two expressions have different dependencies upon the properties
of the galaxies in question. Thus, we should not expect the two to
make equal contributions to the scale-height of galaxies dissimilar to
the Milky Way. This may be seen in Fig.~\ref{fig:scmag}, where it is
clear that the heating by substructures is relatively less important
for lower luminosity galaxies. In conclusion, the fact that the two
heating mechanisms make similar contributions to the scale-heights of
Milky Way-like galaxies appears to be coincidental.

\begin{figure}
\psfig{file=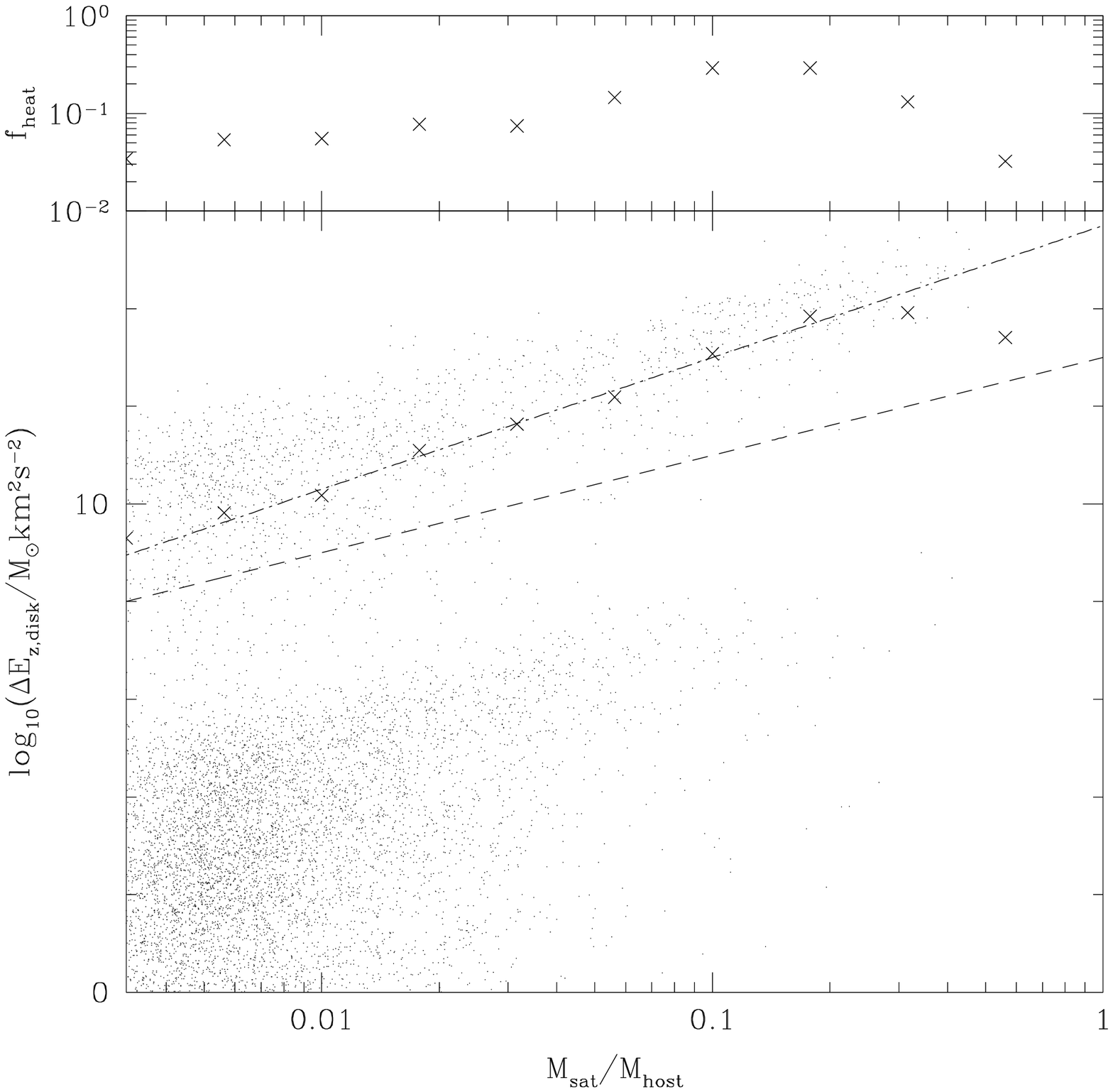,width=80mm}
\caption{The energy contributed to disk heating by satellites as a
function of their mass in shown in the lower panel. Points show the
results for a large sample of satellites, for which we plot the
heating energy supplied ($\Delta E_{\rm z,disk}$) against the mass of
the satellite (expressed in units of the mass of the host system
halo). At fixed mass, the distribution shows a bimodal form, the
dashed line indicates the approximate division between the two peaks
of the distribution. Crosses indicate the mean heating energy per
satellite at each mass, and the dot-dashed line shows an approximate
fit to these points. In the upper panel, we show the fraction of
points at each mass which lie above the dashed line in the lower
panel.}
\label{fig:energyrep}
\end{figure}

It is interesting to compare our conclusions with those of TO, who
found that the Milky Way disk could have accreted only up to 5\% of
its mass within the Solar circle within the past 5\,Gyr without
becoming too thick. Our calculations show that the Milky Way halo in
fact accreted around 25\% of \emph{its} mass (i.e. the total dark mass
of the halo) during this time. This is approximately $100$ times more
than the TO limit. Disks in our model are able to remain fairly thin
despite this substantial accretion for two reasons. Firstly, many of
the accreted subhalos have orbits which do not take them close to the
central galaxy disk, and so they contribute almost nothing to the
heating of the disk. Figure~\ref{fig:energyrep} shows the amount of
energy transferred to the disk of a Milky Way-like galaxy by
individual dark matter satellites as a function of their mass. At a
fixed mass, the distribution of heating energies has a bimodal
distribution. Virtually all of the heating energy is supplied by the
satellites in the upper branch which are those whose orbits take them
close to the central galaxy disk. These satellites are ``trapped'' by
dynamical friction and damage the disk during an extended period;
distant satellites, on the other hand, have a negligible effect.  The
satellites that cause most of the heating amount to only 6\% by mass.
Thus, of the $5\times 10^{11}M_\odot$ infalling, only around $3 \times
10^{10}M_\odot$ contribute to heating the disk. This is still 6 times
larger than the TO limit ($0.5 \times 10^{10}M_\odot$).

The second difference with TO is that we find that tidal mass loss in
subhalos substantially reduces the amount of heating experienced by
the disk. This is in disagreement with TO, who found that tidal mass
loss reduced disk scale-heights by at most a factor of two.  If we do
not allow satellites to lose mass, the peak of the scale-height
distribution is shifted to a value of $h$ which is approximately ten
times larger than our standard result.  This, of course, reflects the
different density profiles that we assign to both the host and
satellite halos (and which are significantly more extended than the
objects considered by TO), and the associated increase in the
dynamical friction timescale in our model. In conclusion, our results
are in partial agreement with those of TO---halos in our
model accrete much more mass in the past 5\,Gyr than the TO limit, but
little of this mass ever contributes to heating the disk.

There is clearly a need for further study of the heating of galactic
disks. In particular, the importance of heating by satellite-triggered
bars and the extent to which heating is local or global are important,
yet poorly understood aspects of the problem. We believe that
analytical modeling of the type developed in this paper provides a
powerful means by which to estimate the degree of heating by
substructures and could easily incorporate any improvements in our
understanding of the physics of the process. Its particular strengths
are the ability to resolve fully all substructures contributing to the
heating and to compute many realizations of the heating process
rapidly, thus allowing the full distribution of scale-heights to be
determined. These features have allowed us to present predicted
distributions of galaxy scale-heights which will be tested by
forthcoming observational data.

In conclusion, the observed thickness of the Milky Way's stellar disk
seems to be entirely consistent with the amount of substructure in
galactic halos expected in a cold dark matter universe. Stars
scattering from giant molecular clouds and substructures passing
through or near the disk produce similar amounts of
heating. Distinguishing these two contributions observationally might
be possible by means of the stellar age-velocity dispersion relation
in the Milky Way disk. An important extension of this work will
therefore be to examine model predictions for heating as a function of
time within individual galaxies. The lowest values of $h$ for disk
galaxies are set by the heating due to star-cloud interactions, while
the highest values are set by the heating due to substructures. Thus,
precise measurements of the disk scale-height distribution can
potentially constrain these two processes.

\section*{Acknowledgments}

We thank Peter Goldreich, Chung-Pei Ma, Milos Milosavljevic and Simon
White for enlightening discussions, James Taylor for providing results
from his satellite orbit calculations in electronic form, Volker
Springel for making his {\sc gadget} N-body code publically available
and the anonymous referee for many valuable suggestions. AJB
acknowledges the hospitality of the University of Durham and the Kavli
Institute for Theoretical Physics where much of this work was
completed. AJB and CMB acknowledge receipt of Royal Society University
Research Fellowships. CGL was supported at Durham by the PPARC rolling
grant in Cosmology and Extragalactic Astronomy. This research was
supported in part by the National Science Foundation under Grant
No.~PHY99-07949.

\appendix

\section{Improvements in the Satellite Evolution Model}
\label{sec:improve}

We detail here the changes in the model of satellite evolution
presented in Paper~I.

i) As before, that mass of a satellite which has become unbound due to
tidal forces is lost gradually over a time comparable to the orbital
period. The fraction of the unbound mass lost in a small timestep of
duration $\delta t$ was chosen to be proportional to $\delta t/t_{\rm
orb}$, where $t_{\rm orb}$ is an estimate of the orbital timescale. In
Paper~I, we chose $t_{\rm orb}=2 \pi/\omega$, where $\omega$ is the
instantaneous angular velocity of the satellite. In the present work, 
we instead take $t_{\rm orb}=f_{\rm orb} 2 \pi /\sqrt{\omega_{\rm
peri}\omega_{\rm apo}}$, where $\omega_{\rm peri}$ and $\omega_{\rm
apo}$ are the angular velocity of the satellite at its most recent
pericentric and apocentric passages, and $f_{\rm orb}$ is an
adjustable parameter which we expect to be of order unity. (Prior to
the first pericentric passage we revert to our previous definition of
mass loss timescale --- this makes little difference to our results as
typically very little mass loss occurs prior to this time.) The
advantage of this choice is that it produces smoother mass loss
histories (as is shown in \S\ref{sec:results}). Furthermore, when
considering cosmological distributions of satellites, we occasionally
find orbits which are near-radial. The above definition then prevents
the mass loss rate from becoming arbitrarily small.

ii) In Chandrasekhar's formula for the dynamical friction force,
\scite{taybab} adopted fixed Coulomb logarithms of $\ln \Lambda = 2.4$
for the dynamical friction force due to the combined halo/bulge system
and $\ln \Lambda = 0.5$ for the force due to the disk. They found that
these values resulted in the best match to the results of the
numerical simulations of \ncite{VW}~(1999; hereafter VW), and we
adopted the same values in Paper~I. Since we will be interested here
in a wide range of satellite and host halo masses, we adopt more
general definitions. For the halo and bulge systems we take
$\Lambda_{\rm h} = f_{\Lambda, \rm h} r (v_{\rm sat}^2+\sigma_{\rm
1D}^2)/\G M_{\rm sat}$, where $r$ is the orbital radius of the
satellite, $M_{\rm sat}$ its mass, $v_{\rm sat}$ the orbital velocity
of the satellite, $\sigma_{\rm 1D}$ the one dimensional velocity
dispersion of the halo at radius $r$, and $f_{\Lambda, \rm h}$ is a
parameter. Since $\Lambda_{\rm h} \leq 1$ is possible with this
definition, we replace the usual $\ln \Lambda_{\rm h}$ term in the
expression for the dynamical friction force (eqn.~20 of Paper~I) with
$\frac{1}{2}\ln (1+\Lambda_{\rm h}^2)$, the correct form for small
$\Lambda_{\rm h}$ \cite{bintrem}. We also account for the finite size
of the satellite as described in Appendix~\ref{app:rcv}. For the disk
we must account for the differing scale-lengths in the radial and vertical
directions. A suitable expression for the Coulomb logarithm is derived
in Appendix~\ref{app:couldisk}, and depends on the disk scale-length
and velocity dispersions, the velocity of the satellite relative to
the disk, the angle this velocity makes with the disk plane, and on
a parameter, $f_{\rm \Lambda ,d}$ which plays a similar role to
$f_{\Lambda, \rm h}$. These forms are used throughout our
calculations.

iii) The disk is now treated as having an anisotropic velocity
dispersion $(\sigma_{\rm R},\sigma_\phi,\sigma_{\rm z})$ in the
radial, azimuthal and vertical directions, and this anisotropy is
included in the calculation of the dynamical friction force due to the
disk (see Appendix~\ref{app:anisdf}). We adopt essentially the same
model for the disk velocity dispersion components as VW. For the
radial velocity dispersions, we set $\sigma_{\rm R}^2 \propto
\exp(-R/R_{\rm d})$ \cite{lewis89}\footnote{Note that VW contains an
error in this equation, although the text of that paper is correct.},
where $R_{\rm d}$ is the disk radial scale-length, and fix the
normalization by assuming the disk to have a Toomre $Q$-parameter of
$1.5$ at its half-mass radius, which results in $Q\approx 1.5$ at the
Solar radius in a Milky Way-like galaxy disk (VW). The azimuthal
velocity dispersion is then determined using the epicyclic
approximation, $\sigma_\phi^2=\sigma_{\rm R}^2\kappa^2/4\Omega^2$
(where $\kappa$ is the epicyclic frequency and $\Omega$ the orbital
frequency of the disk). The vertical velocity dispersion at each
radius is calculated from the vertical scale-height $H_{\rm d}$, assumed
constant with radius, using the expressions in \S\ref{sec:sch} (the
vertical scale-height in turn is related to the disk vertical energy).
In the analytical disk-heating calculation, the radial and azimuthal
velocity dispersions are kept fixed in time, but the vertical velocity
dispersion evolves with the disk vertical energy.

iv) When computing the dynamical friction force due to the disk, we
 smooth the disk density to account for the finite size of the
 satellite halo as did \scite{taybab}. We smooth on a scale
 equal to the current radius of the satellite after tidal limitation
 and gravitational shock-heating.

v) As the disk scale-height will increase as a function of time due to
disk heating, we allow for a variable disk scale-height in our
satellite orbit calculations. This affects both the dynamical friction
force due to the disk and also the gravitational forces exerted by the
disk.

vi) Heating by gravitational shocks causes shells of material within a
satellite to expand before they become completely unbound. Previously,
this effect was included in the calculation of the tidal mass loss,
but not in the calculation of the final internal structure. We now
calculate the evolution of the internal density and circular velocity
profile assuming that the radii of shells of dark matter scale in
inverse proportion to their energy. We have repeated the comparison we
performed in Paper~I of the distribution of peak internal circular
velocities of satellite halos predicted by the semi-analytical model
with the results of cosmological N-body simulations. We find that the
same choice of initial satellite orbital parameters as in Paper~I
still gives the best match to the N-body simulations.

\section{Dynamical Friction Formulae}
\label{sec:dynfric}

In this appendix we derive several formulae related to dynamical
friction which are employed in this work. For completeness, in
\S\ref{sec:sse} and \S\ref{app:rcv} we derive several well-known
relations relevant to dynamical friction. A more complete discussion
of these results can be found in \scite{bintrem} for example. We
consider a mass $M$ moving through an infinite and homogeneous sea of
particles of mass $m(\ll M)$, number density $n$ and density
$\rho=mn$.

\subsection{Single Scattering Events}
\label{sec:sse}

\begin{figure}
\psfig{file=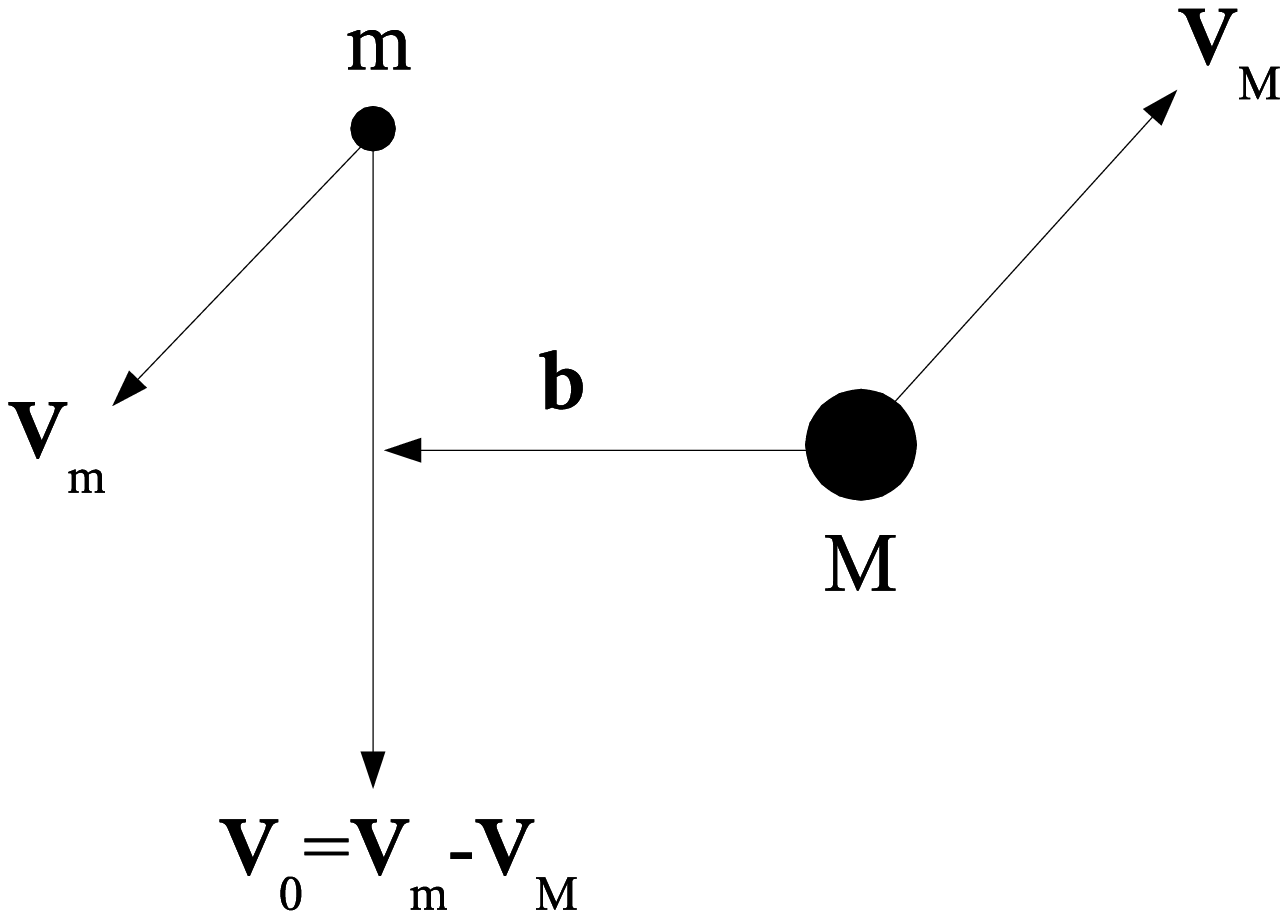,width=80mm}
\caption{The geometry of a scattering event contributing to the
dynamical friction force on mass $M$ (the satellite) due to a particle
of mass $m$ (a background particle). Here, $V_{\rm M}$ and $v_{\rm m}$
are the velocities of $M$ and $m$ respectively, while $V_0$ is the
relative velocity of the two and $b$ is the impact parameter for this
scattering event.}
\label{fig:DFgeom1}
\end{figure}

For a single scattering event we take the results of Binney \&
Tremaine (1987; page~422). The scattering geometry is illustrated in
Fig.~\ref{fig:DFgeom1}. The changes in $M$'s velocity parallel and
perpendicular to the initial relative velocity vector of the $m$ and
$M$, ${\bf V}_0$, are:
\begin{equation}
\Delta {\bf V}_{||} = {2 m \over M}\left[ 1 + {b^2 V_0^4 \over \G^2M^2} \right]^{-1} {\bf V_0},
\end{equation}
and
\begin{equation}
\Delta {\bf V}_\perp = {2 m V_0^3 \over \G M^2}\left[ 1 + {b^2 V_0^4
\over \G^2M^2} \right]^{-1} {\bf b},
\end{equation}
where $b$ is the impact parameter and we have assumed the background
particles to be much less massive than the object for which the force
is being calculated.

\subsection{Rate of Change of Velocity}
\label{app:rcv}

We now envisage a sea of particles $m$ with a distribution of
velocities given by $f({\bf V}_{\rm m})$. The contribution to the rate
of change of velocity in the parallel direction from particles with
velocity ${\bf V}_{\rm m}$ is simply
\begin{equation}
{\d {\bf V}_{||} \over \d t} = f({\bf V}_{\rm m}) \int_0^{b_{\rm max}}
2 \pi b n {\bf V_0} \Delta V_{||} \d b.
\label{eq:bmaxint}
\end{equation}
This gives,
\begin{equation}
{\d V_{||} \over \d t} = 2 \pi \ln (1+\Lambda^2) \rho \G^2 M f({\bf
V}_{\rm m}) {{\bf V_0} \over V_0^3},
\end{equation}
where $\Lambda = b_{\rm max}V_0^2/\G M$. If $M$ has a finite extent
(corresponding to replacing the lower integration limit of $0$ with
$b_{\rm min}$), the above equation still holds with an effective
$\Lambda$ given by
\begin{equation}
\Lambda_{\rm eff} = \left( {1+\Lambda^2 \over 1 + [b_{\rm min}/b_{\rm
max}]^2 \Lambda^2} - 1 \right)^{1/2}.
\end{equation}
Throughout this work, we take $b_{\rm min}$ equal to half the current
tidal radius of the satellite.

Clearly, the net change in the velocity of $M$ perpendicular to ${\bf
V}_0$ is zero by symmetry. Thus, the net rate of change of velocity of
$M$ is
\begin{equation}
{\d {\bf V}_{\rm M} \over \d t} = 2 \pi \ln (1+\Lambda^2) \rho \G^2 M
\int f({\bf V}_{\rm m}) {({\bf V}_{\rm m} - {\bf V}_{\rm M}) \over
({\bf V}_{\rm m} - {\bf V}_{\rm M})^3} \d^3 {\bf V}_{\rm m}.
\label{eq:chandra}
\end{equation}
The integral in the above equation has an identical form to integrals
used to find the gravitational force at position $x_0$ due to a
density distribution, if we identify $f({\bf V}_{\rm m})\equiv \G
\rho({\bf x})$, ${\bf V}_{\rm m}\equiv{\bf x}$ and ${\bf V}_{\rm
M}\equiv {\bf x}_0$.

Thus, the power extracted from the body through dynamical friction is
given by,
\begin{eqnarray}
P_{\rm scat} & = & M {\bf V}_{\rm M}\cdot {\d {\bf V}_{\rm M} \over \d t} \nonumber \\ & = & 2 \pi \ln (1+\Lambda^2) \rho \G^2 M^2 \nonumber \\
 & &  {\bf V}_{\rm M} \cdot \int f({\bf V}_{\rm m}) {({\bf V}_{\rm m} - {\bf V}_{\rm M}) \over ({\bf V}_{\rm m} - {\bf V}_{\rm M})^3} \d^3 {\bf V}_{\rm m}.
\end{eqnarray}

\subsubsection{Application to an Arbitrary Velocity Ellipsoid}
\label{app:anisdf}

\scite{binney77} derives an expression for the dynamical friction
force due to a system of particles with uniform density and Gaussian
velocity distribution with dispersion $\sigma_{\rm \perp}$ in one
direction and $\sigma_{\rm ||}$ in the other two directions. Binney's
equation (A4) is trivially generalized to the case where the velocity
dispersions differ in all three directions. Combining this with his
equation (A3) we find the following expression for the dynamical
friction force:
\begin{eqnarray}
{\bf F}_{\rm df} & = & \sqrt{2\pi} \ln (1+\Lambda^2) \rho \G^2 M^2 {\sqrt{(1-e_\phi^2)(1-e_{\rm z}^2)}\over \sigma_{\rm R}\sigma_\phi \sigma_{\rm z}} \nonumber \\
 & & \times (B_{\rm R}v_{\rm R}{\bf \hat{e}}_{\rm R} + B_\phi v_\phi{\bf \hat{e}}_\phi + B_{\rm z}v_{\rm z}{\bf \hat{e}}_{\rm z})
\end{eqnarray}
where $\rho$ is the background density, $M$ the mass of the orbiting
object, $(v_{\rm R},v_\phi,v_{\rm z})$ is the relative velocity vector
of object and background particles (in cylindrical polar coordinates
since we will apply this expression to a galaxy disk), ${\bf
\hat{e}}_{\rm R}$, ${\bf \hat{e}}_\phi$, ${\bf \hat{e}}_{\rm z}$ are
the basis vectors of the cylindrical polar coordinate system. The
coefficients $B$ are given by,
\begin{eqnarray}
B_{\rm R} & = & \int_0^\infty {\d q \over [(1+q)^3 (1-e^2_\phi+q) (1-e^2_{\rm z}+q)]^{1/2} } \nonumber \\
 & & \times \exp\left( -{1 \over 2} \left[ {v_{\rm R}^2/\sigma_{\rm R}^2 \over (1+q)} + {v_\phi^2/\sigma_{\rm R}^2 \over (1-e^2_\phi+q)} \right. \right. \nonumber \\
 & & \left. \left. + {v_{\rm z}^2/\sigma_{\rm R}^2 \over (1-e^2_{\rm z}+q)} \right] \right), \\
B_\phi & = & \int_0^\infty {\d q \over [(1+q) (1-e^2_\phi+q)^3 (1-e^2_{\rm z}+q)]^{1/2} } \nonumber \\
 & & \times \exp\left( -{1 \over 2} \left[ {v_{\rm R}^2/\sigma_{\rm R}^2 \over (1+q)} + {v_\phi^2/\sigma_{\rm R}^2 \over (1-e^2_\phi+q)} \right. \right. \nonumber \\
 & & \left. \left. + {v_{\rm z}^2/\sigma_{\rm R}^2 \over (1-e^2_{\rm z}+q)} \right] \right), \\
B_{\rm z} & = & \int_0^\infty {\d q \over [(1+q) (1-e^2_\phi+q) (1-e^2_{\rm z}+q)^3]^{1/2} } \nonumber \\
 & & \times \exp\left( -{1 \over 2} \left[ {v_{\rm R}^2/\sigma_{\rm R}^2 \over (1+q)} + {v_\phi^2/\sigma_{\rm R}^2 \over (1-e^2_\phi+q)} \right. \right. \nonumber \\
 & & \left. \left. + {v_{\rm z}^2/\sigma_{\rm R}^2 \over (1-e^2_{\rm z}+q)} \right] \right),
\end{eqnarray}
where $1-e^2_\phi=\sigma^2_\phi/\sigma^2_{\rm R}$ and $1-e^2_{\rm
z}=\sigma^2_{\rm z}/\sigma^2_{\rm R}$.

\subsubsection{Effective Coulomb Logarithm for the Disk}
\label{app:couldisk}

In calculating the dynamical friction force due to the disk we require
the Coulomb logarithm, $\frac{1}{2}\ln(1+\Lambda^2)$, where $\Lambda$
is normally defined as $\Lambda = b_{\rm max} V_0^2/\G M$, where $V_0$
is the typical relative velocity of the satellite and stars in the
disk. We adopt $V_0^2=V_{\rm rel}^2 + (\sigma_{\rm
R}^2+\sigma_\phi^2+\sigma_{\rm z}^2)/3$, where $V_{\rm rel}$ is the
relative velocity of the satellite and the bulk disk motion, and
$\sigma_{\rm R}$, $\sigma_\phi$ and $\sigma_{\rm z}$ are the three
components of the disk velocity dispersion.

\begin{figure}
\psfig{file=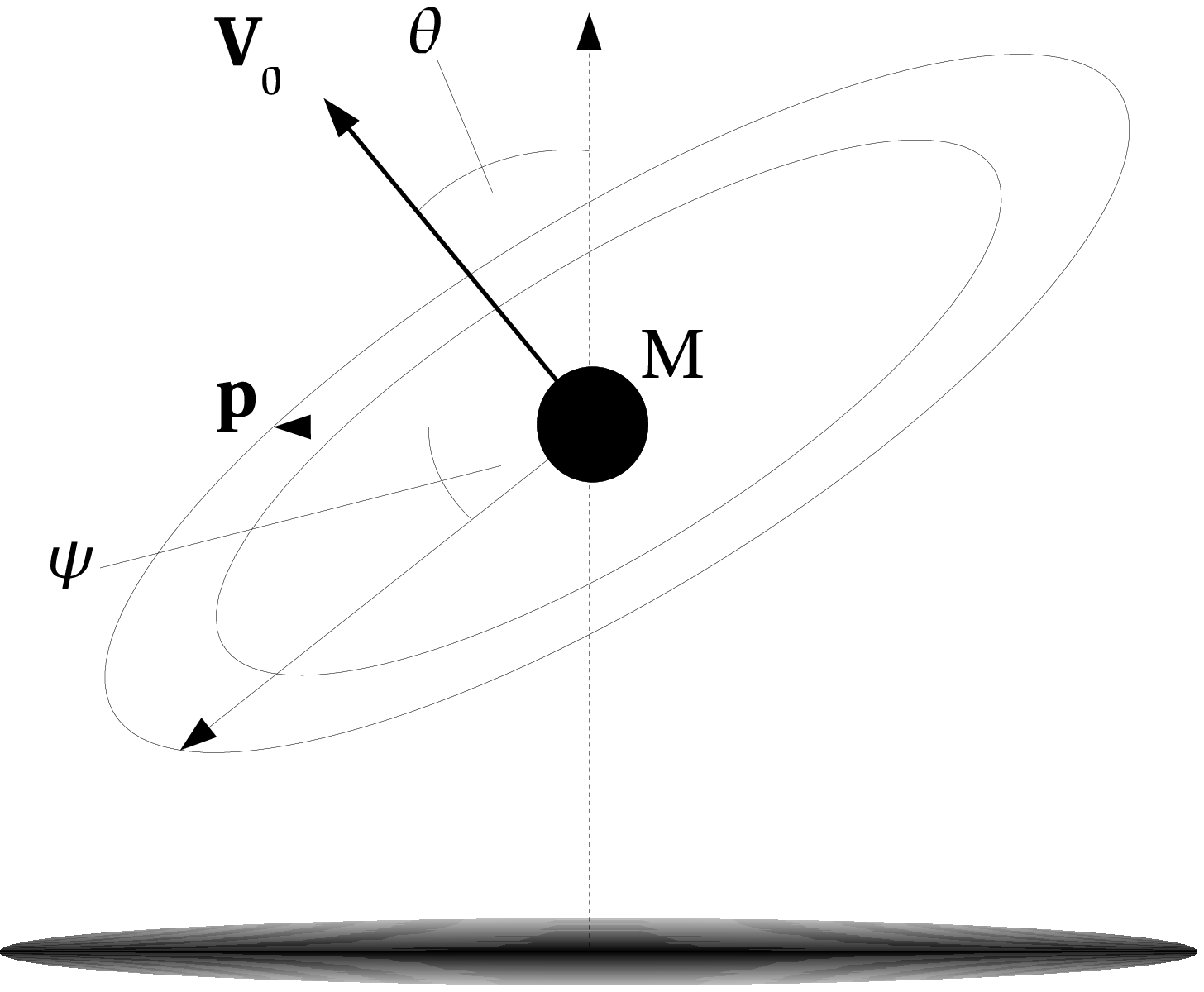,width=80mm}
\caption{The geometry used in calculating the Coulomb logarithm for
the disk. Scatterings through an annulus normal to ${\bf V}_0$ (the
relative velocity of satellite and disk stars) are considered. ${\bf
p}$ is a vector lying in this annulus and parallel to the galaxy
disk. Angles in the annulus, $\psi$, are measured from ${\bf
p}$. Finally, $\theta$ is the angle between ${\bf V}_0$ and the normal
to the disk.}
\label{fig:DFgeomnew}
\end{figure}

When computing the dynamical friction force we sum the contributions
from all particles with impact parameter $b$ by integrating around an
annulus of radius $b$ normal to the relative velocity vector of the
particles and the satellite. We define $\psi$ as the angle of a point
on this annulus measured from a vector, ${\bf p}$, which lies in the
plane of the annulus and which is parallel to the disk plane (see
Fig.~\ref{fig:DFgeomnew}). For the disk, the value of $b_{\rm max}$,
the upper limit of integration in eqn.~(\ref{eq:bmaxint}), will vary
as a function of $\psi$. In the direction corresponding to $\psi=0$
(and $\psi=\pi$) the disk density distribution has a characteristic
length-scale of $R_{\rm d}$ (the exponential scale-length). This will
therefore correspond (approximately) to the largest impact parameter
scatterings occurring in that direction. In perpendicular directions
($\psi=\pi/2$ and $\psi=3\pi/2$) a more appropriate characteristic
length is $r_{\rm eff} = R_{\rm d} (\cos ^2\theta +
h^2\sin^2\theta)^{1/2}$, where $\theta$ is the angle between the
satellite-disk relative velocity vector and the $z$-axis, and $h$ the
ratio of disk scale-height to scale-length. Thus, the effective
$\Lambda$ in direction $\psi$ is
\begin{equation}
\Lambda = {f_{\rm \Lambda,d} R_{\rm d} h^\prime V_0^2 \over \G M},
\end{equation}
where $h^\prime = (\cos^2\psi + \left[ \cos^2\theta + h^2 \sin^2\theta
\right] \sin^2\psi )^{1/2}$. If we account for the finite size of the
satellite then:
\begin{equation}
(1 + \Lambda^2)_{\rm eff} = { 1 + \Lambda^2 \over 1 + (b_{\rm min}
\Lambda / R_{\rm d} h^\prime f_{\rm \Lambda,d})^2}.
\end{equation}
The effective Coulomb logarithm is found by averaging over all $\psi$:
\begin{eqnarray}
\left< {1\over 2} \ln (1+\Lambda^2)_{\rm eff} \right> & = & {1 \over 4 \pi} \int_0^{2 \pi} \ln ( 1 + \Lambda^2 )_{\rm eff} \d \psi .
\end{eqnarray}
This integral is solved numerically.

\subsection{Rate of Increase of Scattered Particle Velocity Dispersion}
\label{app:zeffic}

We now wish to determine the rate of increase of the one-dimensional
velocity dispersion, measured in direction ${\bf \hat{n}}$, of the
particles $m$ due to dynamical friction scatterings.  Since the centre of
mass remains fixed during the scattering, $m\Delta{\bf V}_{\rm m} +
M\Delta{\bf V}_{\rm M}=0$. Therefore, to find the change in velocity
of $m$ we multiply the equations (7-10a) and (7-10b) of Binney \&
Tremaine by $-M/m$. Writing these in a more convenient form:
\begin{equation}
\Delta V_{\rm m \perp} = - 2 V_0 \Lambda {b \over b_{\rm max}} \left[ 1 + \Lambda^2 {b^2 \over b_{\rm max}^2} \right]^{-1},
\label{eq:deltavmperp}
\end{equation}
\begin{equation}
\Delta V_{\rm m ||} = - 2 V_0 \left[ 1 + \Lambda^2 {b^2 \over b_{\rm max}^2} \right]^{-1},
\label{eq:deltavmpara}
\end{equation}
for the components of velocity perpendicular and parallel to the
relative velocity vector ${\bf V}_0$ as measured in the frame in which
the centre of mass of $M$ and $m$ is at rest.

\begin{figure}
\psfig{file=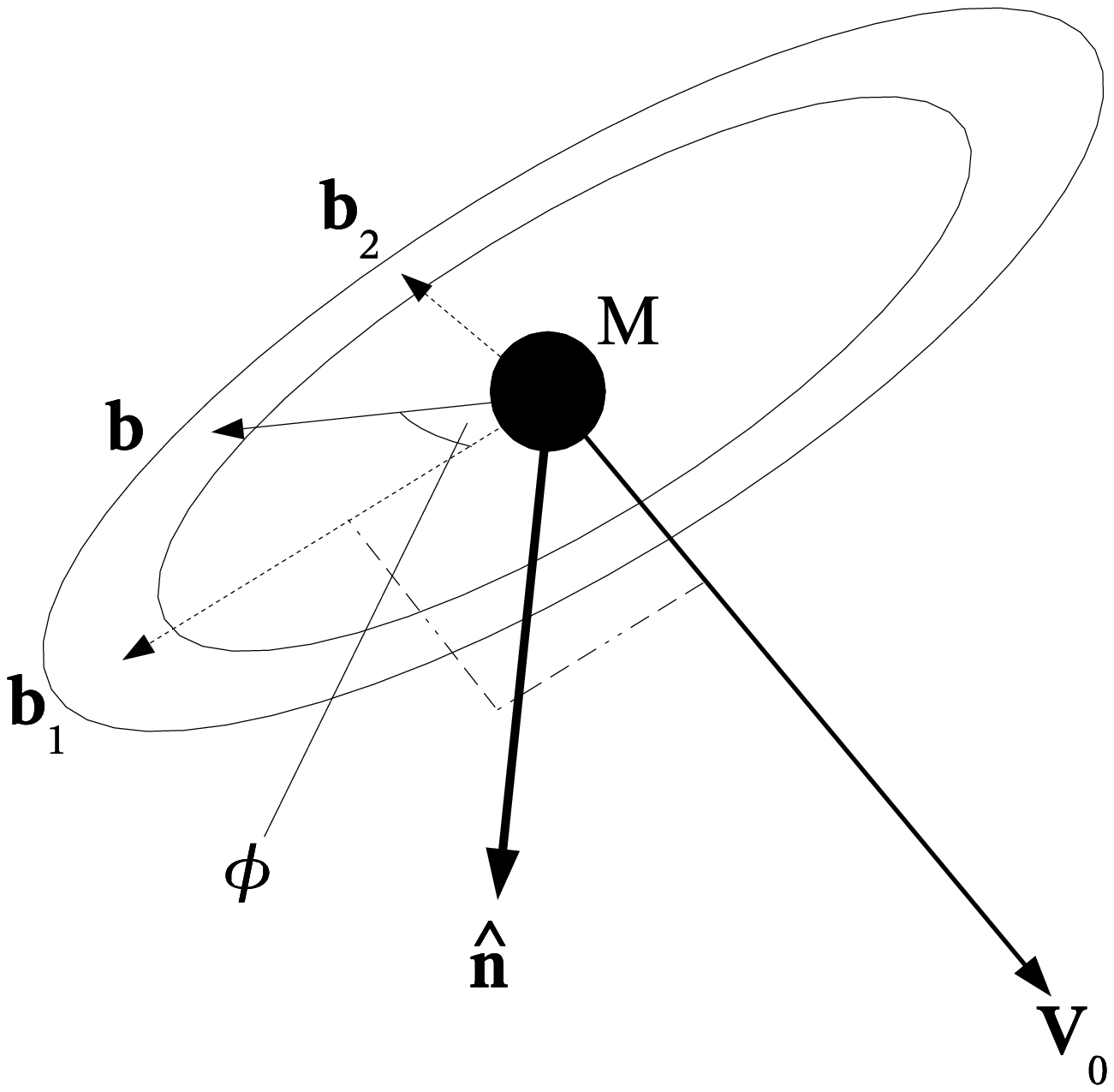,width=80mm}
\caption{Geometry used in computing the rate of increase of velocity
dispersion in direction ${\bf \hat{n}}$. Vectors ${\bf b}$, ${\bf
b}_1$ and ${\bf b}_2$ lie in the plane of the annulus. Vector ${\bf
V}_0$ is normal to the annulus and vector ${\bf \hat{n}}$ lies in the
plane of ${\bf V}_0$ and ${\bf b}_1$.}
\label{fig:geom2}
\end{figure}

Consider now the velocity of $m$ in the frame in which the centre of
mass of the central galaxy and its halo is at rest. The velocity
changes are independent of frame so the final velocity of $m$ in this 
frame is: 
\begin{equation}
{\bf V}_{\rm m}^{\rm (f)} = {\bf V}_{\rm m} + {{\bf V_0}\over V_0} \Delta V_{\rm m ||} + {{\bf b} \over b} \Delta V_{\rm m \perp}.
\end{equation}
We are interested in the velocities in some direction ${\bf
\hat{n}}$. The initial and final velocities of $m$ in this direction
are:
\begin{eqnarray}
{\bf V}^{\rm (i)}_{{\rm m,}{\bf \hat{n}}} & = & {\bf V}_{\rm m}\cdot{\bf \hat{n}}, \\
{\bf V}^{\rm (f)}_{{\rm m,}{\bf \hat{n}}} & = & {\bf V}_{\rm m}\cdot{\bf \hat{n}} + \Delta V_{\rm m ||} \cos \theta_{V_0} + \Delta V_{\rm m \perp} \cos \theta_b,
\end{eqnarray}
where $\theta_{V_0}$ and $\theta_b$ are the angles between ${\bf
\hat{n}}$ and ${\bf V}_0$ and ${\bf \hat{n}}$ and ${\bf b}$
respectively. The change in the component of the kinetic energy in
direction ${\bf \hat{n}}$ is therefore
\begin{eqnarray}
\Delta E_{\bf \hat{n}} & = & {m \over 2} \left\{ \Delta V_{\rm m ||}^2 \cos^2 \theta_{V_0} + \Delta V_{\rm m \perp}^2 \cos^2 \theta_b \right. \nonumber \\
 & & + 2 \Delta V_{\rm m ||} \Delta V_{\rm m \perp}  \cos \theta_{V_0} \cos \theta_b  \nonumber\\
 & & \left. + 2 {\bf V}_{\rm m} \cdot {\bf \hat{n}} \left[  \Delta V_{\rm m ||} \cos \theta_{V_0} + \Delta V_{\rm m \perp} \cos \theta_b \right] \right\}.
\end{eqnarray}
To sum over all particles $m$, we first integrate around an annulus of
constant $|{\bf b}|$. On this annulus, $\theta_{V_0}$ is constant and
we can write the vector ${\bf b}={\bf b}_1 \cos \phi + {\bf b}_2 \sin
\phi$, where ${\bf b}_1 \cdot {\bf b}_2=0$, $|{\bf b}_1|=|{\bf
b}_2|=b$ and $\phi$ is a parameter (see Fig.~\ref{fig:geom2}). We then
note that
\begin{eqnarray}
\int_0^{2 \pi} \cos^2 \theta_b \d \phi & = & \int_0^{2 \pi} \left[{{\bf b}_1\cdot{\bf \hat{n}} \over b}\right]^2 \cos^2 \phi \nonumber \\
 & & + \left[{{\bf b}_2\cdot{\bf \hat{n}} \over b}\right]^2 \sin^2 \phi \nonumber \\
 & & + 2 \left[{({\bf b}_1\cdot{\bf \hat{n}}) ({\bf b}_2\cdot{\bf \hat{n}}) \over b^2}\right] \sin \phi \cos \phi  \d \phi \nonumber \\
 & = & \pi \left({{\bf b}_1\cdot{\bf \hat{n}} \over b}\right)^2 + \pi \left({{\bf b}_2\cdot{\bf \hat{n}} \over b}\right)^2.
\end{eqnarray}
If we choose ${\bf b}_1$ to be parallel to the projection of ${\bf
\hat{n}}$ into the plane of the annulus then ${\bf b}_1\cdot{\bf
\hat{n}}/b=\sin \theta_{V_0}$ and ${\bf b}_2\cdot{\bf \hat{n}}/b=0$,
so
\begin{equation}
\int_0^{2 \pi} \cos^2 \theta_b \d \phi = \pi \sin^2\theta_{V_0}.
\end{equation}
Using a similiar approach, it is simple to show that $\int_0^{2 \pi}
\cos \theta_b \d \phi = 0$. Thus, the change in energy becomes
\begin{eqnarray}
\Delta E_{\bf \hat{n}} & = & {m \over 2} \left\{ 2 \pi \Delta V_{\rm m ||}^2 \cos^2 \theta_{V_0} + \pi \Delta V_{\rm m \perp}^2 \sin^2 \theta_{V_0} \right. \nonumber \\
& & \left.  + 4 \pi {\bf V}_{\rm m} \cdot {\bf \hat{n}} \Delta V_{\rm m ||} \cos \theta_{V_0} \right\}.
\end{eqnarray}
Substituting eqns.~(\ref{eq:deltavmperp}) and (\ref{eq:deltavmpara})
we find
\begin{eqnarray}
\Delta E_{\bf \hat{n}} & = & {m V_0^2 \over 2} \left\{ 8 \pi \left[ 1 + \Lambda^2 {b^2\over b_{\rm max}^2} \right]^{-2} \cos^2 \theta_{V_0} \right. \nonumber \\
 & & + 4 \pi \Lambda^2 {b^2 \over b_{\rm max}^2} \left[ 1 + \Lambda^2 {b^2\over b_{\rm max}^2} \right]^{-2} \sin^2 \theta_{V_0} \nonumber \\
 & & \left. - 8 \pi {{\bf V}_{\rm m} \cdot {\bf \hat{n}}\over V_0} \left[ 1 + \Lambda^2 {b^2\over b_{\rm max}^2} \right]^{-1} \cos \theta_{V_0} \right\}.
\end{eqnarray}
To find the total energy change we multiply by the flux of particles
passing through the annulus, $n V_0 b \d b$, and integrate over $b$
from $0$ to $b_{\rm max}$. This gives
\begin{eqnarray}
{\d E_{\bf \hat{n}} \over \d t} & = & \rho V_0^3 b_{\rm max}^2 \left\{ {2 \pi \cos^2 \theta_{V_0} \over 1+\Lambda^2} \right. \nonumber \\
 & & + \pi {(1+\Lambda^2)\ln(1+\Lambda^2)-\Lambda^2 \over \Lambda^2 (1+\Lambda^2)} \sin^2 \theta_{V_0} \nonumber \\
 & & \left. - 2 \pi {{\bf V}_{\rm m} \cdot {\bf \hat{n}}\over V_0} {\ln(1+\Lambda^2) \over \Lambda^2} \cos \theta_{V_0} \right\}.
\end{eqnarray}
We next average over the velocity distribution of ${\bf V}_{\rm
m}$. The total rate of energy change is then
\begin{eqnarray}
{\d E_{\bf \hat{n}} \over \d t} & = & \int \rho V_0^3 b_{\rm max}^2 \left\{ {2 \pi \cos^2 \theta_{V_0} \over 1+\Lambda^2} \right. \nonumber \\
& & + \pi {(1+\Lambda^2)\ln(1+\Lambda^2)-\Lambda^2 \over \Lambda^2 (1+\Lambda^2)} \sin^2 \theta_{V_0} \nonumber \\
 & & \left. - 2 \pi {{\bf V}_{\rm m} \cdot {\bf \hat{n}}\over V_0} {\ln(1+\Lambda^2) \over \Lambda^2} \cos \theta_{V_0} \right\} f({\bf V}) \d^3{\bf V}.
\end{eqnarray}
In general it seems that this equation is not analytically solvable,
even if $f({\bf V}_{\rm m})$ is an isotropic Gaussian. However, if we
are interested in systems where random motions are much smaller than
the bulk motion (such as galaxy disks), then we can approximate $f({\bf
V}_{\rm m})=\delta({\bf V}_{\rm m}-{\bf V}_{\rm d})$, where ${\bf
V}_{\rm d}$ is the disk bulk velocity and $\delta$ is the Dirac
delta function. Note that ${\bf V}_0 = {\bf V}_{\rm m} - {\bf V}_{\rm
M}$ where ${\bf V}_{\rm M}$ is the velocity of $M$. For this case
\begin{eqnarray}
{\d E_{\bf \hat{n}} \over \d t} & = & \rho V_0^3 b_{\rm max}^2 \left\{ {2 \pi \cos^2 \theta_{V_0} \over 1+\Lambda^2} \right. \nonumber \\
 & & + \pi {(1+\Lambda^2)\ln(1+\Lambda^2)-\Lambda^2 \over \Lambda^2 (1+\Lambda^2)} (1 - \cos^2 \theta_{V_0}) \nonumber \\
 & & \left. - 2 \pi v_{\rm d}\cos \theta_{v_{\rm d}} {\ln(1+\Lambda^2) \over \Lambda^2} \cos \theta_{V_0} \right\},
\end{eqnarray}
where $\cos \theta_{V_0} = v_{\rm d}\cos \theta_{v_{\rm d}} - v_{\rm
M} \cos \theta_{v_{\rm M}}$. Here $v_{\rm d}=V_{\rm d}/V_0$ and
$\theta_{v_{\rm d}}$ is the angle between ${\bf \hat{n}}$ and ${\bf
V}_{\rm d}$, with similar definitions for $v_{\rm M}$ and
$\theta_{v_{\rm M}}$. The efficiency of energy transfer to direction
${\bf \hat{n}}$ is then easily found by dividing the above by the same
expression summed over three orthogonal directions (taking one of
these to be parallel to ${\bf V}_0$ simplifies the summation):
\begin{eqnarray}
\epsilon_{\bf \hat{n}} & = & \left\{ {2  \cos^2 \theta_{V_0} \over 1+\Lambda^2} +  {(1+\Lambda^2)\ln(1+\Lambda^2)-\Lambda^2 \over \Lambda^2 (1+\Lambda^2)} (1 - \cos^2 \theta_{V_0}) \right. \nonumber \\
 & & \left. - 2  v_{\rm d}\cos \theta_{v_{\rm d}} {\ln(1+\Lambda^2) \over \Lambda^2} \cos \theta_{V_0} \right\} \left/ \left\{ {2  \over 1+\Lambda^2} \nonumber \right. \right. \\
 & &  + 2  {(1+\Lambda^2)\ln(1+\Lambda^2)-\Lambda^2 \over \Lambda^2 (1+\Lambda^2)} \nonumber \\
& & \left. - 2  v_{\rm d}\cos \theta_{v_{\rm d}} {\ln(1+\Lambda^2) \over \Lambda^2} \right\} .
\end{eqnarray}

We are interested specifically in the vertical velocity dispersion of
a galactic disk. In this case ${\bf V}_{\rm d}$ lies in the disk
plane, while ${\bf \hat{n}}$ is perpendicular to that
plane. Consequently $\cos \theta_{v_{\rm d}}=0$ and the above
expression simplifies to
\begin{eqnarray}
\epsilon_{\bf \hat{z}} & = & \left\{ {2  \cos^2 \theta_{V_0} \over 1+\Lambda^2} +  {(1+\Lambda^2)\ln(1+\Lambda^2)-\Lambda^2 \over \Lambda^2 (1+\Lambda^2)} (1 - \cos^2 \theta_{V_0}) \right\} \nonumber \\
 & & \left/ \left\{ {2  \over 1+\Lambda^2} + 2  {(1+\Lambda^2)\ln(1+\Lambda^2)-\Lambda^2 \over \Lambda^2 (1+\Lambda^2)} \right\} \right. .
\label{eqn:epsz}
\end{eqnarray}
This expression is then used in eqn.~(\ref{eq:heatz}) to calculate the
vertical heating rate of galaxy disks. Note that $0\leq \epsilon_{\rm
z}\leq 1$, as expected for an efficiency factor.

\section{Disk Surface Energy Densities}
\label{app:enrels}

We here derive expressions for the different contributions to the
surface energy density of the disk. These are used in \S\ref{sec:sch}.

We assume a disk with a density structure
\begin{equation}
\rho_{\rm d}(R,z)=\Sigma(R) {{\rm sech}^2(z/H_{\rm d}) \over 2 H_{\rm d}}.
\end{equation}
with $H_{\rm d}$ constant with radius.  Assuming that the disk is thin,
$H_{\rm d}\ll R$, the potential of the disk can be found by approximating
the density distribution as a set of infinite, homogeneous planes,
such that
\begin{eqnarray}
\phi_{\rm d}(R,z) & = & \int_{-\infty}^\infty 2 \pi \G \rho_{\rm d}(R,z) |z-z^\prime| \d z^\prime + \phi_{\rm d}(R,0) \nonumber \\
 & = & 2 \pi \G \Sigma(R) H_{\rm d} [\ln \cosh(z/H_{\rm d}) +\ln 2 ] + \phi_{\rm d}(R,0).
\end{eqnarray}
Close to the disk plane, the $z$-component of the force due to the
spherical halo plus bulge is
\begin{equation}
F_{\rm h} = - {\G M_{\rm h}(R)\over R^3} z,
\end{equation}
hence the potential due to these components is
\begin{eqnarray}
\phi_{\rm h}(R,z) & \approx & \int_0^z {\G M_{\rm h}(R)\over R^3} z^\prime \d z^\prime + \phi_{\rm h}(R,0) \nonumber \\
 & = & {\G M_{\rm h}(R)\over 2 R^3} z^2 + \phi_{\rm h}(R,0).
\end{eqnarray}
Referencing all energies to $z=0$, we can neglect the final terms in
the above equations. We can now calculate the different contributions
to the disk vertical energy per unit area, where necessary using the
thin disk approximation $H_{\rm d}\ll R$. The gravitational self-energy of
the disk is then
\begin{eqnarray}
w_{\rm dd}(R) & = & {1 \over 2} \int_{-\infty}^\infty \phi_{\rm d}(R,z) \rho_{\rm d}(R,z) \d z \nonumber \\
 & = & \pi \G \Sigma(R)^2 H_{\rm d}.
\label{eq:wdd}
\end{eqnarray}
The disk-halo gravitational potential energy is 
\begin{eqnarray}
w_{\rm dh}(R) & = & \int_{-\infty}^\infty \phi_{\rm h}(R,z) \rho_{\rm d}(R,z) \d z \nonumber \\
 & = & {\pi^2 \over 24} {G M_{\rm h}(R) \over R^3} \Sigma_{\rm d}(R) H_{\rm d}^2,
\label{eq:wdh}
\end{eqnarray}
and the kinetic energy of the disk is
\begin{equation}
t_{\rm z}(R) = {1 \over 2} \Sigma(R) \sigma_{\rm z}^2(R).
\label{eq:td}
\end{equation}

\section{Convergence Tests on N-body Simulations}
\label{sec:convergence}

\begin{figure*}
\begin{tabular}{cc}
\psfig{file=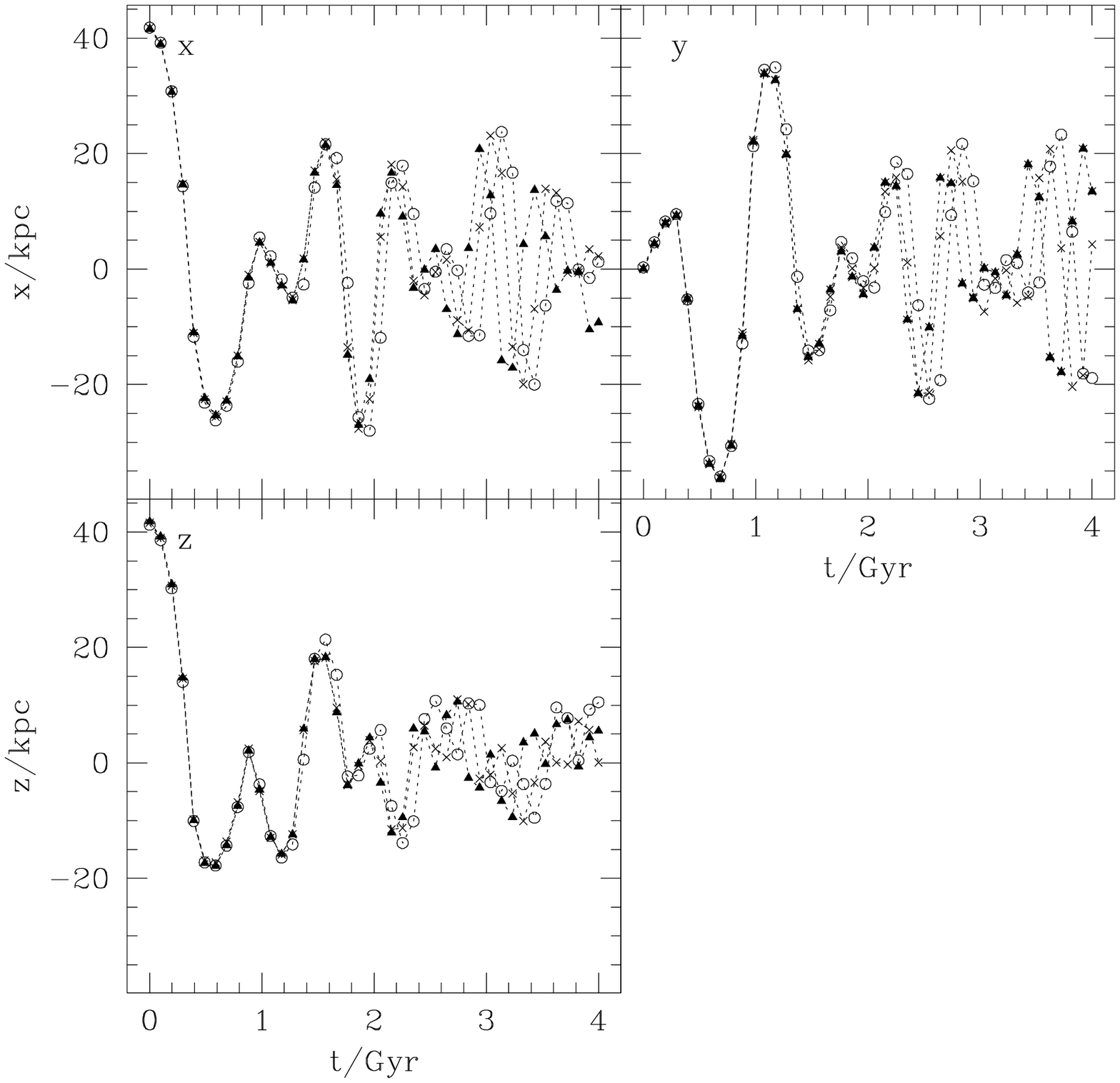,width=75mm} & \psfig{file=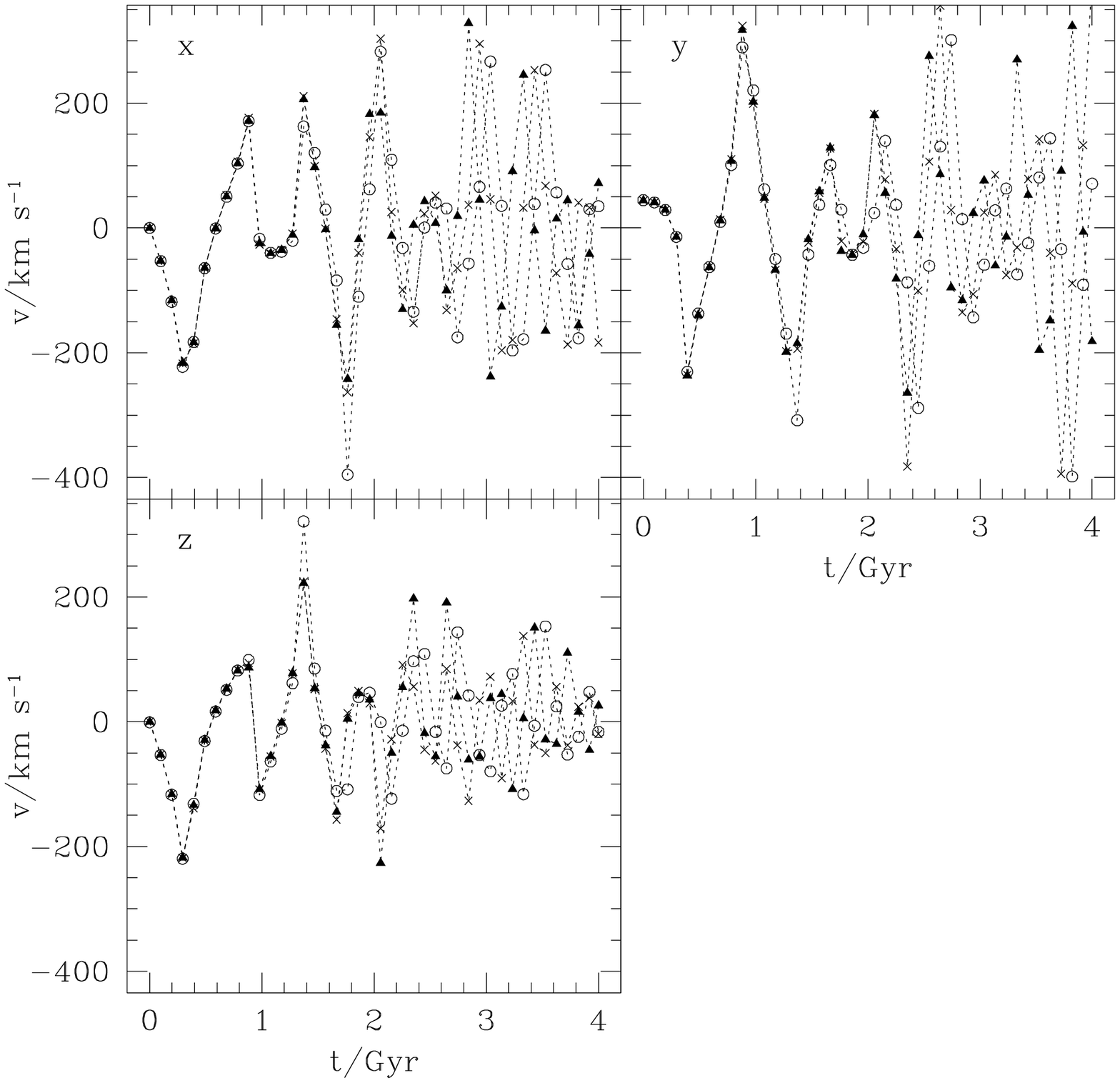,width=75mm} \\
\psfig{file=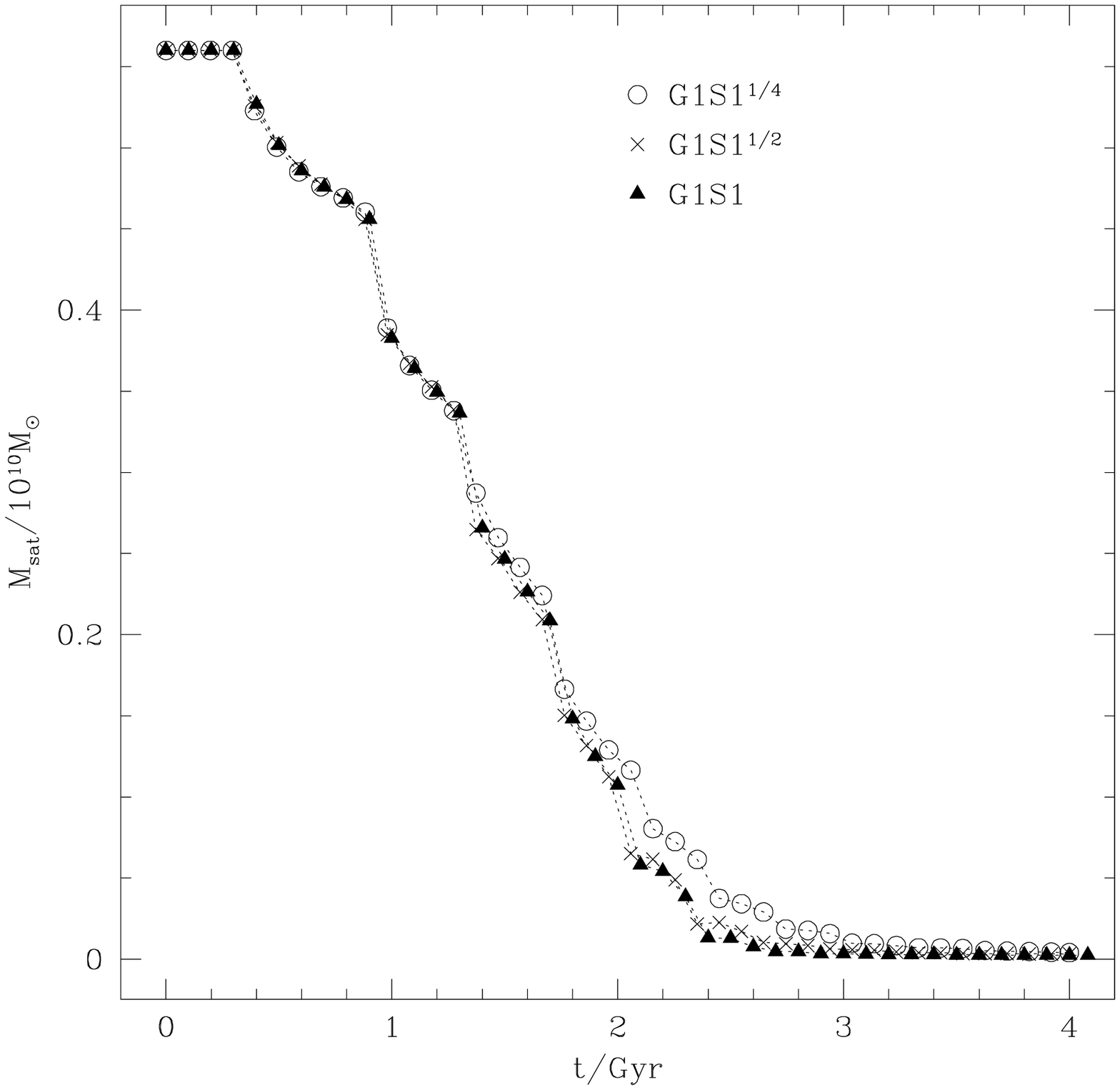,width=75mm} & \psfig{file=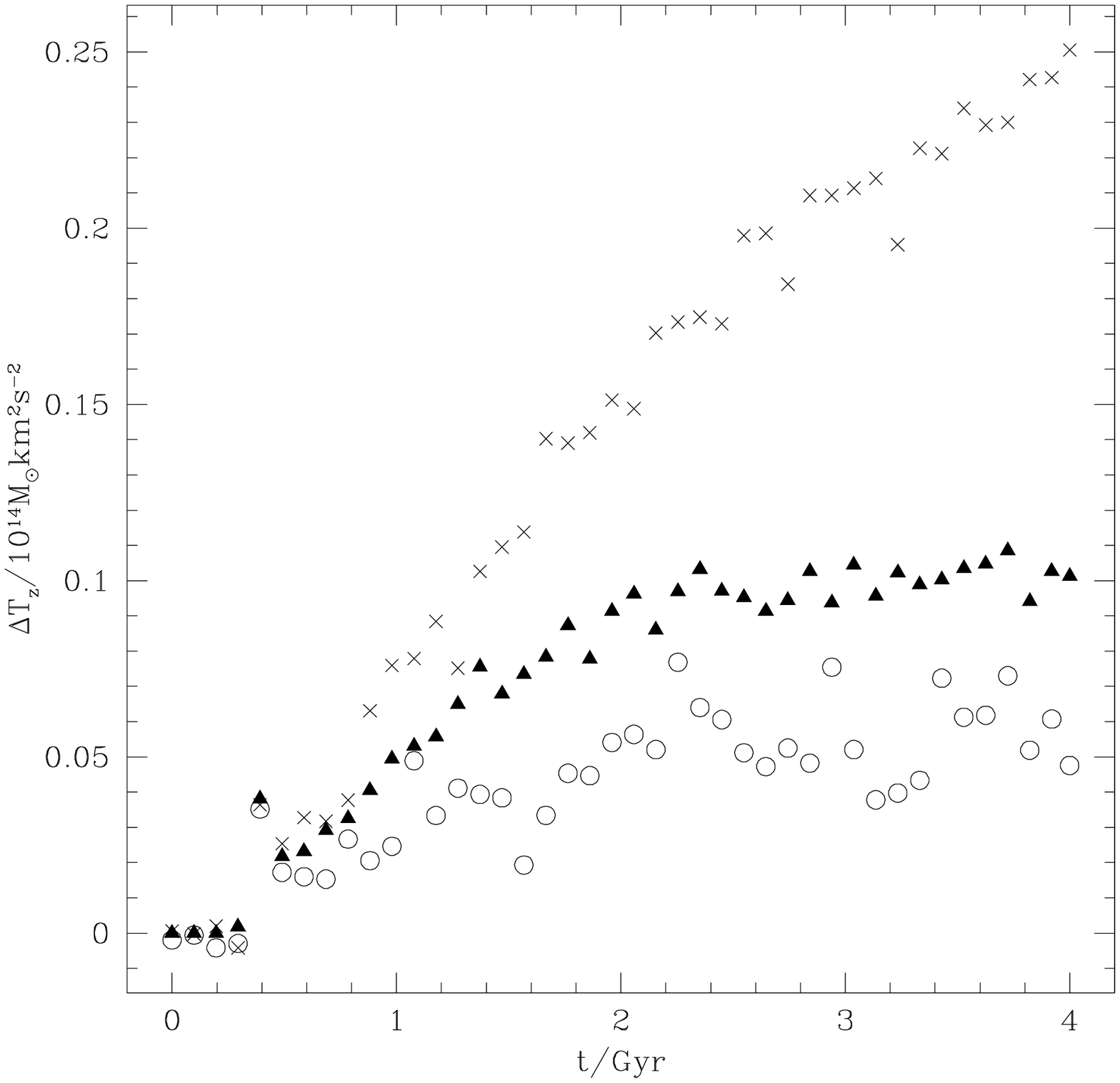,width=75mm}
\end{tabular}
\caption{Convergence tests for model G1S1. Results are shown for the
standard simulation (circles), model G1S1$^{1/2}$ (which has half the
number of particles of  G1S1; crosses) and model G1S1$^{1/4}$ (which
has one quarter the number of particles of G1S1; triangles). Points
are connected by dotted lines to guide the eye only---lines are not
intended as a realistic interpolation of the points. \emph{Top
left-hand panel:} The orbital position of the satellite as a function
of time. \emph{Top right-hand panel:} The orbital velocity of the
satellite as a function of time. \emph{Lower left-hand panel:} The
remaining bound mass of the satellite as a function of
time. \emph{Lower right-hand panel:} The change in the vertical
component of the disk kinetic energy due to heating by the satellite
as a function of time.}
\label{fig:convergence1}
\end{figure*}

\begin{figure*}
\begin{tabular}{cc}
\psfig{file=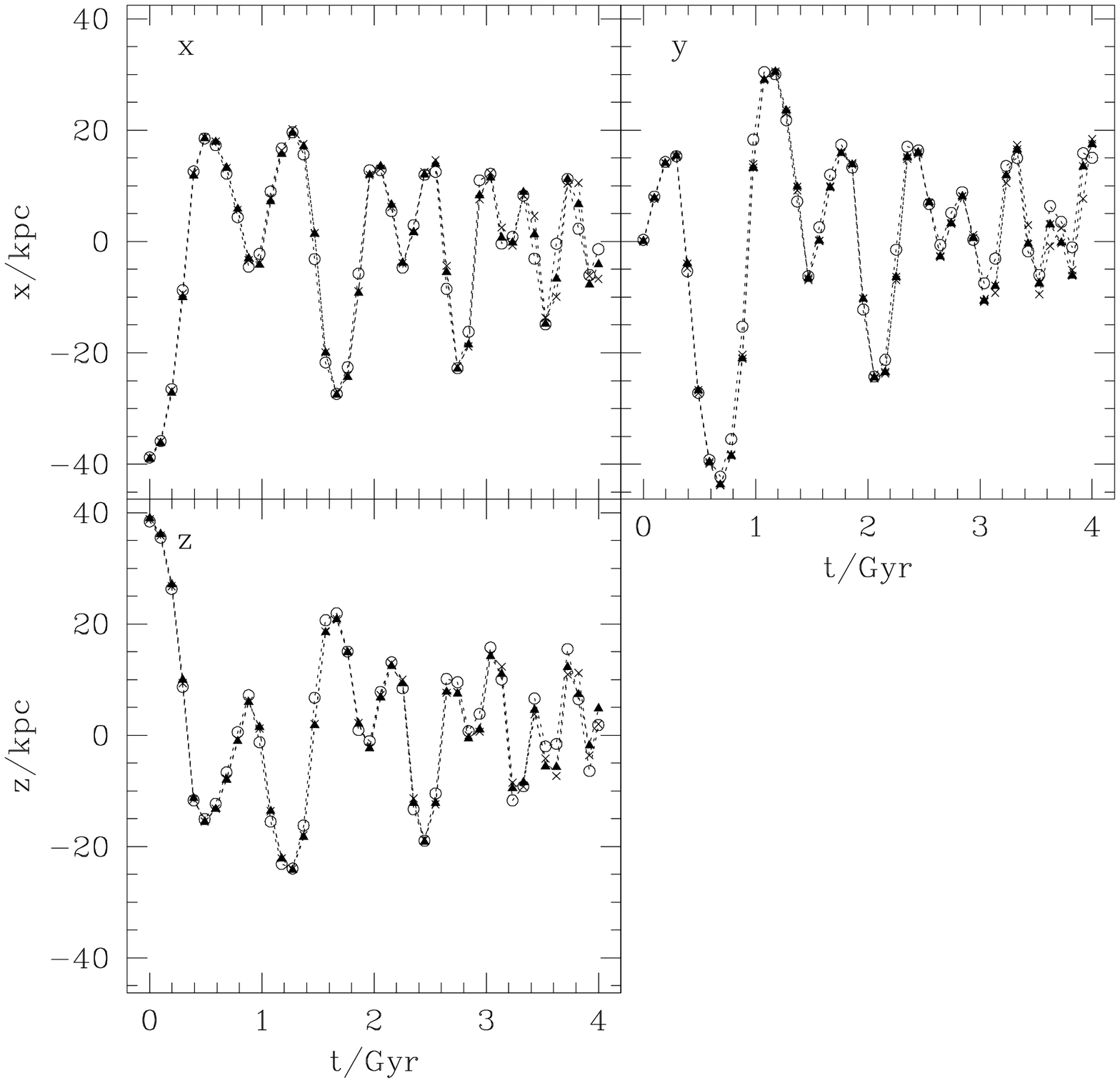,width=75mm} & \psfig{file=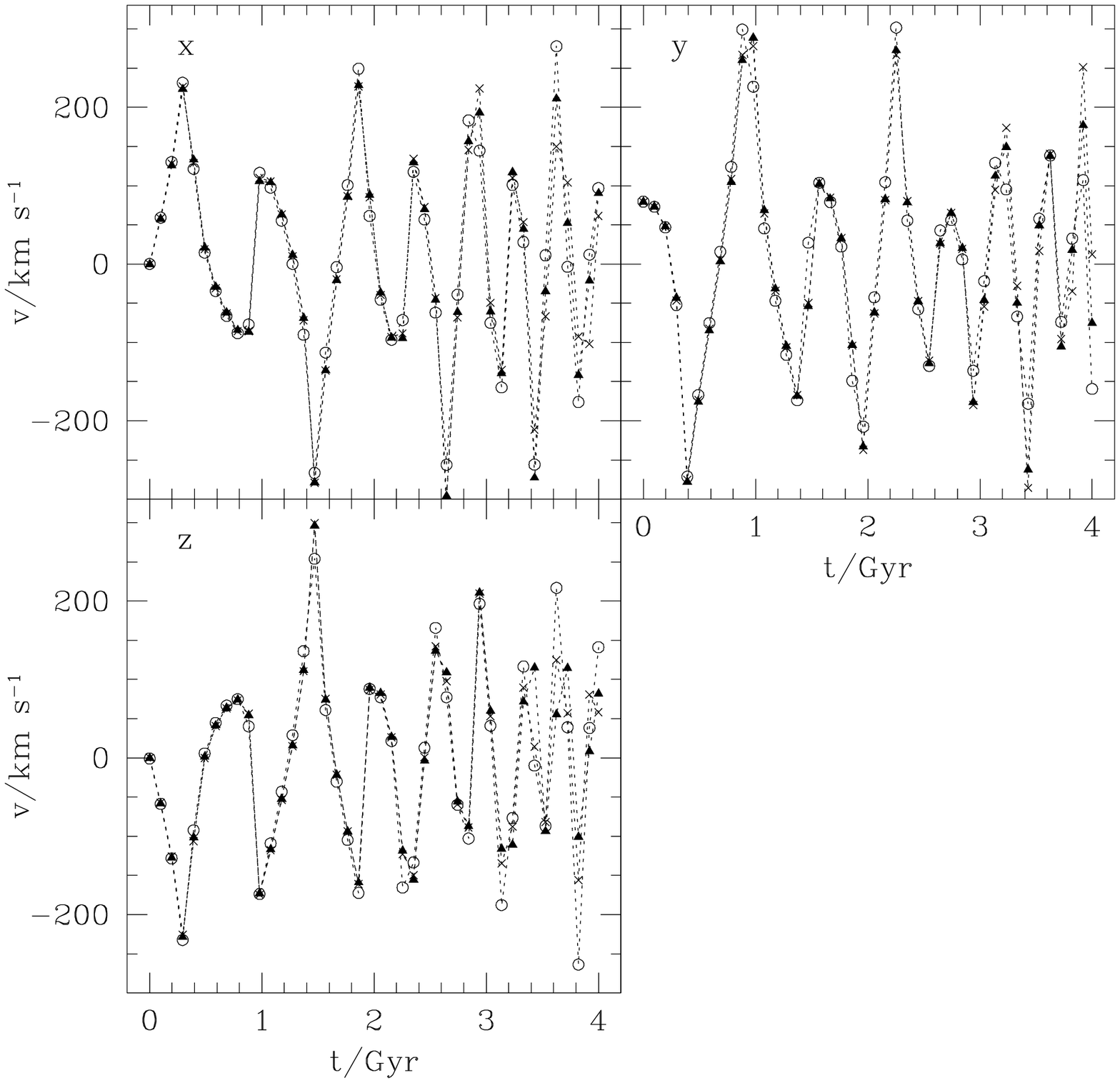,width=75mm} \\
\psfig{file=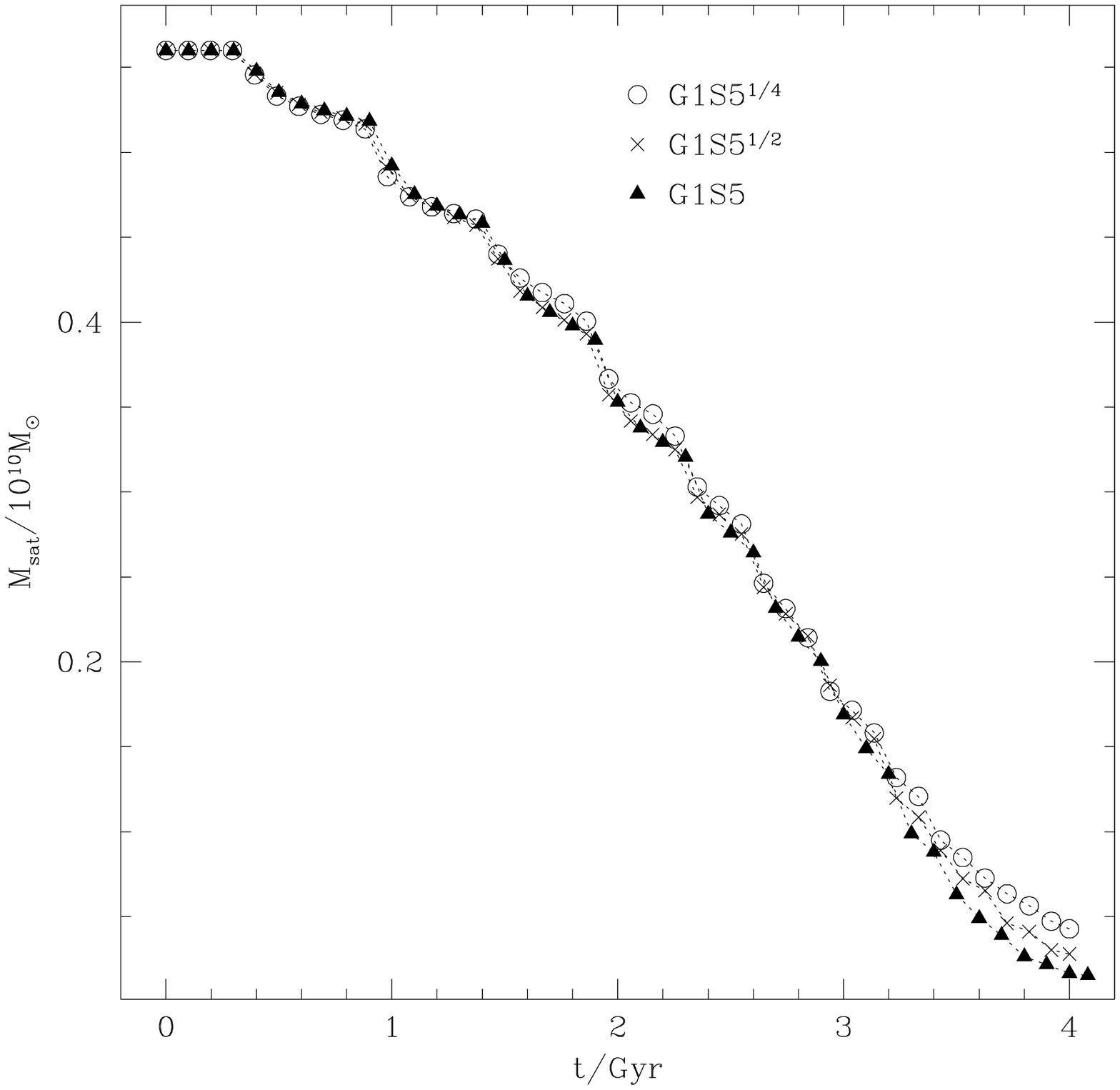,width=75mm} & \psfig{file=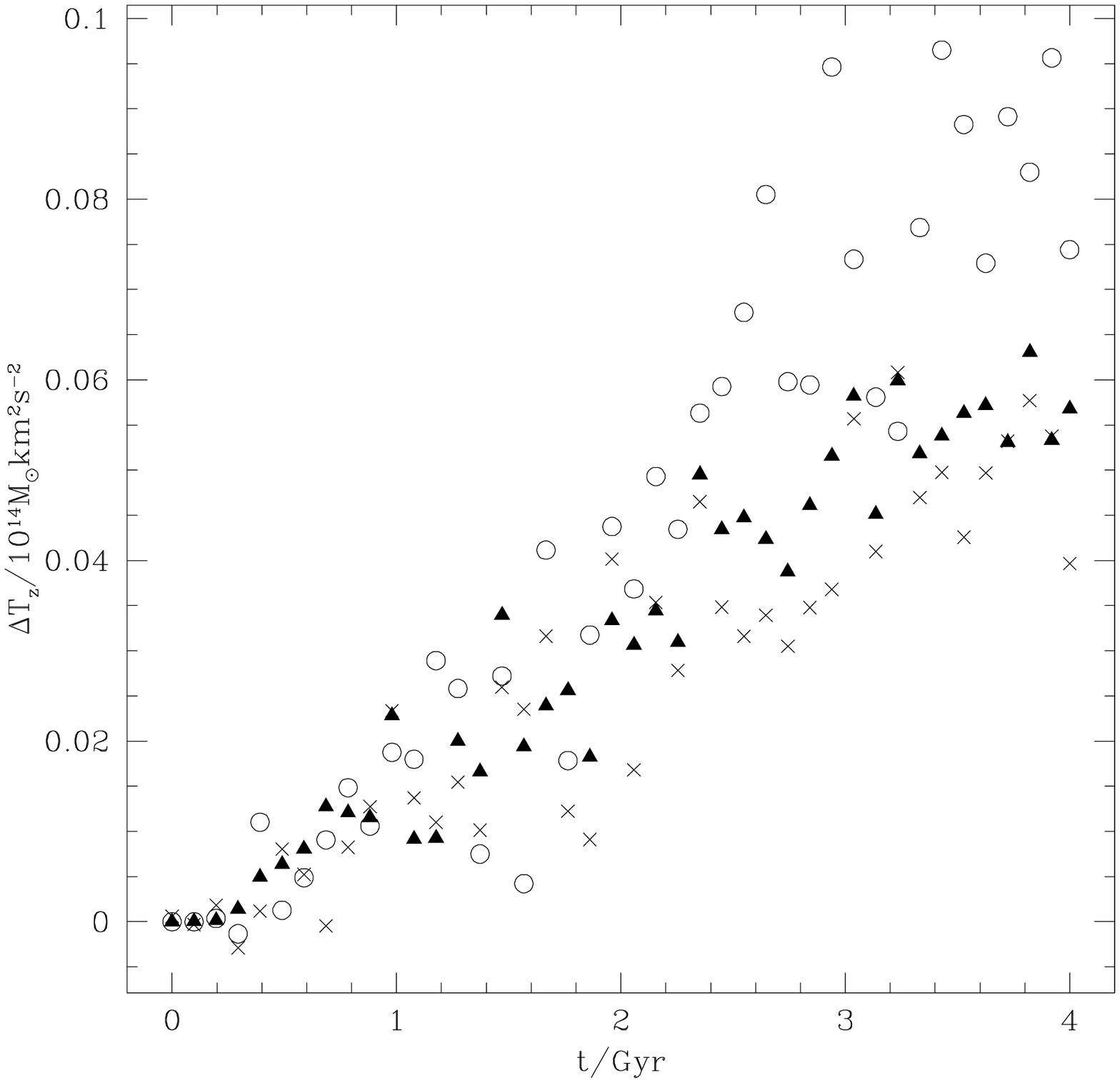,width=75mm}
\end{tabular}
\caption{Convergence tests for model G1S5. Results are shown for the
standard simulation (circles), model G1S5$^{1/2}$ (which has half the
number of particles of G1S5; crosses) and model G1S5$^{1/4}$ (which
has one quarter the number of particles of G1S5; triangles). The
different panels are as for Figure~\protect\ref{fig:convergence1}.}
\label{fig:convergence5}
\end{figure*}

We repeated all the simulations of models G1S1 to G1S15 with
one half and one quarter the number of particles, labelling these runs
G1S1$^{1/2}$, G1S1$^{1/4}$ etc. (In each case, we scaled the softening
in proportion to the mean interparticle separation.)
Figures~\ref{fig:convergence1} and \ref{fig:convergence5} compare the
results for two representative models (G1S1 and G1S5) run with
different numbers of particles. In both cases, the position and velocity
of the satellite are well converged up until the very final stages of
the satellite's life (at which point it becomes difficult to measure
these quantities accurately from the simulation anyway).  The higher
resolution simulations lose mass from the satellite somewhat more
rapidly at late times, but the differences are minor and the mass loss
rate is well determined by the simulations. 

The convergence behaviour seems poorer for the change in vertical
kinetic energy $\Delta T_{\rm z}$ (in which the energy of the
unperturbed disk from model G1S0 has been subtracted off). For the
model G1S5, the difference in the final $\Delta T_{\rm z}$ of
$(0.02-0.03) \times 10^{14}M_\odot$km$^2$s$^{-2}$ between the highest
and lowest resolution runs could result mostly from the error in the
energy of the unperturbed disk that is subtracted off, since the
variation in this value between different realizations is around
$(0.02-0.03)
\times 10^{14}M_\odot$km$^2$s$^{-2}$ in the low resolution
case. However, for model G1S1 the differences in $\Delta T_{\rm z}$
between the high and low resolution runs are much bigger than can be
explained by errors in the subtraction of the unperturbed disk
contribution. In this case, the behaviour of $\Delta T_{\rm z}$ is not
even monotonic as the number of particles is increased. We have
investigated this further by repeating some of the lowest resolution
simulations using a different sequence of random numbers in generating
the initial conditions. We find that this leads to significant
variations in $\Delta T_{\rm z}$, comparable to those seen between the
lowest resolution and higher resolution simulations. It therefore
seems that the amount of disk heating by satellites is sensitive to
stochastic variations in the initial conditions, over and above the two-body
relaxation which heats the unperturbed disk. Comparing the highest and
lowest resolution runs, we find that the error in the low-resolution
estimate of $\Delta T_{\rm z}$ is $\sim 30\%$ for model G1S5, but
$\sim 100\%$ for model G1S1. The convergence of $\Delta T_{\rm z}$
with increasing particle number thus seems to depend on the orbital
properties of the satellite, with different numbers of particles being
required to achieve the same degree of convergence in different
cases. A more comprehensive study of convergence in a variety of
models will be required to address this question fully.

\section{Order of Magnitude Estimates of Disk Thicknesses}
\label{app:ordmag}

In this appendix we make order of magnitude estimates for the
thicknesses of disks resulting from heating by satellites and by stars
scattering from giant molecular clouds.

\subsection{Heating by Satellites}

We assume a distribution of satellite halo masses ${\rm d}N/{\rm d}M =
A M^{-1.7}$, where $A$ is a constant, with a maximum mass of $f_{\rm
max}=0.01$ of the total halo mass and making up a fraction $f{\rm
mass}=0.1$ of the total halo mass (consistent with numerical
simulations; \pcite{springelclus}). Then,
\begin{equation}
A = 0.3 f_{\rm mass} M_{\rm halo}^{0.7}/f_{\rm max}^{0.3}.
\end{equation}
Taking the Chandrasekhar formula for dynamical friction
(eqn.~\ref{eq:chandra}), the power deposited by a satellite is, to
order of magnitude
\begin{equation}
P = 4 \pi {{\rm G}^2 M_{\rm s}^2 \over V_{\rm s}} \ln \Lambda \rho_{\rm d}.
\end{equation}
Assuming that heating is effective primarily within one scale height
of the disk, then heating occurs overly a fraction of roughly $H_{\rm
d}/2 \pi r_{\rm orb}$ of the satellite's orbit, where $r_{\rm orb}$ is
the orbital radius. Allowing for an efficiency of vertical heating
$\epsilon_{\rm z}$ and integrating over masses, the total energy
transfer in time $t$ is given by
\begin{eqnarray}
E & = & 2 t {{\rm G}^2 \over V_{\rm s}} \ln \Lambda {M_{\rm d} \over 4 \pi R_{\rm d}^3 h} \epsilon_{\rm z} {R_{\rm d} h \over r_{\rm orb}} \int_0^{f_{\rm max} M_{\rm halo}} A M_{\rm s}^{-1.7} M_{\rm s}^2 {\rm d}M_{\rm s} \nonumber \\
  & = & 2 {{\rm G}^2 \over V_{\rm s}} \ln \Lambda {M_{\rm d} \over 4 \pi R_{\rm d}^3 h} \epsilon_{\rm z} {R_{\rm d} h \over r_{\rm orb}} 0.3 f_{\rm mass} M_{\rm halo}^2 f_{\rm max}.
\end{eqnarray}
Equating to the energy of the disk, as given by
eqn.~(\ref{eq:eztotal}) after ignoring the contribution proportional
to $h^2$ as this is typically small, and solving for $h$
\begin{equation}
h = {8 \over 3 \pi} 0.3 f_{\rm mass} f_{\rm max} \epsilon_{\rm z} {{\rm G} M_{\rm halo}^2 \ln \Lambda \over V_{\rm s} R_{\rm d} r_{\rm orb} M_{\rm d}} t,
\end{equation}
or
\begin{eqnarray}
h & = & 0.16 \left( {f_{\rm mass} \over 0.1} \right) \left( {f_{\rm max} \over 0.01} \right) \left( {\epsilon_{\rm z} \over 0.3} \right) \left( {M_{\rm halo}\over 10^{12}M_\odot}\right)^2 \left( {\ln \Lambda \over 3}\right) \nonumber \\
 & & \times \left( {V_{\rm s} \over 200\hbox{ km/s}} \right)^{-1} \left( {R_{\rm d}  \over 3.5\hbox{ kpc}} \right)^{-1}  \left( {r_{\rm orb} \over 200\hbox{ kpc}} \right)^{-1} \nonumber \\
 & & \times  \left( {M_{\rm d} \over 5\times 10^{10}M_\odot} \right)^{-1} \left( { t \over \hbox{Gyr}} \right).
\end{eqnarray}

\subsection{Heating by Scatterings from Giant Molecular Clouds}

We begin with eq.~(\ref{eq:cloudheat_tot}), which we approximate to
order of magnitude (replacing $\d \epsilon_{\rm z}/\d t$ with
$\epsilon_{\rm z}/t$ etc.) as
\begin{equation}
E = 2{\G^2 M_{\rm d} M_{\rm c} \Sigma_{\rm c} \over \sigma_{\rm z}^2}  \nu \ln \Lambda_{\rm c} \alpha_{\rm s}^3(\beta) K_{\rm s}(\beta) t.
\end{equation}
Inserting eqn.~(\ref{eq:sigmaz}) to eliminate $\sigma_{\rm z}$ (we
again ignore the contribution proportional to $h^2$) this reduces to
\begin{equation}
E = {2 \over \pi} {\G M_{\rm d} M_{\rm c} \over R_{\rm d}h} {\Sigma_{\rm c} \over \Sigma_{\rm d}} \ln \Lambda_{\rm c} \nu \alpha_{\rm s}^3(\beta) K_{\rm s}(\beta) t.
\end{equation}
Equating to the energy of the disk and solving for $h$ results in
\begin{equation}
h^2 = {32 \over 3 \pi} {M_{\rm c} \over M_{\rm d}} f \ln \Lambda_{\rm c} \nu \alpha_{\rm s}^3(\beta) K_{\rm s}(\beta) t,
\end{equation}
where $f=\Sigma_{\rm c}/\Sigma_{\rm d}$. This can be expressed as
\begin{eqnarray}
h & = & 7.2 \times 10^{-3} \left[ {f \over 0.025} \right]^{1/2} \left[ {M_{\rm c}/M_{\rm d} \over 3 \times 10^{-5}} \right]^{1/2} \left[ {\ln \Lambda \over 3} \right]^{1/2} \nonumber \\
 & &\times \left[ {\nu \over 90\hbox{ Gyr}^{-1}} \right]^{1/2} \left[ {\alpha_{\rm S}(\beta) \over 0.7} \right]^{3/2} \left[ {K_{\rm S}(\beta) \over 0.15} \right]^{1/2} \left[ {t \over \hbox{Gyr}} \right]^{1/2}.
\end{eqnarray}


\begin{thebibliography}{}
\bibitem[Benson et al. <2000a>]{cluster1}Benson~A.~J., Cole~S., Frenk~C.~S., Baugh~C.~M., Lacey~C.~G., 2000a, MNRAS, 311, 793
\bibitem[Benson et al. <2002a>]{benson02a}Benson~A.~J., Lacey~C.~G., Baugh~C.~M., Cole~S., Frenk~C.~S., 2002a, MNRAS, 333, 156
\bibitem[Benson et al. <2002b>]{benson02b}Benson~A.~J., Frenk~C.~S., Lacey~C.~G., Baugh~C.~M., Cole~S., 2002b, MNRAS, 333, 177
\bibitem[Benson et al. <2002c>]{benson02c}Benson~A.~J., Lacey~C.~G., Baugh~C.~M., Cole~S., Frenk~C.~S., 2002c, MNRAS, 343, 679
\bibitem[Binney <1977>]{binney77}Binney~J., 1977, MNRAS, 181, 735
\bibitem[Binney \& Tremaine <1987>]{bintrem}Binney~J., Tremaine~S., 1987, ``Galactic Dynamics'', Princeton University Press, Princeton
\bibitem[Bizyaev \& Mitronova <2002>]{bizyaev02}Bizyaev~D., Mitronova~S., 2002, A\&A, 389, 795
\bibitem[Bontekoe \& van Albada <1987>]{bont87}Bontekoe~T.~R., van Albada~T.~S., 1987, MNRAS, 224, 349
\bibitem[Bullock, Kravtsov \& Weinberg <2000>]{bullock00}Bullock~J.~S., Kravtsov~A.~V., Weinberg~D.~H., 2000, ApJ, 539, 517
\bibitem[Carlberg \& Sellwood <1985>]{carl85}Carlberg~R.~G., Sellwood~J.~A., 1985, ApJ, 292, 79
\bibitem[Chiba <2002>]{chiba01}Chiba~M., 2002, ApJ, 565, 17
\bibitem[Cole et al. <2000>]{cole00}Cole~S., Lacey~C.~G., Baugh~C.~M., Frenk~C.~S., 2000, MNRAS, 319, 168
\bibitem[Colpi, Mayer \& Governato <1999>]{colpi99}Colpi~M., Mayer~L., Governato~F., 1999, ApJ, 525, 720
\bibitem[Core, Muzzio \& Vergne <1997>]{cora97}Cora~S.~A., Muzzio~J.~C., Vergne~M.~M., 1997, MNRAS, 289, 253
\bibitem[Dalal \& Kochanek <2002a>]{dal01}Dalal~N., Kochanek~C.~S., 2002a, ApJ, 572, 25 
\bibitem[Dalal \& Kochanek <2002b>]{dal02}Dalal~N., Kochanek~C.~S., 2002b, submitted to ApJ (astro-ph/0202290)
\bibitem[de Grijs \& Peletier <1997>]{reynier}de Grijs~R.,
  Peletier~R.~F., 1997, A\&A, 320, L21
\bibitem[Dehnen \& Binney <1998>]{dehnen}Dehnen~W. \& Binney~J., 1998,
  MNRAS, 294, 429.
\bibitem[Donner \& Sundelius <1993>]{donner93}Donner~K.~J., Sundelius~B., 1993, MNRAS, 265, 88
\bibitem[Font et al. <2001>]{font01}Font~A.~S., Navarro~J.~F., Stadel~J., Quinn~T., 2001, ApJ, 563, L1
\bibitem[Fontaine, Brassard \& Bergeron <2001>]{fontaine01}Fontaine~G., Brassard~P., Bergeron~P., 2001, PASP, 113, 409
\bibitem[Ghigna et al. <1998.>]{ghigna98}Ghigna~S., Moore~B., Governato~F., Lake~G., Quinn~T., Stadel~J., 1998, MNRAS, 300, 146
\bibitem[Goldreich \& Tremaine <1979>]{goldtre79}Goldreich~P., Tremaine~S., 1979, ApJ, 233, 857
\bibitem[Granato et al. <2000>]{granato00}Granato~G.~L., Lacey~C.~G.,
  Silva~L., Bressan~A., Baugh~C.~M., Cole~S., Frenk~C.~S., 2000, ApJ,
  542, 710
\bibitem[Hanninen \& Flynn <2002>]{hanninen}Hanninen~J. \& Flynn~C.,
  2002, MNRAS, 337, 731
\bibitem[Hernquist <1993>]{hernquist93}Hernquist~L., 1993, ApJS, 86, 389
\bibitem[Huang \& Carlberg <1997>]{huang97}Huang~S., Carlberg~R.~G., 1997, ApJ, 480, 503
\bibitem[King <1966>]{king66}King~I.~R., 1966, AJ, 71, 64
\bibitem[Klypin et al. <1999>]{klypin99}Klypin~A.~A., Kratsov~A.~V., Valenzuela~O., Prada~F., 1999, ApJ, 522, 82
\bibitem[Lacey <1984>]{lacey84}Lacey~C.~G., 1984, MNRAS, 208, 687
\bibitem[Lewis \& Freeman <1989>]{lewis89}Lewis~J.~R., Freeman~K.~C., 1989, AJ, 97, 139
\bibitem[Mao \& Schneider <1998>]{mao98}Mao~S., Schneider~P., 1998,
MNRAS, 295, 587 
\bibitem[Mendez \& Guzman <1998>]{guzman} Mendez~R.~A.\& Guzman~R., 1998, A\&A,
333, 106
\bibitem[Metcalf \& Madau <2001>]{metmad01}Metcalf~R.~B., Madau~P., 2001, ApJ, 563, 9
\bibitem[Moore <2001>]{moore01}Moore~B., 2001, in J.~C.~Wheeler and H.~Martel eds. ``20$^{\rm th}$ Texas Symposium on Relativistic Astrophysics'', AIP Conference Proceedings, Vol~586, p.~73
\bibitem[Moore etal <1999>]{moore99} Moore~B., Ghigna~S.,
Governato~F., Lake~G., Quinn~T., Stadel~J., Tozzi~P., 1999, ApJ,  
524, L19 
\bibitem[Navarro, Frenk \& White <1994>]{NFW94} Navarro~J.,
Frenk~C.S. \& White~S.D.M., 1994, MNRAS, 267, L1
\bibitem[Siegel et al. <2002>]{siegel02}Siegel~M.~H., Majewski~S.~R., Reid~I.~N., Thompson~I.~B., 2002, ApJ, 578, 151
\bibitem[Somerville <2002>]{somerville02}Somerville~R.~S., 2002, ApJ, 572, 597
\bibitem[Spitzer \& Schwarzchild <1953>]{spitschwar53}Spitzer~L., Schwarzchild~M., 1953, ApJ, 118, 106
\bibitem[Springel, Yoshida \& White <2001>]{springel01}Springel~V., Yoshida~N., White~S.~D.~M., 2001, NewA, 6, 79
\bibitem[Springel et al. <2001>]{springelclus}Springel~V., White~S.~D.~M., Tormen~G., Kauffmann~G., 2001, MNRAS< 328, 726
\bibitem[Taffoni et al. <2003>]{taff03}Taffoni~G., Mayer~L., Colpi~M., Governato~F., 2003, MNRAS, 341, 434
\bibitem[Taylor \& Babul <2001>]{taybab}Taylor~J.~E., Babul~A., 2001, ApJ, 559, 716
\bibitem[Taylor \& Babul <2003>]{taybab03}Taylor~J.~E., Babul~A., 2003, submitted to MNRAS (astro-ph/0301612)
\bibitem[T\'oth \& Ostriker <1992>]{TO}T\'oth~G., Ostriker~J.~P., 1992, ApJ, 389, 5
\bibitem[van den Bosch <1999>]{vdb99}van den Bosch~F.~C., Lewis~G.~F., Lake~G., Stadel~J., 1999, ApJ, 515, 50
\bibitem[Vel\'azquez \& White <1999>]{VW}Vel\'azquez~H.,
  White~S.~D.~M., 1999, MNRAS, 304, 254
\bibitem[Villumsen <1985>]{villumsen}Villumsen~J.~V., 1985, ApJ, 290, 75
\bibitem[Wahde, Donner \& Sundelius <1996>]{wahde96}Wahde~M., Donner~K.~J., Sundelius~B., 1996, MNRAS, 281, 1165
\bibitem[Weinberg <1986>]{weinberg86}Weinberg~M.~D., 1986, ApJ, 300, 93
\bibitem[Weinberg \& Katz <2002>]{weinberg02}Weinberg~M.~D., Katz~N., 2002, ApJ, 580, 627
\end{thebibliography}
\end{document}